\documentclass[reprint,superscriptaddress,aps,prd,nofootinbib]{revtex4-2}

\usepackage{graphicx}
\usepackage{dcolumn}
\usepackage{bm}
\usepackage{amsmath}
\usepackage{graphicx}

\usepackage{amsthm,amssymb}
\usepackage{mathrsfs,subfigure}
\usepackage{color,multirow,hyperref}
\usepackage{verbatim}
\usepackage{booktabs}
\usepackage{diagbox}
\usepackage[normalem]{ulem}

\newcommand \p {\partial}

\allowdisplaybreaks[3]

\def \bal#1\eal  {\begin{align} #1 \end{align}}
\def\({\left(}
\def\){\right)}
\def\[{\left[}
\def\]{\right]}
\def\<{\langle}
\def\>{\rangle}

\newcommand{\bim} {\begin{itemize}[noitemsep]}

\newcommand{\eim}{\end{itemize}}
\newcommand{\be} {\begin{equation}}
\newcommand{\ee} {\end{equation}}
\newcommand{\bc}{\begin{center}}
\newcommand{\ec}{\end{center}}

\newcommand{\gi}{{\gamma}}

\newcommand{\oi}{\omega}

\begin{document}

\hfill {{\footnotesize USTC-ICTS/PCFT-25-11}}

\title{Superradiance of Friedberg-Lee-Sirlin Solitons}

\author{Guo-Dong Zhang}
\email[]{guodongz@mail.ustc.edu.cn}
\affiliation{Interdisciplinary Center for Theoretical Study, University of Science and Technology of China, Hefei, Anhui 230026, China}
\author{Cheng-Hao Li}
\email[]{hzpeterli@mail.ustc.edu.cn}
\affiliation{School of Physical Sciences, University of Science and Technology of China, Hefei, Anhui 230026, China}
\author{Qi-Xin Xie}
\email[]{qixin.xie@nottingham.ac.uk}
\affiliation{School of Physics and Astronomy, University of Nottingham, University Park, Nottingham NG7 2RD, United Kingdom}
\author{Shuang-Yong Zhou}
\email[]{zhoushy@ustc.edu.cn}
\affiliation{Interdisciplinary Center for Theoretical Study, University of Science and Technology of China, Hefei, Anhui 230026, China}
\affiliation{Peng Huanwu Center for Fundamental Theory, Hefei, Anhui 230026, China}

\date{\today}

\begin{abstract}

It has recently been pointed out that rotation in internal space can induce superradiance. We explore this effect in non-topological solitons of the two-field Friedberg-Lee-Sirlin model. This renormalizable model admits very large solitons, making the perturbative scattering equations highly sensitive to boundary conditions and requiring a relaxation method for their solution. We find that the energy extraction rate is strongly influenced by the mass hierarchy of the two scalars, and solitons with lower internal frequencies lead to more peaks in the spectra of the amplification factors. Additionally, we derive absolute bounds on the amplification factors for general ingoing modes using a linear fractional optimization algorithm and establish analytical bounds near the mass gap.

\end{abstract}

\maketitle

\tableofcontents

\section{Introduction}
\label{sec:level1}


Non-topological solitons are stable, localized and time-dependent solutions in certain field theories \cite{Friedberg:1976me,Coleman:1985ki} (see \cite{Zhou:2024mea} and references therein for a recent review). Unlike topological defects, the stability of non-topological solitons does not rely on topological features of the solutions but rather on the presence of conserved Noether charges. The attractive nature of the interactions in these models ensures that the energy of a unit-charge in a soliton is lower than that of a free particle, causing charges to naturally coalesce and form a spherical configuration.

$Q$-balls are one of the simplest kinds of non-topological solitons, typically involving only a complex scalar field\,\footnote{Sometimes the term $Q$-ball is used interchangeably with non-topological soliton.}, in which case they must be sustained by sizable higher-dimensional effective operators in 4D spacetime \cite{Coleman:1985ki}.  Apart from the internal field-space rotation, a $Q$-ball can possess angular momentum in real space, which must be an integer multiples of the $Q$-ball charge $Q$ for the strictly stationary case \cite{Volkov:2002aj,Kleihaus:2005me}, but can take non-integer values for the generic quasi-stable case \cite{Almumin:2023wwi}. Another type of long-lived, composite structures of $Q$-balls, known as charge-swapping $Q$-balls \cite{Copeland:2014qra, Xie:2021glp, Hou:2022jcd, Xie:2023psz}, has also been identified recently, in which positive and negative charges within a ball swap quasi-periodically.  Although $Q$-balls are often justifiably studied in the classical limit, thanks to the large occupation numbers for most of the relevant modes in it, their quantum surface evaporation effects were uncovered with a perturbative approach \cite{Cohen:1986ct} and their quantum dynamics have also been explored with non-perturbative lattice simulations at leading order in the 2PI expansion \cite{Tranberg:2013cka, Xie:2023psz, Kovtun:2020udn}. $Q$-balls also exist when there are multiple fields \cite{Alonso-Izquierdo:2023hrr}. The complex scalar can be coupled to a gauge field as well if the gauge coupling is not too large. In this case, gauge interactions introduce repulsive forces, which can significantly alter the energy and charge distribution within a $Q$-ball, as well as other properties of $Q$-balls \cite{Friedberg:1976az,Friedberg:1976ay,Lee:1988ag,Kusenko:1997vi,Benci:2010cs,Gulamov:2013cra,Gulamov:2015fya,Loiko:2019gwk,Nugaev:2019vru,Heeck:2021zvk,Heeck:2021bce,Kinach:2024qzc}. $Q$-balls naturally arise in supersymmetric extensions of the Standard Model and may play important roles in various cosmological scenarios such as in the Affleck-Dine baryogenesis ({\it e.g.}, \cite{Kusenko:1997si, Enqvist:1997si, Fujii:2002kr, Enqvist:2003gh, Roszkowski:2006kw, Shoemaker:2009kg, Zhou:2015yfa, Kawasaki:2019ywz, Gouttenoire:2021jhk, Kasai:2022vhq, ElBourakadi:2023pue} and see \cite{Zhou:2024mea} for more details). $Q$-balls are closely related to oscillons \cite{Blaschke:2024dlt}, another type of non-topological soliton.


More recently, it is found that the internal field-space rotation of a $Q$-ball can enhance the energy of incident waves, which is referred to as $Q$-ball superradiance \cite{Saffin:2022tub}.
Superradiance was originally coined by Dicke in his work on radiation enhancement in coherent media \cite{Dicke:1954zz}. Many phenomena such as Cherenkov radiation, Mach shocks, and the critical speed for superfluidity can also be viewed as some forms of superradiance \cite{Bekenstein:1998nt}. Due to its relevance to current observations in astroparticle physics (see for example \cite{Brito:2015oca,Teukolsky:1974yv,Cardoso:2004hs,Dolan:2007mj,Arvanitaki:2009fg,Bredberg:2009pv,Arvanitaki:2010sy,Pani:2012vp,Witek:2012tr,Brito:2013wya,Brito:2014wla,Berti:2015itd,Marsh:2015xka,East:2017ovw,Baryakhtar:2017ngi,Baumann:2018vus,Zhu:2020tht,Zhang:2020sjh,Stott:2020gjj,Baryakhtar:2020gao,Mehta:2021pwf,Roy:2021uye,Chen:2022nbb,Siemonsen:2022yyf}), an important type of superradiance that has been under extensive study is Zel’dovich's rotational superradiance, where energy extraction occurs through wave scattering around rotating bodies such as Kerr black holes \cite{Zeld1,Zeld2}. 


For $Q$-ball superradiance, Ref \cite{Saffin:2022tub} analyzed the scattering of perturbative waves off a $Q$-ball and computed amplification factors for energy, angular momentum, and charge, pointing out the particle number conservation in the process, which helps establish the relevant amplification criteria. The distinction between amplification factors for energy/angular momentum and energy/angular momentum flux in such a multi-mode system was subsequently clarified in \cite{Cardoso:2023dtm}. This energy enhancement mechanism arises because the complex scalar contains two degrees of freedom and the $Q$-ball rotation in field space introduces coupled perturbation modes modulated by the $Q$-ball background; A key feature of the scattering process is that the number of particles in the ingoing and outgoing modes remains precisely the same and the energy enhancement comes purely from a redistribution among the different modes. 

The more computationally challenging case of 3+1D spinning 
$Q$-balls was examined in \cite{Zhang:2024ufh}. 
The case with strong gravity, {\it i.e.}, boson star superradiance, was studied in both the Newtonian limit \cite{Cardoso:2023dtm} and the relativistic regime \cite{Gao:2023gof}. The relativistic case was further generalized to include rotational effects in real space \cite{Chang:2024xjp}. Non-perturbative time-domain analyses with Gaussian initial wave packets were also carried out \cite{Cardoso:2023dtm} and extended to incorporate strong gravity effects \cite{Chang:2024xjp}, confirming the validity of perturbative results. 


Rather than relying on high dimensional operators, a renormalizable model to support non-topological solitons can be found by simply introducing an additional real scalar field \cite{Friedberg:1976me}. Historically, the Friedberg-Lee-Sirlin (FLS) solitons were constructed and thoroughly studied much earlier, including its quantum stability, and found their applications in phenomenologically modeling hadrons \cite{Lee:1991ax,Lee:1978yu,Friedberg:1978sc,Goldflam:1981tg,Cahill:1985mh}. Indeed, many properties of FLS solitons closely parallel those of $Q$-balls, while also displaying distinct new features \cite{Loiko:2018mhb,Zhong:2018hwm,Loiko:2022noq,Heeck:2023idx,Kim:2023zvf, Kim:2024vam, Hamada:2024pbs}. In particular, the radial profile of the complex field of a large FLS soliton does not exhibit the same thin-wall limit characteristic of a large $Q$-ball, making the solution extremely sensitive to the boundary conditions near the origin. With gravitational effects included, FLS boson stars have been shown to exhibit a variety of interesting features and dynamics \cite{Kunz:2019sgn,Kunz:2021mbm,Herdeiro:2023lze,Kunz:2023qfg,deSa:2024dhj, Kunz:2024uux, Jaramillo:2024cus}.

 
In this paper, we investigate superradiant amplification of scattering waves off non-topological solitons in the FLS model. We will see that FLS soliton superradiance operates similarly to that of $Q$-balls. However, due to the presence of the extra scalar, the superradiance patterns are more diverse, heavily influenced by the soliton internal rotation frequency, the mass ratio of the two scalar fields and the combinations of the ingoing modes. We will use the relaxation method to obtain both the background soliton solutions as well as solutions of the perturbative scatterings. 
An effective implementation of this method is crucial to achieve reasonable accuracy in the amplification factors for very large solitons, which exist in the FLS model and will induce extreme sensitivity for the perturbative solutions near the center. We explicitly compute a variety of choices of model parameters and combinations of ingoing modes. Additionally, we derive absolute bounds on the amplification factors for arbitrary ingoing modes by leveraging a linear fractional optimization procedure that makes use of particle number conservation. For the amplification factors near the mass gap, we are also able to establish some analytical bounds.

Note that when we were finalizing this work, \cite{Azatov:2024npx} appeared, which contains some overlap with our paper in numerically evaluating perturbative energy extractions from FLS solitons.


The paper is organized as follows. 
In Section~\ref{sec:level2}, we first introduce the FLS model and its formulation with dimensionless variables, which only contains one free parameter---the mass ratio of the two scalar fields. Using the relaxation method (see \cite{NRC,Zhang:2024ufh} for more details) with suitable boundary conditions, we then construct solutions of the FLS solitons for various frequencies and mass ratios. In Section~\ref{sec:level3}, we solve the linear perturbation equations on the soliton background again with the relaxation method, which is efficient in handling the case of large solitons with small $\omega_Q$ and large $\gamma$, and derive the formulas to compute the energy and energy flux amplification factors. We also extract the bounds on possible amplification factors with numerical optimization as well as analytically near the mass gap.
The generic bounds on the amplification factors for the $Q$-ball case with a single complex field are supplemented in Appendix~\ref{sec:levela} for completeness. Section~\ref{sec:level4} presents numerical results for superradiant scatterings of FLS solitons, explores strategies to enhance amplification factors, and investigates the amplification peaks associated with background fields. In Appendix~\ref{sec:levelb}, we display the numerical accuracy for the amplification factors. Finally, we summarize our findings in Section~\ref{sec:level5}.

\section{FLS solitons}
\label{sec:level2}

In this section, we will construct the FLS solitons in $3+1$D, focusing on spherical symmetry and a non-spinning configuration. These solutions are obtained using the relaxation method, and its solution will serve as the background for the perturbative waves to scatter off in the next section.

The FLS model consists of one complex scalar field $\widetilde{\Phi}$ and one real mediator scalar $\widetilde{\chi}$, given by \cite{Friedberg:1976me} \footnote{We use a mostly positive signature for the spacetime metric throughout and the natural units $\hbar=c=1$. } 
\begin{align}
 & \widetilde{\mathcal{L}}  = - \left( \widetilde{\p}_{\mu} \widetilde{\Phi}^{\dagger} \widetilde{\p}^{\mu} \widetilde{\Phi} + \frac{1}{2} \widetilde{\p}_{\mu} \widetilde{\chi} \widetilde{\p}^{\mu} \widetilde{\chi} \right) - \widetilde{U}( |\widetilde{\Phi}| , \widetilde{\chi} ) , \\ 
 & \widetilde{U}( |\widetilde{\Phi}| , \widetilde{\chi} )  = e^2 \widetilde{\chi}^2 \widetilde{\Phi}^{\dagger} \widetilde{\Phi} + \frac{1}{8} g^2 \left( \widetilde{\chi}^2 - \chi_{vac}^2 \right)^2 ,
\end{align}
which has a ${\rm U(1)}$ symmetry under the transformation $\widetilde{\Phi} \to \exp(i \beta) \widetilde{\Phi}$, as well as a discrete $\mathbb{Z}_2$ symmetry $\widetilde{\chi} \to -\widetilde{\chi}$. The parameters are chosen such that $\widetilde{\Phi} = 0$ and $ \widetilde{\chi} = \chi_{vac}$ represent the true vacuum.
Due to the vacuum solution occurring at $\widetilde{\chi}=\chi_{vac}$, it is more convenient to redefine the field by shifting the field $\widetilde{\chi} = \chi_{vac} - \bar{\chi}$. The potential then takes the form: 
\begin{align}
\widetilde{U}( |\widetilde{\Phi}| , \bar{\chi} )  = ~& 
m^2_{\Phi} \widetilde{\Phi}^\dagger \widetilde{\Phi} + \left( - 2 e m_{\Phi} \bar{\chi} + e^2 \bar{\chi}^2  \right) \widetilde{\Phi}^\dagger \widetilde{\Phi}  \notag \\
& + \frac{1}{2} m^2_{\chi} \bar{\chi}^2 - \frac{e m^2_{\chi}}{2 m_{\Phi}} \bar{\chi}^3 + \frac{e^2 m^2_{\chi}}{8 m_{\Phi}^2} \bar{\chi}^4 , 
\end{align}
where $m_{\Phi} = e \chi_{vac}$ and $m_{\chi} = g \chi_{vac}$ are the masses of the fields $\widetilde{\Phi}$ and $\bar{\chi}$ respectively in the true vacuum outside the soliton.
It is also convenient to introduce the following dimensionless variables:
\begin{align}
\widetilde{\Phi} = \frac{m_{\Phi}}{\sqrt{2}e} \Phi, \quad \bar{\chi} = \frac{m_{\Phi}}{e} \chi, \quad \widetilde{x}_{\mu} = \frac{ x_{\mu}}{m_{\Phi}} . 
\end{align}
Additionally, we define the dimensionless parameter 
\be
\gamma = m^2_{\chi} / m^2_{\Phi} ,
\ee
representing the mass ratio squared of the particles $\bar{\chi}$ and $\widetilde{\Phi}$.
With these variables, we can get the equivalent dimensionless Lagrangian $\mathcal{L}$:
\begin{align}
\label{dimlessL}
 \! \mathcal{L} = \frac{\widetilde{\mathcal{L}}e^2}{m^4_{\Phi}}  =  - \frac{1}{2} \left( \p_{\mu} \Phi^{\dagger} \p^{\mu} \Phi +  \p_{\mu} \chi \p^{\mu} \chi \right) - V( |\Phi| , \chi ) ,\! 
\end{align}
where the potential $V(|\Phi|,\chi)$ is defined as:
\begin{align}
 V( |\Phi| , \chi )  = ~& \frac{1}{2} \Phi^\dagger \Phi - \chi \Phi^\dagger \Phi + \frac{1}{2} \chi^2 \Phi^\dagger \Phi  \notag \\
&  + \frac{1}{2} \gamma \chi^2 - \frac{1}{2} \gamma \chi^3 + \frac{1}{8} \gamma \chi^4 . 
\end{align}
We shall focus exclusively on the dimensionless FLS model for the remainder of this work, without any loss of generality at the classical level.

The global $U(1)$ symmetry of the FLS model ensures the conservation of the following charge: 
\begin{align}
 Q = \frac{i}{2} \int {\rm d}^3 {\mathbf x} \left( \Phi^\dagger \dot{\Phi} - \dot{\Phi}^\dagger \Phi \right),
\end{align} 
where the dot denotes the time derivative, $\dot{\Phi}=\p \Phi / \p t$. The energy-momentum tensor for the FLS model is given by:
\begin{align}
T_{\mu \nu} = \frac{1}{2} \left( \p_\mu \Phi^\dagger \p_\nu \Phi + \p_\nu \Phi^\dagger \p_\mu \Phi \right) + \p_\mu \chi \p_\nu \chi + g_{\mu \nu} \mathcal{L} ,  
\end{align}
where $g_{\mu\nu}$ represents the Minkowski metric. Using the dimensionless Lagrangian \eqref{dimlessL}, we derive the field equations:
\begin{align}
\label{equ::1}
\Box \Phi & = 2 \frac{\p V}{\p \Phi^\dagger}=  \Phi - 2 \chi \Phi + \chi^2 \Phi , \\ 
\Box \chi & = \frac{\p V}{\p \chi} = \gamma \chi - \Phi^\dagger \Phi + \chi \Phi^\dagger \Phi - \frac{3}{2} \gamma \chi^2 + \frac{1}{2} \gamma \chi^3 , \label{equ::2}
\end{align}
where $\Box$ is the Minkowski d'Alembertian. 

To find the soliton solutions in the FLS model, we adopt a non-spinning ansatz that satisfies spherical symmetry. The ansatz takes the form: 
\begin{align}
\label{ans_FLS}
\Phi(t,r) = \Phi_Q = f_Q(r) e^{-i \omega_Q t}, \quad \chi(t,r) = \chi_Q (r) , 
\end{align}
where $f_Q$ and $\chi_Q$ are real functions, and $\omega_Q$ is the internal rotation frequency of the field. The existence of the FLS soliton requires the modulus of frequency of the complex field to be smaller than the mass of the complex field but greater than zero, which can be expressed as \cite{Friedberg:1976me}:
\begin{align}
 1 > |\omega_Q| > 0.
\end{align}
Without loss of generality, we focus on the case where $\omega_Q>0$ in this paper. If $\omega_Q<0$, we can perform the transformations $\omega_Q \to -\omega_Q$ and $t \to -t$ to map the negative frequency scenario back to the positive frequency case. Substituting the non-spinning ansatz Eq.~\eqref{ans_FLS} into the equations of motion Eqs.~\eqref{equ::1} and \eqref{equ::2}, we obtain the explicit field equations in $3+1$D spacetime: 
\begin{align}
\label{eom::1}
 \! 0 =& \left( \p_r^2 + \frac{2}{r} \p_r + \omega_Q^2 -1 + 2 \chi_Q - \chi_Q^2 \right) f_Q   ,\! \\
 \! 0 =& \left( \p_r^2 + \frac{2}{r} \p_r + \frac{\gamma}{2} \left( 3\chi_Q - \chi_Q^2 - 2 \right) \right) \chi_Q  \! \notag \\ 
& + f_Q^2 (1-\chi_Q) . 
\label{eom::2}
\end{align} 
These equations describe the radial behavior of the fields $f_Q$ and $\chi_Q$ under the spherically symmetric and non-spinning ansatz.
The resulting system consists of two coupled, nonlinear  differential equations, which are challenging to solve analytically. Therefore, it is necessary to consider numerical methods to obtain solutions. In Section~\ref{sec:level2:2}, we will employ the relaxation method to solve these equations, subject to appropriate boundary conditions.

For given profiles $\Phi_Q$ and $\chi_Q$, the $U(1)$ charge and the energy of the FLS soliton are expressed as:
\begin{align}
Q & = 4 \pi \omega_Q  \int {\rm d}r r^{2} f_Q^2,  \\
E & = 4 \pi \int {\rm d}r r^{2} T_{tt} .
\end{align}
Using the virial theorem \cite{Friedberg:1976me}, we can derive a simplified expression for the energy:
\begin{align}
E = \omega_Q Q + \frac{4 \pi}{3} \int {\rm d}r r^2 \left(  (\nabla f_Q)^2 + (\nabla \chi_Q)^2  \right).
\end{align}

It is crucial to highlight the stability of the FLS solitons. Specifically, if $E<Q$ (with the mass of the field $\Phi_Q$ normalized), the system is in a stable configuration. On the other hand, if $E>Q$, the system corresponds to an unstable case. However, even in the unstable regime, the solitons can persist for a long time at the classical level, provided that the condition ${\rm d}Q/{\rm d}\omega_Q < 0$ holds \cite{Friedberg:1976me}.  

\begin{figure*}
	\centering
		\includegraphics[height=5.3cm]{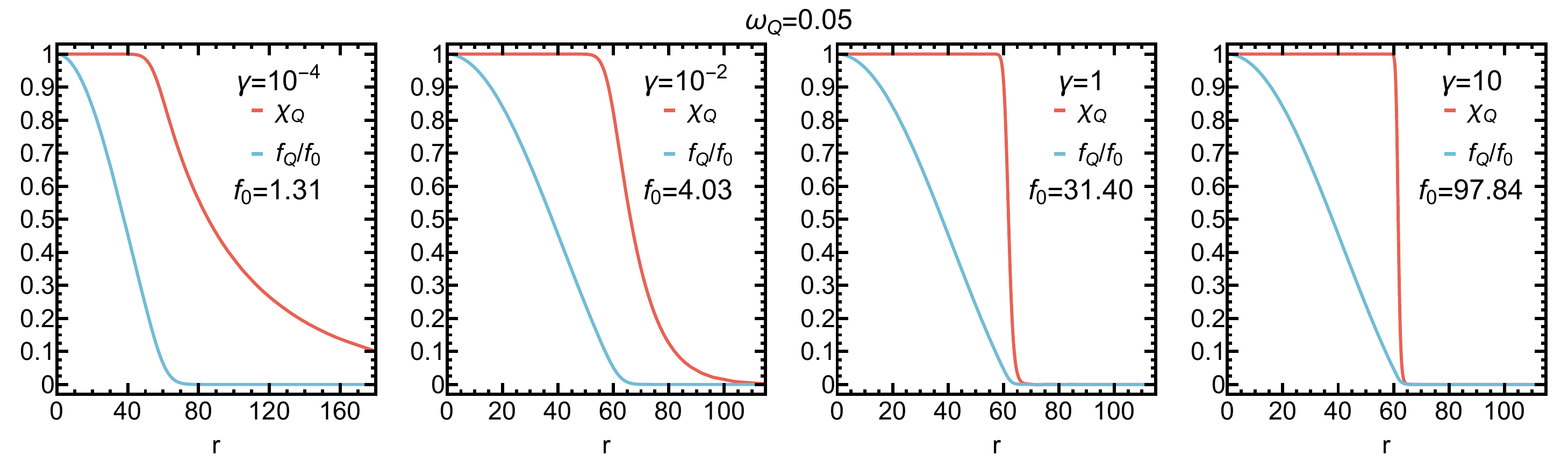}
      \includegraphics[height=5.3cm]{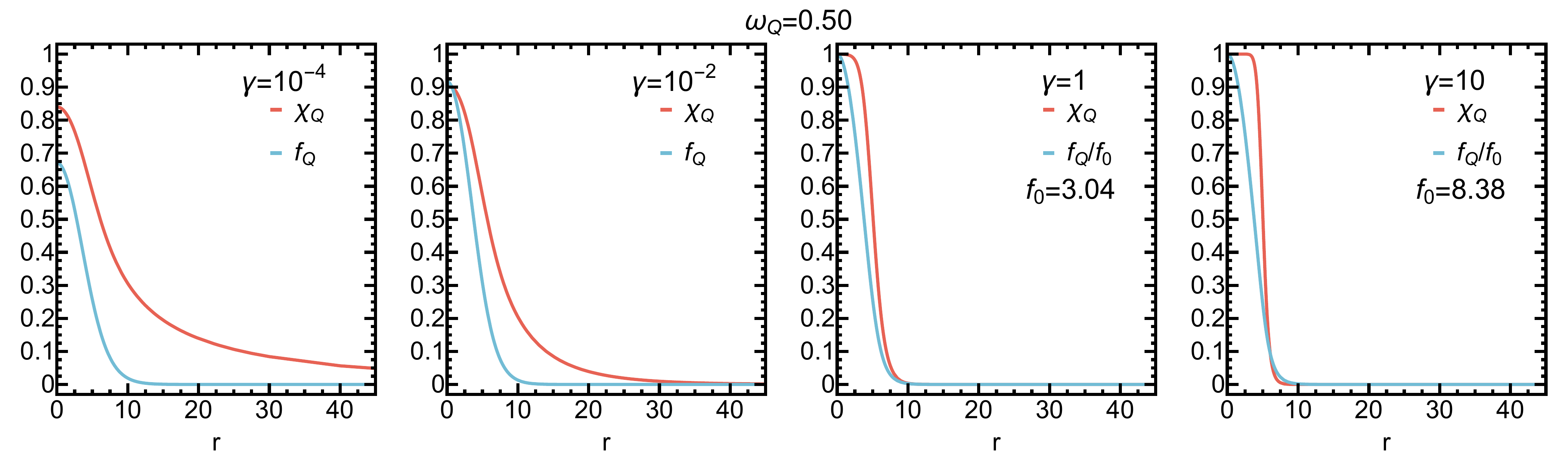}
      \includegraphics[height=5.3cm]{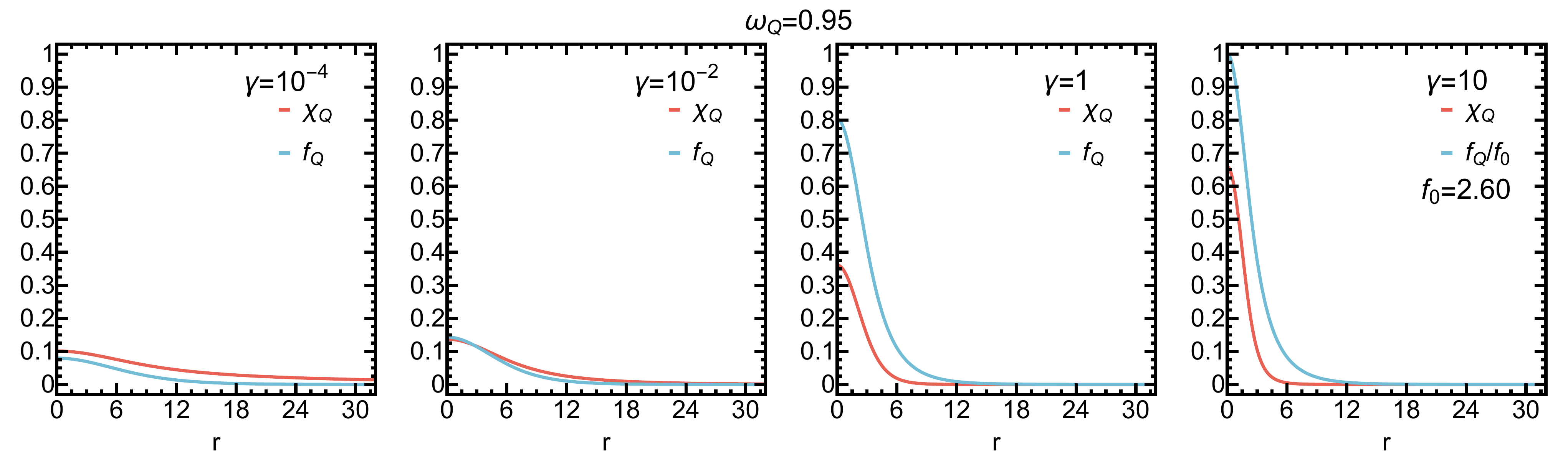}
	\caption{Radial amplitudes $f_Q$ and $\chi_Q$ for the solutions of Eqs.~\eqref{eom::1} and \eqref{eom::2} obtained via the relaxation method. The corresponding values of charge and energy are summarized in Table~\ref{Tab:QE}.   
    }
 \label{fig:FLSmode1}
\end{figure*}

\begin{table*}[ht]
    \centering
    \caption{Charge and energy values corresponding to the solutions in Fig.~\ref{fig:FLSmode1} for different configurations. All solutions are stable, except those with $\omega_Q=0.95$ and $\gamma=1$ or $10$.}
\begin{tabular}{c|c|c|c|c}
\hline \hline
(charge, energy) & $\gamma=10^{-4}$  & $\gamma=10^{-2}$  & $\gamma=1$ & $\gamma=10$ \\
\hline 
\rule{0pt}{10pt}
$\omega_Q=0.05$ & $(1.360\times10^{4},1142)$  & $(1.286\times10^{5},9080)$  & $(7.786\times10^{6},5.191\times10^{5})$  & $(7.559\times10^{7},5.021\times10^{6})$ \\ 
\hline
\rule{0pt}{10pt}
$\omega_Q=0.50$ & $(59.88,42.75)$  & $(91.59,64.00)$  & $(778.3,507.0)$  & $(5727,3651)$ \\ 
\hline
\rule{0pt}{10pt}
$\omega_Q=0.95$ & $(7.818,7.688)$  & $(11.89,11.76)$  & $(76.43,77.34)$  & $(429.7,441.8)$ \\ 
\hline \hline        
\end{tabular}
    \label{Tab:QE}
\end{table*}

\subsection{Boundary conditions}
\label{sec:level2:1}

We now proceed to consider the appropriate boundary conditions for the system.
It is well-established that, outside the soliton, the system naturally tends toward the true vacuum state, where $f_Q=0$ and $\chi_Q=0$. To prevent the divergence of the terms $f_Q'/r$ and $\chi_Q'/r$ as $r\to 0$, the boundary conditions should be given by:
\begin{align}
\p_r f_Q \to 0 ,~ \p_r \chi_Q \to 0, ~~\text{ for }~ r \to 0.
\end{align}
As $r \to \infty$, both $f_Q$ and $\chi_Q$ asymptotically approach zero:
\begin{align}
    f_Q \to 0 ,~ \chi_Q \to 0, ~~\text{ for }~ r \to \infty.
\end{align}
The asymptotic behaviors can be derived from the governing equations of the system, yielding the following expressions \cite{Heeck:2023idx}:
\begin{align}
&\begin{cases}
  f_Q  \to f_0 \frac{\sin(\omega_Q r)}{\omega_Q r} , \\
  \chi_Q \to \chi_0 , 
\end{cases}
\quad r \to 0 ,\\
&\begin{cases}
  f_Q  \to \frac{f_\infty}{r}e^{-\sqrt{1-\omega_Q^2}r} , \\
  \chi_Q \to \frac{\chi_\infty}{r} e^{-\sqrt{\gamma}r} ,
\end{cases}
\quad r \to \infty ,
\end{align}
where $f_0,\chi_0,f_\infty,$ and $\chi_\infty$ are constants that depend on the specific parameters of the system. 

It is evident that four parameters $f_0,\chi_0,f_\infty,$ and $\chi_\infty$ need to be determined. These parameters depend on the system, in other words, on $(\omega_Q,\gamma)$. To eliminate these undetermined parameters, we instead impose the following boundary conditions when we numerically solve Eqs.~\eqref{eom::1} and \eqref{eom::2}:
\begin{align}
\label{bd1}
&\begin{cases}
 f'_Q  = 0 , \\
 \chi_Q' = 0, 
\end{cases}
\quad r \to 0 ,\\
&\begin{cases}
 f_Q' + \left( \frac{1}{r} + \sqrt{1-\omega_Q^2} \right) f_Q  = 0 , \\
 \chi_Q' + \left( \frac{1}{r} + \sqrt{\gamma} \right) \chi_Q = 0 ,
\end{cases}
\quad r \to \infty .
\label{bd2}
\end{align}

\subsection{Numerical results}
\label{sec:level2:2}

Two commonly used methods for solving a system of ODEs such as Eqs.~\eqref{eom::1} and \eqref{eom::2}, along with their boundary conditions, are the high-dimensional shooting method and the relaxation method. In this work, we mainly adopt the relaxation method, with some results also crossed checked with the shooting method. The relaxation method is particularly advantageous for systems that require stringent accuracy. This approach iteratively refines the solution while controlling numerical errors, ensuring that both absolute and relative errors remain below $10^{-10}$. Indeed, as we will see later, it is rather challenging to solve the perturbation equations around large solitons with the shooting method. 

Based on the above discussion, the solutions are shown in Fig.~\ref{fig:FLSmode1}, with the corresponding charge and energy values summarized in Table~\ref{Tab:QE}. Here, $f_0$ and $\chi_0$ are the limiting constants that $f_Q$ and $\chi_Q$ approach as $r\to0$. The figures present various parameter combinations of $\omega_Q$ and $\gamma$, where the top-right corner is close to the large soliton limit, and the bottom-left corner corresponds to the small soliton limit. This layout allows for a comprehensive comparison of the solutions across varying $\omega_Q$ and $\gamma$ values, highlighting the transition between the large and small soliton limits. For low $\omega_Q$ values and heavier $\chi_Q$ (near the top-right corner of the figure), the behavior of the real scalar field resembles a step function, with a larger value of $f_0$. In contrast, the bottom-left corner corresponds to the smaller value of $f_0$. The bottom-right corner, representing $\omega_Q=0.95$ and $\gamma=10$, falls within the unstable region. The adjacent panel to the left, representing $\omega_Q=0.95$ and $\gamma=1$, lies in the metastable region (see \cite{Friedberg:1976me,Heeck:2023idx} for further details and the definition of the metastable state). 

For the background soliton solutions, we can also obtain them with a standard shooting method, even for the very large solitons. However, as we will demonstrate later, when solving the perturbative modes for very large solitons, the shooting method fails while the relaxation method remains effective.

\section{Perturbative scattering}
\label{sec:level3}

In this section, we will investigate the scattering of waves on the background of the FLS solitons, building upon the solutions derived in the previous section. To explore this phenomenon, we employ a linear perturbation expansion to describe the scattering waves. The perturbative fields resulting from this expansion provide essential information regarding the energy and energy flux associated with both incoming and outgoing waves. After that, we examine the enhancements in energy and energy flux that arise from the scattering process.

\subsection{Perturbative waves}
\label{sec:pertwaves}

To study the waves scattering on the background of the FLS solitons, we introduce small linear perturbations, denoted by $\phi$ and $\xi$, on top of the FLS soliton solutions $\Phi_Q $ and $\chi_Q$. Specifically, we express the perturbative fields as:
\begin{align}
\Phi & = \Phi_Q + \phi , \\
\chi & = \chi_Q + \xi , 
\end{align}
where $\Phi_Q$ and $\chi_Q$ are the FLS soliton solutions that satisfy the field equations given by Eq.~\eqref{equ::1} and Eq.~\eqref{equ::2}. 
The linearized equations of motion for the scattering fields $\phi$ and $\xi$, treated as small perturbations, take the following form:
\begin{align}
\label{equ::phi}
\Box \phi & = 2 \left. \left(  \frac{\p^2 V}{\p \Phi^\dagger \p \Phi} \phi +  \frac{\p^2 V}{(\p \Phi^\dagger)^2 }  \phi^\dagger +  \frac{\p^2 V}{\p \Phi^\dagger \p \chi}  \xi \right) \right|_{\Phi_Q,\chi_Q} , \notag \\ 
& = (1 + U_\phi ) \phi + 2 e^{-i\omega_Q t} W \xi , \\  
\Box \xi & = \left. \left( \frac{\p^2 V}{(\p \chi)^2} \xi + \frac{\p^2 V}{\p \chi \p \Phi}  \phi +  \frac{\p^2 V}{\p \chi \p \Phi^\dagger }  \phi^\dagger \right) \right|_{\Phi_Q,\chi_Q}   , \notag \\ 
& =  ( \gamma + U_\xi ) \xi + e^{i\omega_Q t} W \phi + e^{-i\omega_Q t} W \phi^\dagger , 
\label{equ::xi}  
\end{align}
where
\begin{align}
U_\phi & = \chi_Q^2 - 2 \chi_Q , \\
U_\xi & = f_Q^2 + \frac{3}{2} \gamma \chi_Q^2 - 3 \gamma \chi_Q , \\
W & = f_Q (\chi_Q-1). \label{equ::w}
\end{align}
The relative masses of the two fields, $\phi$ and $\xi$, are identified as $1$ and $\gamma$, respectively. The potentials $U_\phi,U_\xi,$ and the background coefficient $W$ depend solely on the background FLS soliton solution. Due to the asymptotic behavior of the FLS soliton fields $f_Q$ and $\chi_Q$, both the potentials and the background coefficient vanish as $r \to \infty$.
  
The perturbative equations of motion are more conveniently solved by performing a Fourier transformation into the frequency domain, ``factoring out'' the time dependence. In this context, we consider non-spinning scattering waves in a spherically symmetric, $3+1$ dimensional spacetime. The ansatzes for the perturbative fields take the following forms: 
\begin{align}
\phi(t,r) & = \eta_{+}(r) e^{ - i \omega_+ t} + \eta_{-}(r) e^{ - i \omega_- t} , \notag \\
& = \left( \eta_{+}(r) e^{ - i \omega t} + \eta_{-}(r) e^{ i \omega t} \right) e^{-i\omega_Q t},
\label{phi::a1} \\
\xi(t,r) & = \rho_{+}(r) e^{ - i \omega t} + \rho_{-}(r) e^{ i \omega t}, \text{ and } \rho_{+} = \rho_{-}^\dagger,
\label{xi::a1}
\end{align}
where $\omega_{\pm} = \omega_Q \pm \omega$. 
Since the background field $\chi_Q$ is real, this property is naturally inherited by its perturbative counterpart, $\xi$. Therefore, the spectral decomposition of the perturbative field imposes a conjugate relationship between $\rho_+$ and $\rho_-$. This constraint effectively reduces the degrees of freedom, allowing only one independent field in the spectral decomposition, as opposed to the two independent fields ($\eta_{\pm}$) that would arise in the case of a complex scalar field. Consequently, in the following analysis, we will focus on a single field for the real field perturbation, with the other component determined by the conjugate relationship.


Substituting the ansatzes Eq.~\eqref{phi::a1} and Eq.~\eqref{xi::a1} into Eq.~\eqref{equ::phi} and Eq.~\eqref{equ::xi}, we have
\begin{align}
\label{equ::ptphi}
\left( \p_r^2 + \frac{2}{r}\p_r + k_{\pm}^2  \right)\eta_{\pm} & = U_\phi\eta_{\pm} + 2 W \rho_{\pm},  \\ 
\left( \p_r^2 + \frac{2}{r}\p_r + k_\xi^2 \right)  \rho_{\pm} & =  U_\xi \rho_{\pm} + W \left( \eta_{\pm} + \eta_{\mp}^\dagger \right),
\label{equ::ptxi}
\end{align}
where the wave numbers $k_{\pm}^2 = \omega_{\pm}^2 - 1$ and $k_\xi^2 = \omega^2 - \gamma$. After applying the transformation $\omega \to -\omega$, we can simply exchange the subscripts  $+ \leftrightarrow -$ to return to the original case. Therefore, without loss of generality, we focus on the case $\omega>0$. For the case $\omega<0$, we can perform the transformation $\omega \to -\omega$, exchange the subscripts of $+ \leftrightarrow -$, and then redefine the $\omega$ to return to the case $\omega>0$. 

To obtain a physically propagating solution for the three fields $\eta_\pm$ and $\rho_+$, we impose conditions on the wave number to guarantee a validity of the solution. Specifically, the conditions are given by:
\begin{align}
|\omega_Q \pm \omega| > 1, \text{ and } |\omega| > \sqrt{\gamma},
\end{align}
such that all the wavenumbers $k_{\pm}$ and $k_{\xi}$ are real, representing propagating waves. It is evident that for a light field $\xi$ (or $\rho_+$), where $\gamma<1$, the second term does not significantly constrain the frequency range of $\omega$. In contrast, for a heavy field $\xi$, where $\gamma>1$, the second term becomes the dominant factor in determining the allowed frequency range for $\omega$. In the extreme case of a very heavy field $\xi$, where $\gamma \gg 1$, the frequency satisfies $|\omega| > \sqrt{\gamma} \gg 1$ and the wave numbers approximate $k_+^2 \sim k_-^2 \sim \omega^2 - 1$. 

We observe that the perturbation preserves both a $U(1)$ symmetry and a scaling symmetry. Consequently, the solutions remain invariant under the following transformation: 
\begin{align}
\left\{\begin{matrix}
(\eta_{+},\rho_{+}) \\
(\eta_{-},\rho_{-})
\end{matrix}\right. \to 
\left\{\begin{array}{l}
\alpha (\eta_{+},\rho_{+}) e^{-i \beta} \\
\alpha (\eta_{-},\rho_{-}) e^{i \beta}
\label{equ::scal}
\end{array}\right.  ,
\end{align}
where $\alpha,\beta \in \mathbb{R}$ are constants. This $U(1)$ symmetry associated with the perturbation allows us to construct the corresponding Lagrangian for the perturbative fields, which is expressed as:
\begin{align}
\label{lag::er}
& \mathcal{L}(\eta_{\pm},\rho_{\pm})  = \sum_{s=\pm} \left( - \eta_{s}^{\dagger} (\nabla^2 + k_s^2) \eta_{s} - \rho_{s}^{\dagger} (\nabla^2 + k_\xi^2) \rho_{s} \right) \notag \\
& + \sum_{s=\pm} \left( U_\phi (\eta_{s}^{\dagger} \eta_s) + U_\xi (\rho_{s}^{\dagger} \rho_s) \right) \notag \\ 
& + W \left( (\rho_+ + \rho_{-}^{\dagger}) (\eta_- + \eta_{+}^{\dagger}) + h.c. \right), 
\end{align}
where $\nabla^2$ represents the Laplace operator, and $h.c.$ denotes the Hermitian conjugate of the previous term. It is the $U(1)$ symmetry of the perturbation that enables the establishment of a conserved particle between the ingoing and outgoing modes, as demonstrated in Section~\ref{sec:ampf}.

\subsection{Boundary conditions}
\label{sec:bdcond}

As discussed in the previous subsection, the perturbative equations of motion, given by Eq.~\eqref{equ::ptphi} and Eq.~\eqref{equ::ptxi}, have been derived.  In this subsection, we will focus on constructing the solutions to these perturbations. This task presents significant challenges for a large FLS soliton, which will be addressed in the subsequent analysis.

When constructing the perturbation solution formulated as a boundary value problem, it is crucial to carefully consider the asymptotic behaviors near the boundaries. To avoid the divergence of the terms $\p_r\eta_{\pm}/r$ and $\p_r\rho_{+}/r$ as $r\to 0$, the derivatives of $\eta_{\pm}$ and $\rho_+$ must either vanish in this limit. Solving the perturbations for a large FLS soliton is challenging. In the following, we will illustrate this in detail.  As $r \to 0$, the perturbation equations for a large FLS soliton reduce to the following asymptotic forms:
\begin{align}
\left( \p_r^2 + \frac{2}{r}\p_r + \omega_{\pm}^2  \right)\eta_{\pm} & = 0 , \\
\left( \p_r^2 + \frac{2}{r}\p_r + \omega^2  \right)  \rho_{+}  & = \omega_0^2 \rho_{+}   ,
\end{align}
where $ \omega_0^2 \equiv { f^2_0 - \frac{1}{2}\gamma }$. For a light field $\rho_+$, where $\gamma<1$, the term $\omega_0^2$ remains relatively small. However, for a heavy real field, where $\gamma \ge 1$, the term $\omega_0^2$ becomes large, resulting in a significant increase in $\rho_+$ as the evolution in the radial direction proceeds. For example, if we neglect the frequency term $\omega^2$,  which is sufficiently small relative to $\omega_0^2$ for low frequencies $\omega$ and large $\gamma$, and also disregard the background coefficient related to $W$--noting that the condition $(\chi_Q - 1) = 0$ still holds in the initial region--we treat the background field $f_Q$ as a step function, defined as:
\begin{align}
f_Q = \left\{\begin{array}{ll}
 \frac{f_0}{2} &  \text{ for } r_0 \ge r \ge 0 \\
 0 &  \text{ for } r>r_0
\end{array}\right. ,
\end{align}
where $r_0$ represents the position where the field $f_Q$ attains half of its maximum value, {\it i.e.}, $ f_Q(r_0) = f_0/2$. Under this assumption, the solution for $\eta_{\pm}$ and $\rho_+$ are given by:
\begin{align}
\eta_{\pm} = F_{\pm} \frac{\sin (\omega_{\pm} r)}{\omega_{\pm} r}, \\
\rho_+ = F_\xi \frac{\sinh (\omega_0 r)}{\omega_0 r},
\end{align}
where $F_{\pm}$ and $F_\xi$ are tunable complex constants, and adjusting them can alter the amplitudes of the ingoing and outgoing modes. A more detailed explanation will be provided later. 

In the case $\gamma=1$ and $\omega_Q=0.05$, we obtain $\omega_0 = 31.38$ and $r_0=37.91$. The solution for $\rho_+$ at the point $ r= r_0$ takes the following form:
\begin{align}
\rho_+(r_0) = F_\xi \frac{\sinh (1189.62)}{1189.62} \approx   1.86 \times 10^{513} F_\xi.
\end{align}
To obtain a physically feasible solution, it is necessary to set $F_\xi \sim 10^{-513}$ at a minimum to counterbalance the extremely large value. When $\gamma=10$ and $\omega_Q=0.05$, the required value for $F_\xi$ becomes $ \sim 10^{-1606}$. This makes it impractical to consider this case using the shooting method. However, the relaxation method can still be applied to solve the system in this scenario. This also suggests that the influence of perturbations in $\xi$ in the initial region, at least for $r\le r_0$, can be largely neglected for a large FLS soliton with a heavy field $\xi$ or $\rho_{+}$ and low frequencies $\omega$. Therefore, as $r\to 0$, to eliminate these adjustable parameters $F_{\pm}$ and $F_\xi$, we impose the following boundary condition:
\begin{align}
\p_r \eta_{\pm} = 0 ,\quad \p_r \rho_{+} = 0 ,\quad  \text{as } r \to 0. 
\label{bc::1}
\end{align}

As $r \to \infty$, the asymptotic equations are obtained:
\begin{align}
\left( \p_r^2 + \frac{2}{r}\p_r + k_{\pm}^2 \right)\eta_{\pm} & = 0 , \\
\left( \p_r^2 + \frac{2}{r}\p_r + k_\xi^2 \right) \rho_{+} & = 0 .
\end{align}
The asymptotic behaviors are:
\begin{align}
\label{equ::asymphi}
\eta_{\pm} & \to \frac{A_{\pm}}{k_\pm r} e^{i k_{\pm} r} + \frac{B_{\pm}}{k_\pm r} e^{-i k_{\pm} r}, \\
\rho_{+} & \to  \frac{C_{+}}{k_\xi r} e^{i k_\xi r}   + \frac{D_{+}}{k_\xi r} e^{-i k_\xi r} ,
\label{equ::asymxi} 
\end{align}
where $A_\pm, B_\pm, C_+,$ and $ D_+$ are complex constants. By substituting the asymptotic behavior into the ansatz equations (Eq.~\eqref{phi::a1} and Eq.~\eqref{xi::a1}), we can distinguish the ingoing and outgoing modes for each case. With a positive frequency, where $\omega>0$, the ingoing modes are $A_-,B_+$ and $D_+$, while the outgoing modes are $A_+,B_-$ and $C_+$. This results in three ingoing and three outgoing modes for the FLS soliton perturbations, which represents a more complex structure than the case of the single-scalar $Q$-ball model, where only two ingoing and two outgoing modes are present \cite{Zhang:2024ufh}. 

If a set of data for $(F_+,F_-,F_\xi)$ is provided, we can evolve Eq.~\eqref{equ::ptphi} and Eq.~\eqref{equ::ptxi} to a large $r$, which will yield a set of complex constants $A_\pm, B_\pm, C_+,$ and $ D_+$ at the large $r$. This system consists of six ODEs and, therefore, requires six boundary conditions to be solved. Three of these boundary conditions are obtained from Eq.~\eqref{bc::1}. Therefore, by providing $(F_+,F_-,F_\xi)$, we can fully determine the solution for the ODEs. However, for large FLS solitons, the value of $F_\xi$ needs to be extremely small to counterbalance the rapid growth of the hyperbolic sine function. Thus, we are left with the option of imposing constraints on $A_\pm, B_\pm, C_+, D_+$ to provide the remaining three boundary conditions. 

The case of a single ingoing mode is a clear scenario worth considering in detail. Since there are three ingoing modes available, we can designate each ingoing mode as the sole ingoing mode while setting the others to zero. These scenarios can be expressed as:
\begin{align}
\left\{\begin{matrix}
 A_-=1, & B_+=D_+=0 , \\
 B_+=1, & A_-=D_+=0 , \\
 D_+=1, & A_-=B_+=0 .
\end{matrix}\right.
\end{align}
The first two cases correspond to a single complex perturbation serving as the ingoing mode, while the third case involves a single real perturbation serving as the ingoing mode. Notably, the non-zero term's value can be treated as an arbitrary constant. However, due to the system's scaling and $U(1)$ symmetry described in Eq.~\eqref{equ::scal}, this constant can be normalized to 1 for simplicity. In case of no ambiguity, we adopt the following convention: if not explicitly stated, only the non-zero incident mode is specified, and any unmentioned modes are considered zero. For example, $A_- = 1$ represents the case where the single ingoing mode for $A_-$ is set to 1, while $B_+$ and $D_+$ are set to 0. 

It is also possible to consider two or all three ingoing modes, allowing two or three parameters $A_-, B_+,$ and $D_+$ to be non-zero. This scenario will be discussed in Section~\ref{sec:level4}.

After that, we need to consider how to incorporate the remaining three boundary conditions into the relaxation method, which only involves the calculation of position points $r$ and the corresponding functions $\eta_{\pm}$ and $\rho_+$.  The asymptotic behaviors are employed to derive the suitable boundary conditions, which are expressed as follows:
\begin{align}
i (r\eta_-' + \eta_-) - k_- r\eta_- &= -2 A_- e^{i k_- r}, \notag \\
i (r\eta_+' + \eta_+) + k_+ r\eta_+ &= 2 B_+ e^{-i k_+ r}, \notag \\ 
i (r\rho_+' + \rho_+) + k_\xi r\rho_+ &= 2 D_+ e^{-i k_\xi r}.
\label{bc::2}
\end{align}

Up to this point, the relaxation method (see \cite{NRC,Zhang:2024ufh} for more details) can be applied using the obtained six boundary conditions, including those given by Eqs.~\eqref{bc::1} and Eqs.~\eqref{bc::2}, to fully solve the system given by Eqs.~\eqref{equ::ptphi} and Eq.~\eqref{equ::ptxi}. The set of equations is solved iteratively on a discretized lattice, and the derivatives are replaced by finite differences. To achieve higher accuracy, additional grids are introduced in each iteration. The absolute and relative errors are maintained below $10^{-10}$, ensuring high accuracy in the solution.

\subsection{Amplification factors}
\label{sec:ampf}

In the previous subsection, we derived the suitable boundary conditions required to solve the system, and the wave amplitudes, denoted as $A_{\pm},B_{\pm},C_+,D_+$, can be determined via the perturbative solutions. These amplitudes contain the physical information of the ingoing and outgoing modes, and they can be used to derive relevant physical quantities, such as energy and energy flux. With these, we can define the amplification factors between the ingoing and outgoing scattering waves.

To this end, note that the system possesses a $U(1)$ symmetry, which is reflected in the ansatz given by Eq.~\eqref{phi::a1} and Eq.~\eqref{xi::a1}.  First, let us define the Noether charge as follows:
\begin{align}
M_{\eta\rho} =  i r^2 \left(  \eta_{+}^{\dagger} \overleftrightarrow{\p_r} \eta_+  -  \eta_{-}^{\dagger} \overleftrightarrow{\p_r} \eta_-  + 2 \rho_{-} \overleftrightarrow{\p_r} \rho_+ \right), 
\end{align}
where $\rho_{-} \overleftrightarrow{\p_r} \rho_+ = \rho_{-} \p_r \rho_+ - \p_r \rho_{-} \rho_+$. This term satisfies $\p_r M_{\eta\rho} =0$, which follows from the equation $0 = \delta (4\pi\int{\rm d}r r^2\mathcal{L})/\delta \beta$, 
where the Lagrangian corresponds to Eq.~\eqref{lag::er} and the U(1) transformation Eq.~\eqref{equ::scal} with the parameter $\beta$, after neglecting the equation of motion. Consequently, $M_{\eta\rho}$ is independent of $r$. Furthermore, the regularities of $\eta_{\pm}$ and $\rho_+$ at $r=0$ ensures $M_{\eta\rho}\equiv0$ which is guaranteed by the boundary conditions Eq.~\eqref{bc::1} \cite{Saffin:2022tub}. This symmetry implies the conservation of the particle number, where the combination of one positive charge and one negative charge gives rise to a single particle number. At large $r$, plugging in the asymptotic behavior Eq.~\eqref{equ::asymphi} and Eq.~\eqref{equ::asymxi}, and integrating over a spherical shell region, the conservation of the particle number between ingoing and outgoing modes is expressed as: 
\begin{align}
\! \frac{|A_-|^2}{k_-} + \frac{|B_+|^2}{k_+} + \frac{ 2 |D_+|^2}{k_\xi} = \frac{|B_-|^2}{k_-} + \frac{ |A_+|^2}{k_+}  + \frac{ 2 |C_+|^2 }{k_\xi} . \!
\end{align}

Since the amplitude of the background FLS soliton solution decays to zero at large $r$, the contributions from the background fields become negligible as $r\to \infty$. Therefore, the energy density of the scattering waves can be approximated by
\begin{align}
 T_{tt} & = \frac{1}{2} \left(\left| \p_t \phi \right|^2 +  \left| \nabla \phi \right|^2 + |\phi|^2  \right) \notag \\
& + \frac{1}{2} \left(  (\p_t \xi)^2 +  ( \nabla \xi )^2 + \gamma \xi^2 \right). 
\end{align}
Considering only the leading order $O(r^{-2})$ terms, the potential term simplifies to the mass term, which yields the following expression:
\begin{align}
V(|\phi|^2, \xi) \sim \frac{1}{2} |\phi|^2 + \frac{1}{2} \gamma \xi^2.
\end{align}
The averaged energy density over a spherical shell region from $r_1$ to $r_2$ as $r_1, r_2 \to \infty$, is given by:
\begin{align}
 E_\circledcirc  = ~& \frac{1}{r_2-r_1} \int_{r_1}^{r_2} {\rm d}r r^2 \left \langle {T_{tt}}  \right \rangle_{T\Omega}, \notag \\
  = ~& \frac{\omega_{+}^2}{k_+^2} \left( |A_+|^2 + |B_+|^2 \right) +  \frac{\omega_{-}^2}{k_-^2} \left( |A_-|^2 + |B_-|^2 \right) \notag \\
& + 2  \frac{\omega^2}{k_\xi^2} \left( |C_+|^2 + |D_+|^2 \right). 
\end{align}
Here the shell region from $r_1$ to $r_2$ includes at least one full spatial oscillation of the longest wavelength. The notation $\langle \cdot \rangle_{T\Omega}$ denotes the average over several temporal oscillations and over the entire 2-sphere. We explicitly define this as:
\begin{align}
\langle f \rangle_{T\Omega} \equiv \frac{1}{4\pi(t_2-t_1)} \int_{t_1}^{t_2} {\rm d}t \int {\rm d}\Omega f(t,r,\theta,\varphi).
\end{align}

On the other hand, the energy flux of the scattering waves is:
\begin{align}
T_{rt} & =  \frac{1}{2} \left( \p_r \phi^\dagger \p_t \phi + \p_t \phi^\dagger \p_r \phi \right) +  \p_r \xi \p_t \xi.
\end{align}
The averaged energy flux over a spherical shell region is:
\begin{align}
P_{rt}  = ~& \frac{-1}{r_2-r_1} \int_{r_1}^{r_2} {\rm d}r r^2 \left \langle {T_{rt}}  \right \rangle_{T\Omega}, \notag \\
= ~& \frac{\omega_+}{k_+} \left( - |A_+|^2 + |B_+|^2 \right) +  \frac{\omega_-}{k_-} \left( - |A_-|^2 + |B_-|^2 \right) \notag \\ 
& + 2 \frac{ \omega}{k_\xi} \left( -|C_+|^2 + |D_+|^2 \right). 
\end{align}
Due to the case $\omega>0$, the ingoing modes are $A_-,B_+,$ and $D_+$, while the outgoing modes are $A_+,B_-,$ and $C_+$. A natural definition of the amplification factors is the ratio of all outgoing particles, weighted by their corresponding physical quantities, to the ingoing particles. Thus, the amplification factors are defined as follows:
\begin{align}
\label{Ampc1}
 \mathcal{A}_{N} & = \frac{ \frac{ |A_+|^2}{k_+} + \frac{|B_-|^2}{k_-} + 2 \frac{ |C_+|^2 }{k_\xi} }{ \frac{|A_-|^2}{k_-} + \frac{|B_+|^2}{k_+} + 2 \frac{ |D_+|^2}{k_\xi} } , \\
\! \mathcal{A}_{tt} & =  \frac{ \frac{\omega_{+}^2}{k_+^2}  |A_+|^2 + \frac{\omega_{-}^2}{k_-^2} |B_-|^2 + 2  \frac{\omega^2}{k_\xi^2} |C_+|^2 }{ \frac{\omega_{-}^2}{k_-^2}  |A_-|^2 +  \frac{\omega_{+}^2}{k_+^2} |B_+|^2 + 2  \frac{\omega^2}{k_\xi^2} |D_+|^2 }  , \! \label{Ampc2} \\ 
\mathcal{A}_{rt} & =  \frac{ \frac{\omega_+}{k_+}  |A_+|^2 + \frac{-\omega_-}{k_-} |B_-|^2 + 2  \frac{\omega}{k_\xi} |C_+|^2 }{  \frac{-\omega_-}{k_-}  |A_-|^2 +  \frac{\omega_+}{k_+} |B_+|^2 + 2  \frac{\omega}{k_\xi} |D_+|^2 }  .
\label{Ampc3}
\end{align}
In other words, $\mathcal{A}_{N}$ is the ratio of outgoing particle number to the ingoing particle number, $\mathcal{A}_{tt}$ is the ratio of energy density, and $\mathcal{A}_{rt}$ is the ratio of energy flux density. Due to the vector nature of the energy current, it is clear that there is a sign difference between the ingoing and outgoing modes. Therefore, a negative sign has been included by default when defining the energy flux amplification factors, meaning that only the magnitude of the energy flow is considered. We can verify that each term defined in this manner is positive for $\omega>0$.

\subsection{Amplification limits}
\label{sec:ampfcons}

In the previous subsection, we derived the formulas of the amplification factors for both energy and energy flux. Exploiting the $U(1)$ symmetry for the system, we can deduce the conservation of particle number, which serves as a constraint. In this subsection, we will see that this constraint allows us to put upper bounds on the amplification factors, which facilitates the refinement of these enhancement factors and provides a framework for verifying the consistency of the numerical results. Note that these bounds are derived by leveraging the asymptotic behavior of the perturbative scattering equations, which do not depend on the specifics of the nonlinear interactions. Thus, these bounds apply to generic two-scalar systems beyond the FLS model.

\subsubsection{Absolute bounds via optimization}
\label{sec:amplim}

If the amplitudes of the ingoing modes ($A_-,B_+$ and $D_+$) are known, the amplification factor for energy can be formulated as a linear fractional optimization (LFO) problem. Below is a description of the LFO problem:

\textbf{Linear Fractional Optimization:}
\begin{align}
\begin{aligned}
\text{Min (or Max)}: \quad & \frac{a^T x + c}{b^T x + d} \\
\text{subject to} \quad & e x \succeq f, \\
& g x = h, 
\end{aligned}
\end{align}
where $x\in\mathbb{R}^n$ is the vector of decision variables, $a,b\in \mathbb{R}^n$ are vectors of constants, and $c,d\in\mathbb{R}$ are scalar constants. The matrices $e\in \mathbb{R}^{p \times n}$ and $g\in \mathbb{R}^{q \times n}$ represent the coefficients of the inequality and equality constraints, respectively, while the vectors $f\in \mathbb{R}^{p}$ and $h\in \mathbb{R}^{q}$ correspond to the constants associated with the inequality and equality constraints, respectively.
Here, the problem involves $p$ inequality constraints and $q$ equality constraints. The problem can be efficiently solved using the \texttt{LinearFractionalOptimization} function in {Mathematica}.

We calculate the limits of the amplification factor for energy within the context of the LFO problem. Specifically, we define
\begin{align}
x &\equiv (|A_-|^2,|B_-|^2,|A_+|^2,|B_+|^2,|C_+|^2,|D_+|^2),\\
a^T &= (0,\frac{\omega_-^2}{k_-^2},\frac{\omega_+^2}{k_+^2},0,2\frac{\omega^2}{k_\xi^2},0), \\
b^T &= (\frac{\omega_-^2}{k_-^2},0,0,\frac{\omega_+^2}{k_+^2},0,2\frac{\omega^2}{k_\xi^2}), \\
g &= (-\frac{1}{k_-},\frac{1}{k_-},\frac{1}{k_+},-\frac{1}{k_+},\frac{2}{k_\xi},-\frac{2}{k_\xi}). 
\end{align}
$e$ is a sixth order identity matrix, and $f$ is a zero vector. This implies that the modulus of the amplitude is non-negative. The parameters $c,d$ and $h$ are zero. This formulation enables us to derive the general limits of the amplification factor for energy under arbitrary ingoing modes.  We can also investigate the limit of the energy flux amplification factor by modifying the vectors $a$ and $b$ as follows:
\begin{align}
a^T &= (0,-\frac{\omega_-}{k_-},\frac{\omega_+}{k_+},0,2\frac{\omega}{k_\xi},0), \\
b^T &= (-\frac{\omega_-}{k_-},0,0,\frac{\omega_+}{k_+},0,2\frac{\omega}{k_\xi}).
\end{align}

Furthermore, by introducing additional constraints on the amplitude—such as considering single ingoing mode, double ingoing mode, or even triple ingoing mode—we can impose more restrictive bounds on the amplification factor, thus refining our analysis for specific physical scenarios.

\begin{figure}
	\centering
		\includegraphics[height=5.0cm]{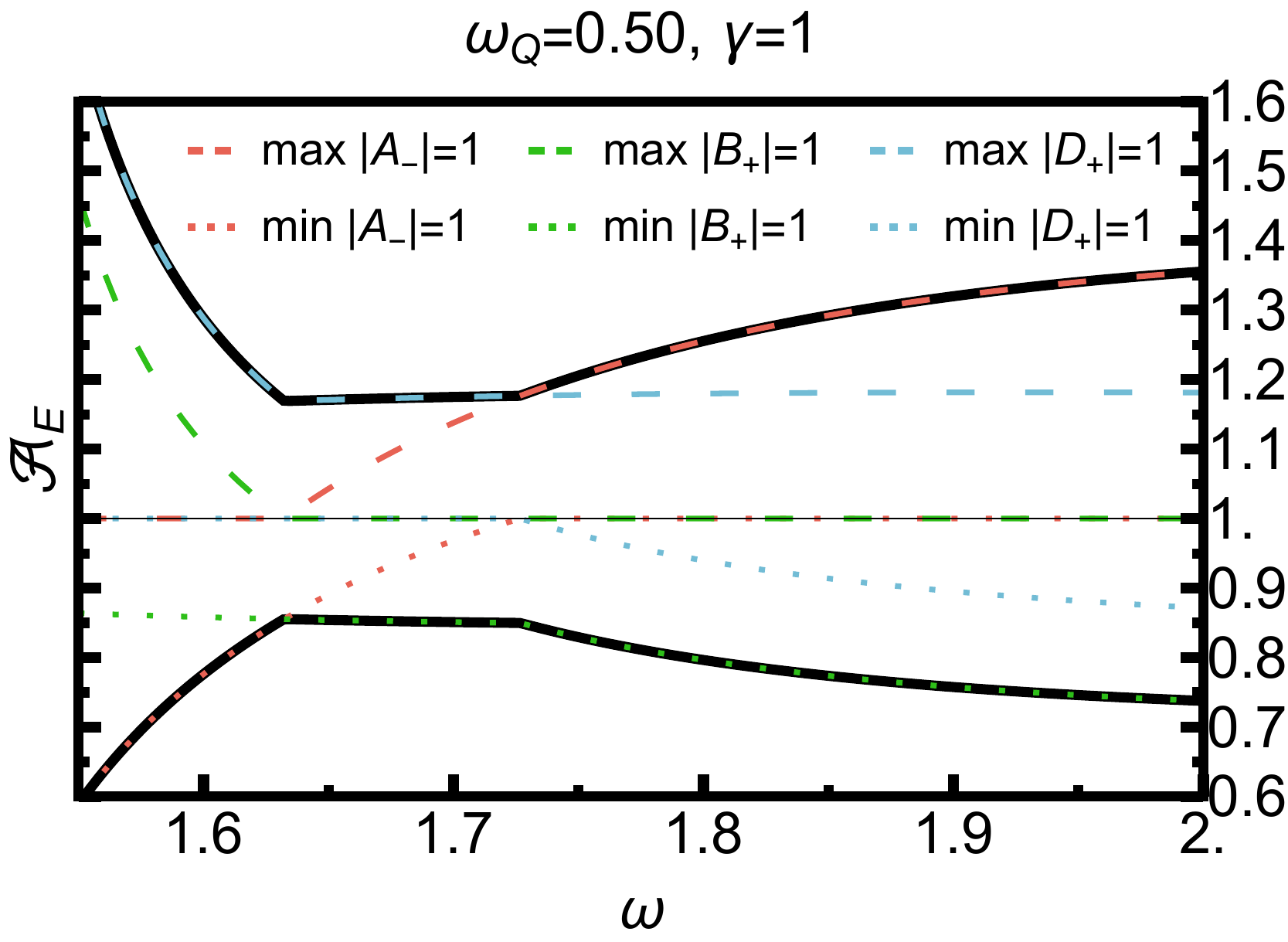}
            \includegraphics[height=5.0cm]{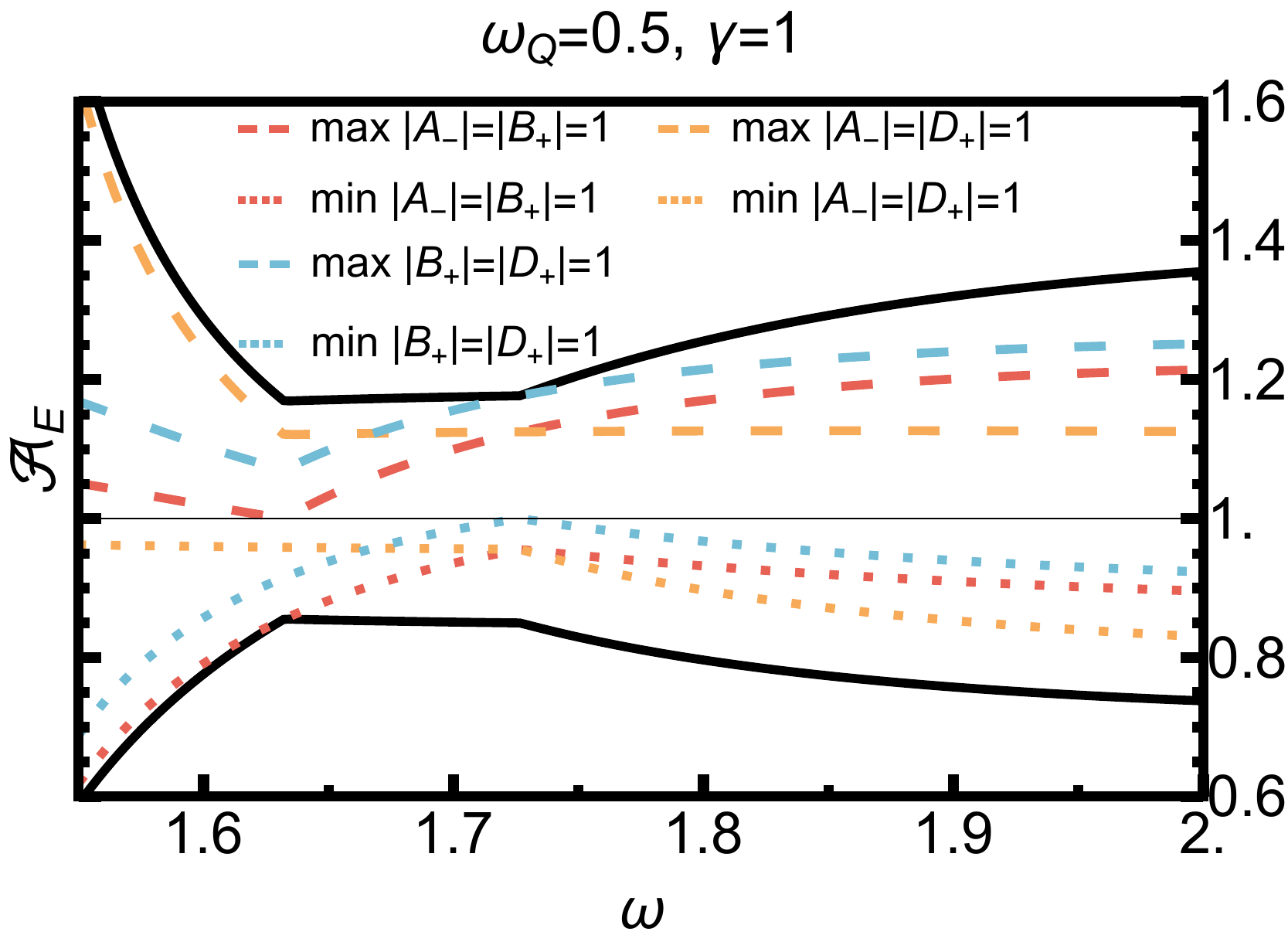}
		\includegraphics[height=5.0cm]{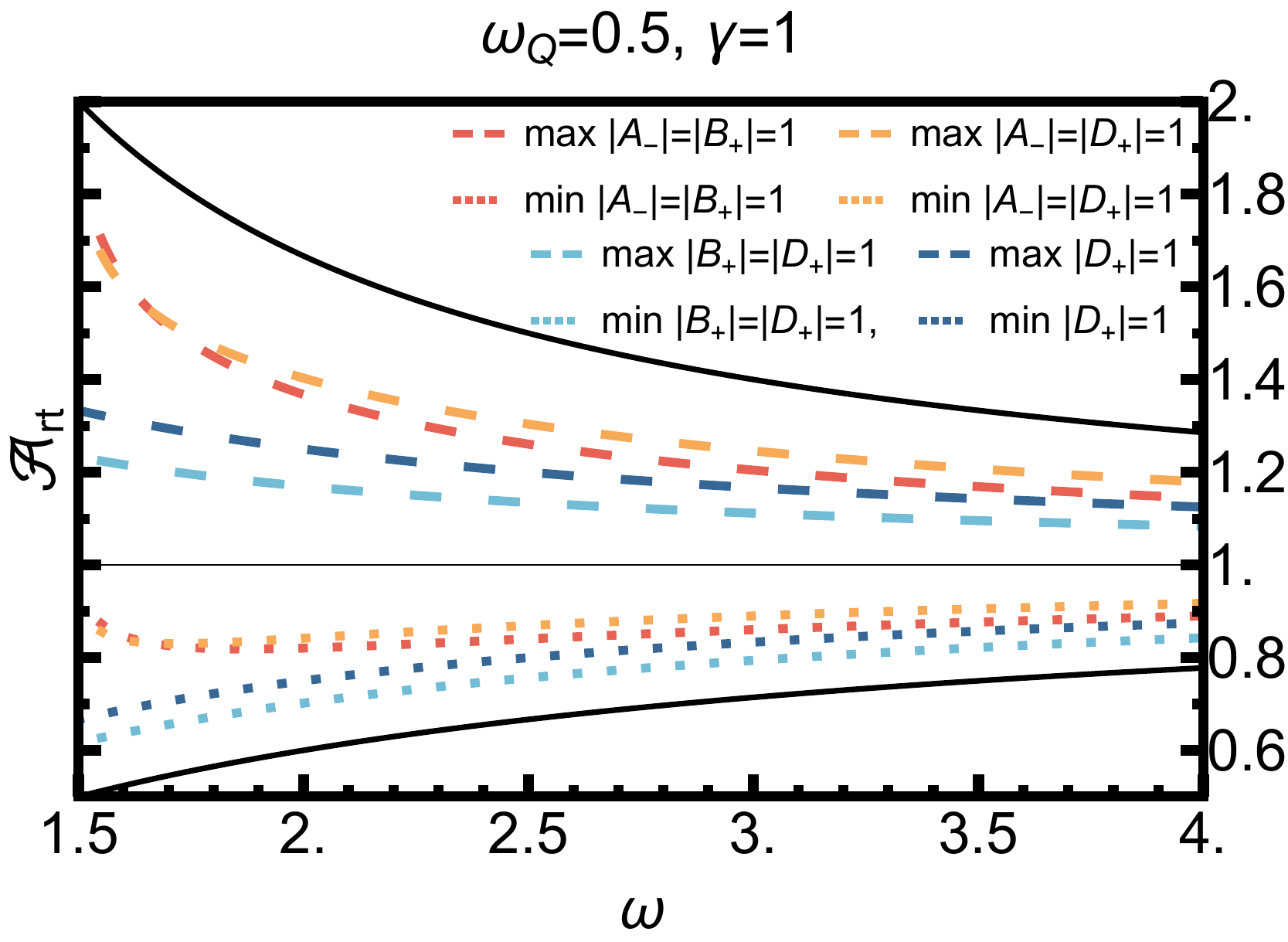}
	\caption{Limits of the energy amplification factors for $\omega_Q=0.5$ and $\gamma=1$. 
    The black lines show the limits of the amplification factors without any amplitude constraints. The top figure shows the bounds on $\mathcal{A}_E$ for the single ingoing mode case, while the middle figure corresponds to the double ingoing mode case. The bottom figure presents the bounds on $\mathcal{A}_{rt}$. 
    }
 \label{fig:aer}
\end{figure}

The results are presented in Fig.~\ref{fig:aer}. In the plot, we choose the parameters: $\omega_Q=0.5$ and $\gamma=1$. The black lines illustrate the limit of the energy amplification factor without any amplitude constraints. The top and middle figures show the bounds in $\mathcal{A}_E$, corresponding to single and double ingoing modes, respectively. The bottom figure presents the bound for $\mathcal{A}_{rt}$. In the bottom figure, the region between the upper black line and $\mathcal{A}_{rt}=1$ corresponds to $|A_-|=1$ for the single ingoing mode, while the region between the lower black line and $\mathcal{A}_{rt}=1$ represents $|B_+|=1$. 
It is evident that the upper and lower bounds of the energy amplification factor, when no amplitude constraints are imposed, are constructed from the bounds for single ingoing modes. Intuitively, the black lines can be viewed as compositions of the colored lines. For the double and triple ingoing modes, the limits lie within the region bounded by the black lines, aligning with our expectations.
We also apply this method to obtain the bounds for the single-scalar $Q$-ball model, as presented in Appendix \ref{sec:levela}.

For clarity, we can interpret these limits from a physical perspective to facilitate understanding. Both ingoing and outgoing modes can be classified into three types based on the frequencies $\omega_+,\omega_-,$ and $\omega$. For each frequency type, we can derive the corresponding particle energy and energy flux, as described by the following equations:
\begin{align}
E_+ &= \frac{\omega_+^2}{k_+},\quad E_- = \frac{\omega_-^2}{k_-},\quad E_\xi = \frac{\omega^2}{k_\xi}, \\
P_+ &= \omega_+,\quad P_- = -\omega_-,\quad P_\xi = \omega.
\end{align}
Next, we define the particle number for each ingoing and outgoing mode corresponding to each frequency:
\begin{align}
N^{in}_+ &= \frac{|B_+|^2}{k_+}, \; 
N^{in}_- = \frac{|A_-|^2}{k_-}, \;
N^{in}_\xi = 2\frac{|D_+|^2}{k_\xi}, \\
N^{out}_+ &= \frac{|A_+|^2}{k_+}, \;
N^{out}_- = \frac{|B_-|^2}{k_-}, \;
N^{out}_\xi = 2\frac{|C_+|^2}{k_\xi}.
\end{align}
The conservation of particle number is expressed as:
\begin{align}
N_c \equiv N^{in}_+ + N^{in}_- + N^{in}_\xi = N^{out}_+ + N^{out}_- + N^{out}_\xi.
\end{align}
The scattering process can be regarded as:
\begin{align}
(N^{in}_+ , N^{in}_- , N^{in}_\xi) \xrightarrow{\text{scattering}}(N^{out}_+ , N^{out}_- , N^{out}_\xi). 
\end{align}
Thus, the scattering process redistributes the particle numbers among the ingoing and outgoing modes. Based on this understanding, it is straightforward to derive the bound of the energy and energy flux amplification factor. Specifically, we consider the single ingoing mode with the lowest energy and the outgoing mode with the highest energy. Due to the conservation of particle number, the bound of the energy amplification factor can be expressed as:
\begin{align}
\max (A_E) = \frac{1}{\min (A_E)}= \frac{\max(E_+,E_-,E_\xi)}{\min(E_+,E_-,E_\xi)}. 
\end{align}
Similarly, the bound for the energy flux amplification factor follows the same pattern:
\begin{align}
 \max (A_{rt}) = \frac{1}{\min (A_{rt})} & = \frac{\max(P_+,P_-,P_\xi)}{\min(P_+,P_-,P_\xi)}, \notag  \\
& = \frac{\omega_+}{-\omega_-}, 
\end{align}
where $\omega_+,-\omega_-,$ and $\omega$ are all greater than zero, due to $\omega>\omega_Q>0$ (the rationale for choosing this case is explained in Section~\ref{sec:pertwaves}). These two bounds can be interpreted as the case without any amplitude constraints. In fact, the bound for the single ingoing mode contributes the bound without any amplitude constraint. This is because the upper (lower) bound for the amplification factor corresponds to the case where the single ingoing mode is in the lowest (highest) particle energy state, while the single outgoing mode is in the highest (lowest) particle energy state. It is evident that the limits on energy flux are exclusively determined by the frequencies of the complex scattering waves. For each single ingoing mode, the limits of the energy flux are given by: 
\begin{align}
\left\{\begin{matrix}
A_{rt} \in (1,\frac{\omega_+}{-\omega_-}),  & \text{ for } |A_-|=1, \\
A_{rt} \in (\frac{-\omega_-}{\omega_+},1),  &  \text{ for } |B_+|=1,\\
A_{rt} \in (\frac{-\omega_-}{\omega},\frac{\omega_+}{\omega}),  &  \text{ for } |D_+|=1.    
\end{matrix}\right.
\label{range::Art}
\end{align} 
These relations indicate that when $|A_-|$ serves as the single ingoing mode, energy flux superradiance must occur. In contrast, when $|B_+|$ is the single ingoing mode, superradiance is not achievable. However, when $|D_+|$ is the single ingoing mode, energy flux can either be amplified above or reduced below $1$. When $\omega \gg \omega_Q $, implying that $\omega_Q$ can be neglected, the limits on the energy flux approach $1$. For a very heavy real scalar field, when all perturbation modes correspond to propagating solutions ($\omega>\sqrt{\gamma}\gg 1$), this condition is naturally satisfied, leading to an energy flux amplification close to $1$. Therefore, we do not consider the case where $\gamma\gg1$.

If the number of ingoing (outgoing) modes for each frequency is fixed, corresponding to the amplitude constraints, the upper and lower bounds of the energy amplification factor can also be determined. The upper bound is obtained by adjusting the outgoing (ingoing) mode with the highest (lowest) single-particle energy, while the lower bound is determined by adjusting the outgoing (ingoing) mode with the lowest (highest) single-particle energy. A similar procedure applies to determining the bounds for the energy flux amplification factor in this case.

From the results of the calculation, we observe that a higher value of $\omega_Q$ leads to an increase in the upper bounds of both the energy and energy flux amplification factors. However, this does not necessarily imply that the amplification factors themselves are greater in the scattering case. Specifically, when $\omega_Q$
approaches its upper limit of 1, the numerical results for the perturbation equations indicate that the energy and energy flux amplification factors approach to $1$. This will be further discussed in Section~\ref{sec:level4}, where we will explain the underlying reasons. Nonetheless, it is important to note that increasing $\omega_Q$ appropriately will yield larger upper bounds for the amplification factors.

\begin{figure*}
	\centering
		\includegraphics[height=3.8cm]{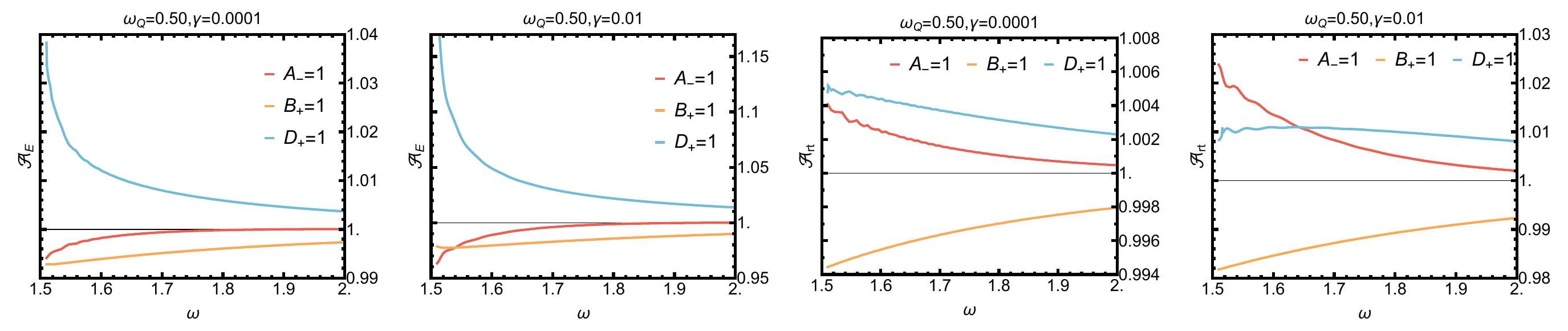}
          \includegraphics[height=3.8cm]{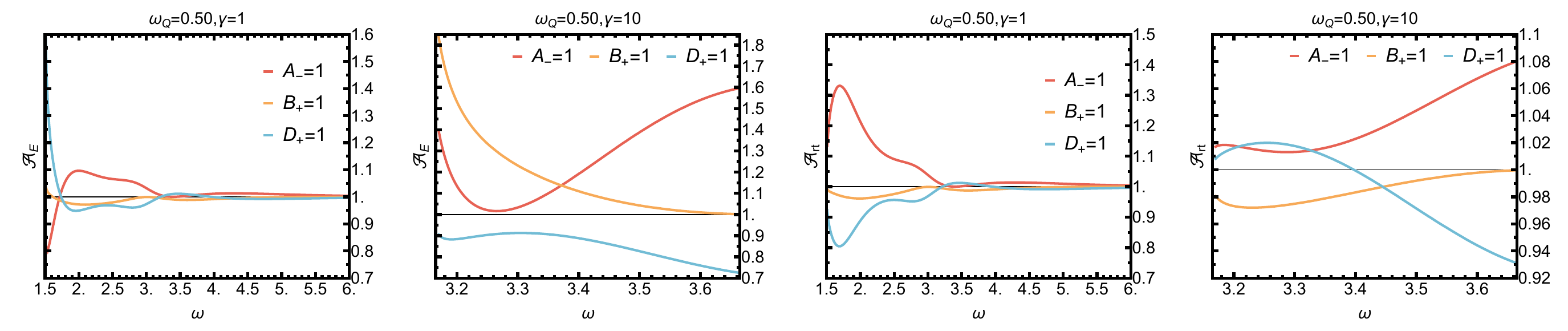}
		\includegraphics[height=3.9cm]{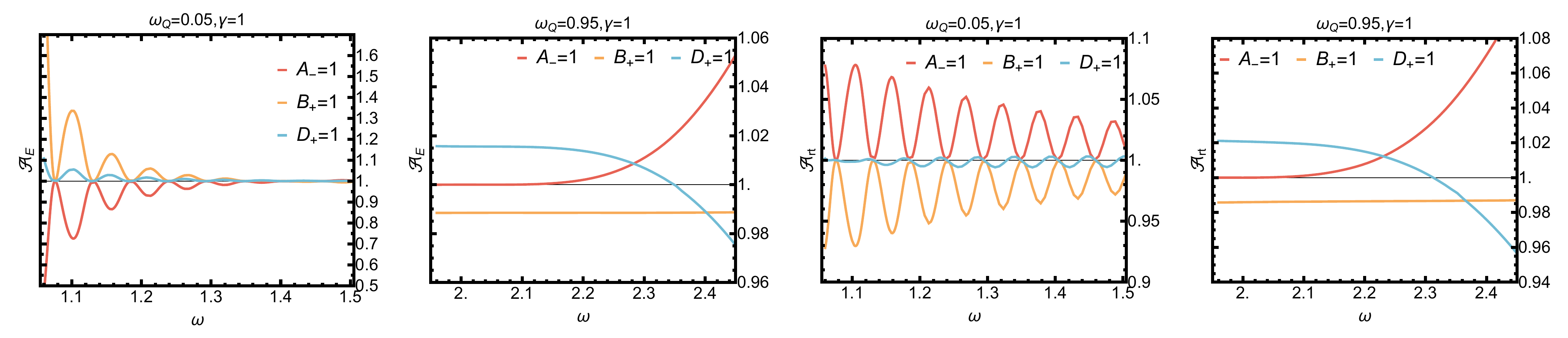}
	\caption{Spectra of the energy amplification factor $\mathcal{A}_E$ and the energy flux amplification factor $\mathcal{A}_{rt}$ for various $\oi_Q$ and $\gi$ for the case of a single ingoing mode. For example, in the top left subplot, $A_-=1$ means that only $A_-$ is nonzero and the amplitudes of all the other ingoing modes ($B_+$ and $D_+$) vanish. }
 \label{fig:sging}
\end{figure*}

\begin{figure*}
	\centering
		\includegraphics[height=4.55cm]{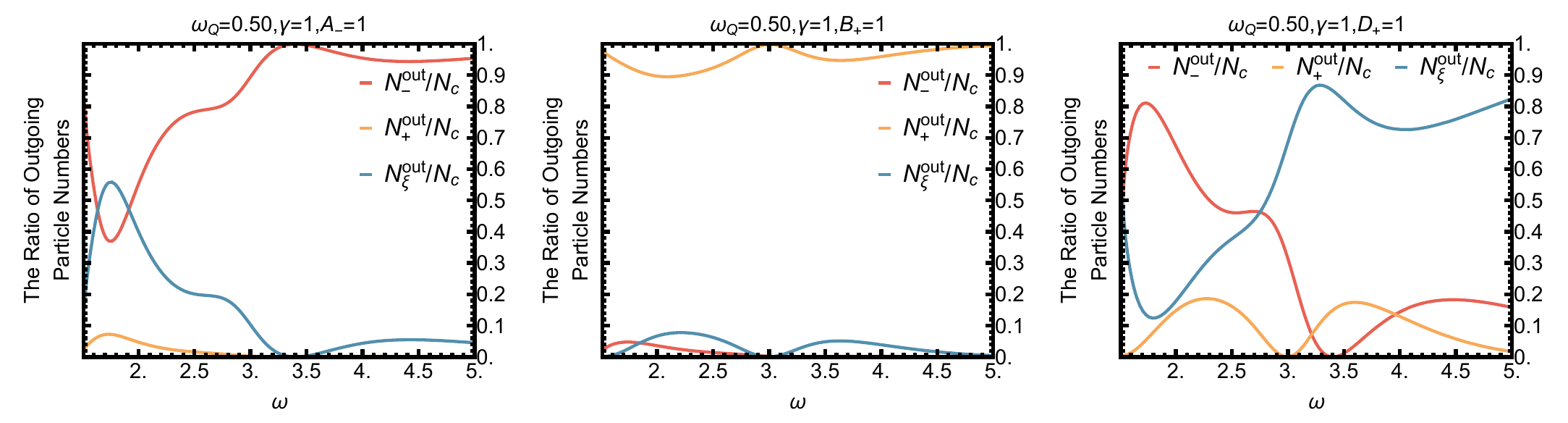}
	\caption{Ratios of outgoing particle numbers to the total particle number for the case of a single ingoing mode. 
}
 \label{fig:rat}
\end{figure*}

\subsubsection{Bounds near the mass gap}
\label{sec:asymp}

We now turn to the asymptotic behavior of the amplification factors and the distribution of particles as the frequency $\omega$ approaches the mass gap or the high-frequency limit $\omega \to \infty$. 

We shall first establish the condition for the frequency $\omega$ to approach the mass gap. Depending on the mass ratio, there are three distinct cases: $\gamma=(1+\omega_Q)^2$, $\gamma<(1+\omega_Q)^2$ and $\gamma>(1+\omega_Q)^2$.

When $\gamma=(1+\omega_Q)^2$, the selected frequency range is given by $\omega \in (1+\omega_Q,\infty)$, where $\omega>0$ (the rationale for choosing this case is explained in Section~\ref{sec:pertwaves}). To facilitate analysis, we express the frequency as $\omega=1+\omega_Q+\epsilon$, where $0<\epsilon \ll 1$. Under this approximation, the relevant quantities can be written as: 
\begin{align}
\! \omega_+ &= 1+2\omega_Q+\epsilon, \quad -\omega_- = 1+\epsilon, \!  \notag \\
\! k_+ &= 2\sqrt{\omega_Q+\omega_Q^2}, ~ k_-=\sqrt{2\epsilon},~k_\xi = \sqrt{2(1+\omega_Q)\epsilon}, \!
\end{align}
where higher-order terms in $\epsilon$ have been neglected. It is obvious that both the particle numbers and energies are closely associated with the wave numbers, which may lead to divergent results in this limits. To properly understand the asymptotic behavior of the amplification factors, it is essential to separately consider the contributions from particle numbers and particle energies. 

To avoid divergence in the particle number, it is essential to impose appropriate constraints on the amplitude. These constraints can be expressed as:
\begin{align}
|A_-|^2, |D_+|^2, |B_-|^2, \text{ and } |C_+|^2  \propto \epsilon^n , \text{ for } n \ge \frac{1}{2}.
\end{align}
If we select the single ingoing mode, specifically  $|A_-|$ or $|D_+|$, we can scale the solutions $(\eta_{\pm},\xi)$ by a factor of $\epsilon^{-n}$. This scaling adjustment ensures that the conditions $|A_-|=1$ or $|D_+|=1$, are satisfied. Additionally, using the  $U(1)$ symmetry of the solutions,  we can normalize the amplitude factor of the single incoming mode to 1, {\it i.e.}, $A_-=1$ or $D_+=1$. For double and triple incoming modes, this method usually can only be used to normalize one of the ingoing amplitudes to 1.

In this limit, the particle energies $E_-$ and $E_\xi$ exhibit a tendency to diverge as the frequency $\omega$ approaches the mass gap. However, the energy fluxes, which are not directly determined by the wave numbers, do not exhibit such divergent behavior. As a result, the non-divergence of the particle number ensures that the energy flux amplification factors provide a well-defined and finite bound in this asymptotic regime.

For each single ingoing mode, if the parameter satisfies $n > 1$, the amplitude terms are sufficient to counterbalance the divergence of the particle energy. Consequently, the energy amplification factor remains finite and does not exhibit divergence. For $n>1$, the particle number and energy for $\eta_-$ and $\xi$ tend to zero, leaving only $\eta_+$, so the energy and energy flux amplification factors tend to $1$. However, when $1 \ge n \ge 1/2$, the behavior becomes more intricate. Below, we outline different scenarios corresponding to distinct single ingoing modes (with all unmentioned ingoing modes set to zero in each case):
\begin{itemize}
\item {\it $A_-=1$}: The energy amplification factor is bounded as:
\begin{align}
\lim_{\epsilon \to 0} \mathcal{A}_E \in (0,(1+\omega_Q)^{3/2}).
\end{align} 
Although both the energies $E_-$ and $E_\xi$  exhibit divergent behavior, their ratio remains constant:
\begin{align}
E_\xi/E_-=(1+\omega_Q)^{3/2}.
\end{align}
This ratio represents the upper bound of the energy amplification factor, corresponding to the scenario where there is only one ingoing mode associated with $A_-$ and one outgoing mode associated with $C_+$. On the other hand, the energy $E_+$ for ougoing mode associated with $A_+$, contributes a finite value, ensuring that the lower bound of the energy amplification factor is $0$.

\item {\it $B_+=1$}: The energy amplification factor satisfies:
\begin{align}
\lim_{\epsilon \to 0} \mathcal{A}_E \in (1,\infty).
\end{align} 
The upper bound arises from the divergence of the outgoing particle energies $E_-$ and $E_\xi$, while the lower bound corresponds to a scenario where no transformation into other particles occurs..

\item {\it $D_+=1$}: The energy amplification factor is bounded by:
\begin{align}
\lim_{\epsilon \to 0} \mathcal{A}_E \in (0,(1+\omega_Q)^{-3/2}).
\end{align} 
Similar to the first scenario, although the outgoing particle energies exhibit divergence, their ratio remains constant, determining the range of the energy amplification factor.
\end{itemize}

When $\gamma<(1+\omega_Q)^2$, the allowed frequency range is $(1+\omega_Q,\infty)$, with divergences occurring only at $k_-$ and $E_-$. For $n>1$, the previous conclusions still hold. When $1 \ge n \ge 1/2$, the results are summarized as follows:
\begin{itemize}
\item {\it $A_-=1$}: $ \lim_{\epsilon \to 0} \mathcal{A}_E \in (0,1)$. 

\item {\it $B_+=1$}: $\lim_{\epsilon \to 0} \mathcal{A}_E \in (\min(1,E_\xi/E_+),\infty)$.

\item {\it $D_+=1$}: $\lim_{\epsilon \to 0} \mathcal{A}_E \in (\min(1,E_+/E_\xi),\infty)$.
\end{itemize}

When $\gamma>(1+\omega_Q)^2$, the allowed frequency range is $(\sqrt{\gamma},\infty)$, with divergences occurring only at $k_\xi$ and $E_\xi$. For $n>1$, the previous results are still applicable. When $1 \ge n \ge 1/2$, the conclusions are as follows:
\begin{itemize}
\item {\it $A_-=1$}: $\lim_{\epsilon \to 0} \mathcal{A}_E \in (\min(1,E_+/E_-),\infty)$. 

\item {\it $B_+=1$}: $\lim_{\epsilon \to 0} \mathcal{A}_E \in (\min(1,E_-/E_+),\infty)$.

\item {\it $D_+=1$}: $\lim_{\epsilon \to 0} \mathcal{A}_E \in (0,1)$.
\end{itemize}

These results highlight the dependence of the allowed frequency range on the mass ratio $\gamma$, and the associated divergences at specific energy and wave number values. The behavior is consistent across different values of $n$, which characterizes the asymptotic amplitude behavior near the mass gap to avoid particle number divergence for $n\ge1/2$. Special attention is given to the boundary cases where $1 \ge n \ge 1/2$.

In the high-frequency limit ($\omega \to \infty$), it becomes evident that the frequencies and wave numbers of the scattering wave exhibit similar asymptotic behavior. Specifically, $\omega_+,-\omega_-,k_+,k_-,$ and $k_\xi$ all converge to $\omega$ as $\omega \to \infty$. Consequently, the energy and energy flux carried by the scattering wave also approach similarity. This convergence implies that the energy and energy flux amplification factors asymptotically approach a value of $1$. Therefore, to achieve a large energy flux amplification factor, it is not necessary to require a large $\gamma$,  which corresponds to a heavier field $\rho_+$.

\section{Numerical results}
\label{sec:level4}

In this section, we will present the numerical results for waves scattering around the FLS soliton, focusing on the energy and energy flux enhancements during the process, key features of the superradiance phenomenon, and methods for achieving higher amplification factors. As previously mentioned, we will employ the relaxation method to solve the perturbative equations. The shooting method can be used to compute the perturbation equations around small solitons, so  the cases of $\omega_Q = 0.05, \gamma = 10^{-4}$ and $\omega_Q = 0.50, \gamma = 10^{-2}$ are solved with the shooting method, which serves as a consistency check.

As discussed at the end of Section~\ref{sec:bdcond}, the system contains three ingoing modes, expressed as $A_-,B_+,D_+$ for the $\omega>0$ case (the rationale for choosing this case being explained in Section~\ref{sec:pertwaves}). This choice ensures that we can systematically determine the amplification factors for each single, double, or triple ingoing modes configuration. In the following, we will analyze each case individually and further explore the strategies for enhancing the amplification factors.

\begin{figure*}
	\centering
		\includegraphics[height=4.29cm]{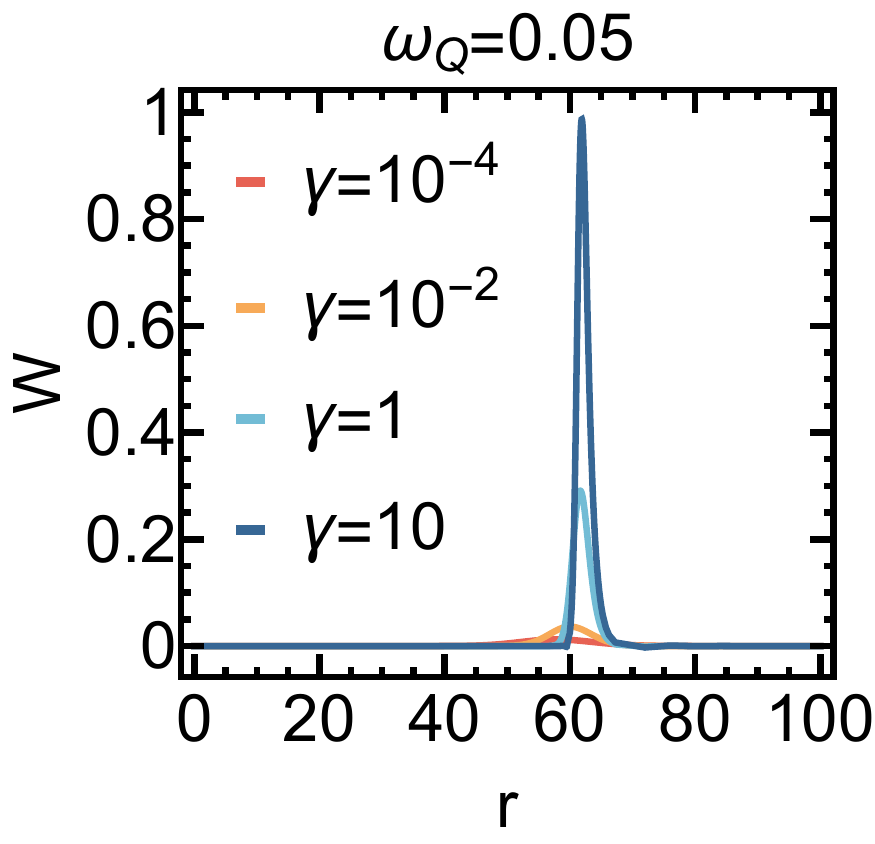}
            \includegraphics[height=4.2cm]{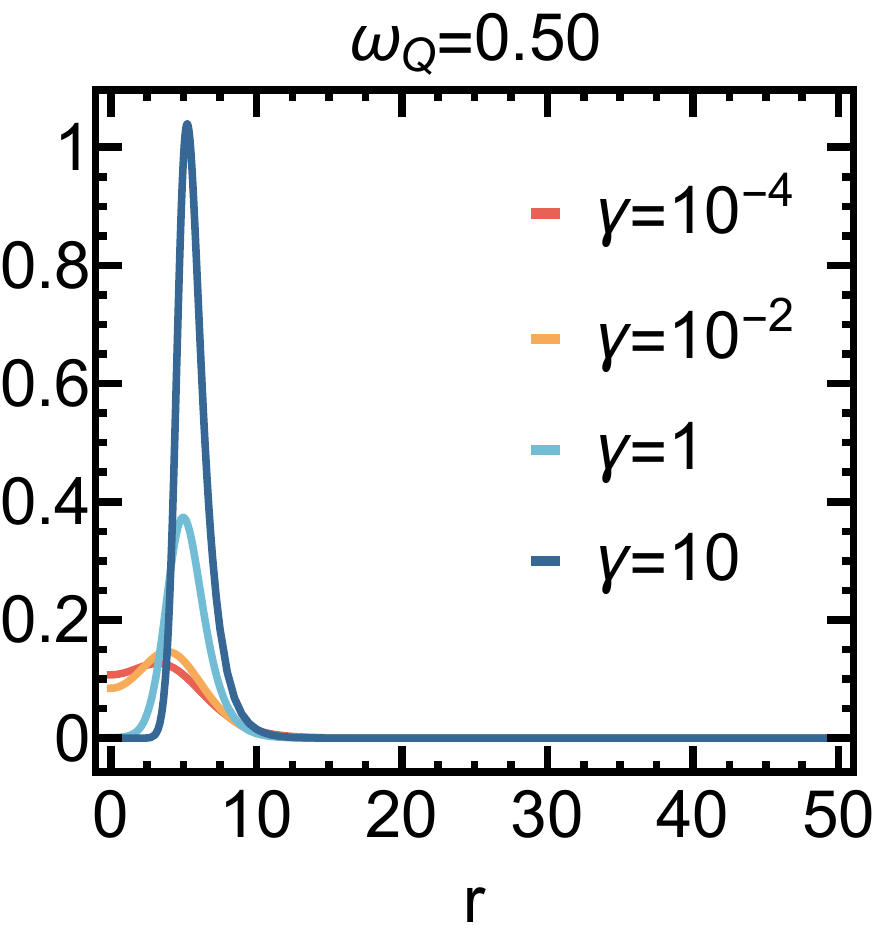}
		\includegraphics[height=4.2cm]{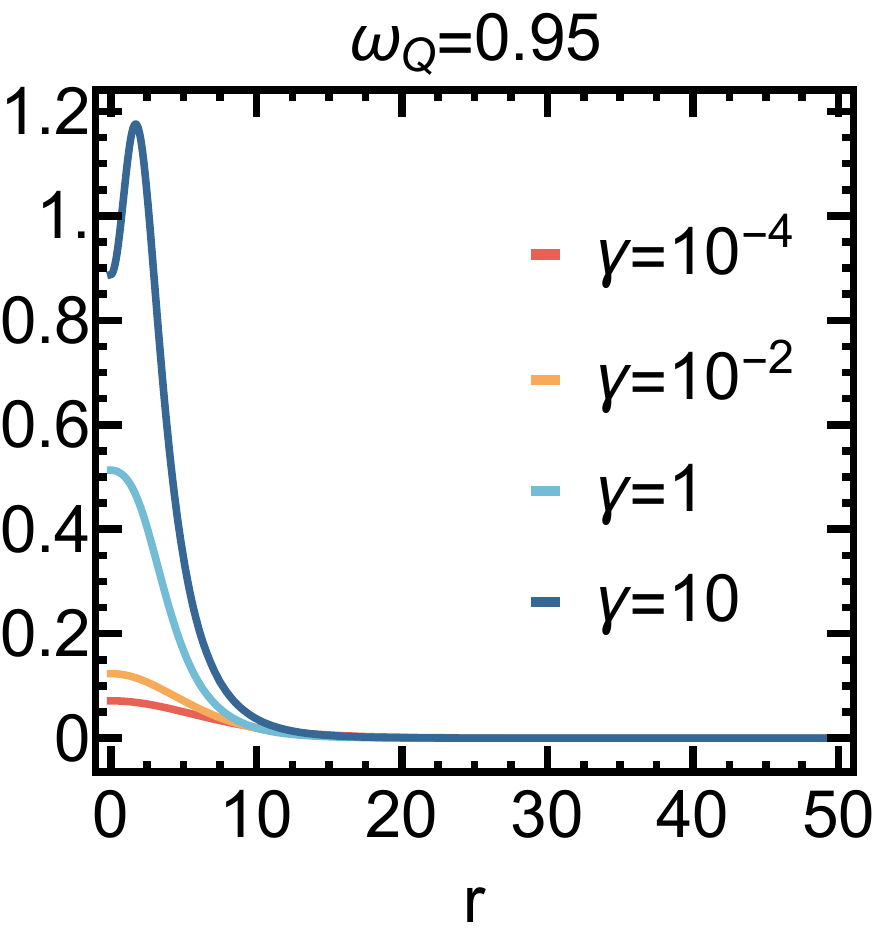}
	\caption{Background coefficient $W$ for different $\omega_Q$ and $\gamma$. 
}
 \label{fig:W}
\end{figure*}

\begin{figure}
	\centering
		\includegraphics[height=3.7cm]{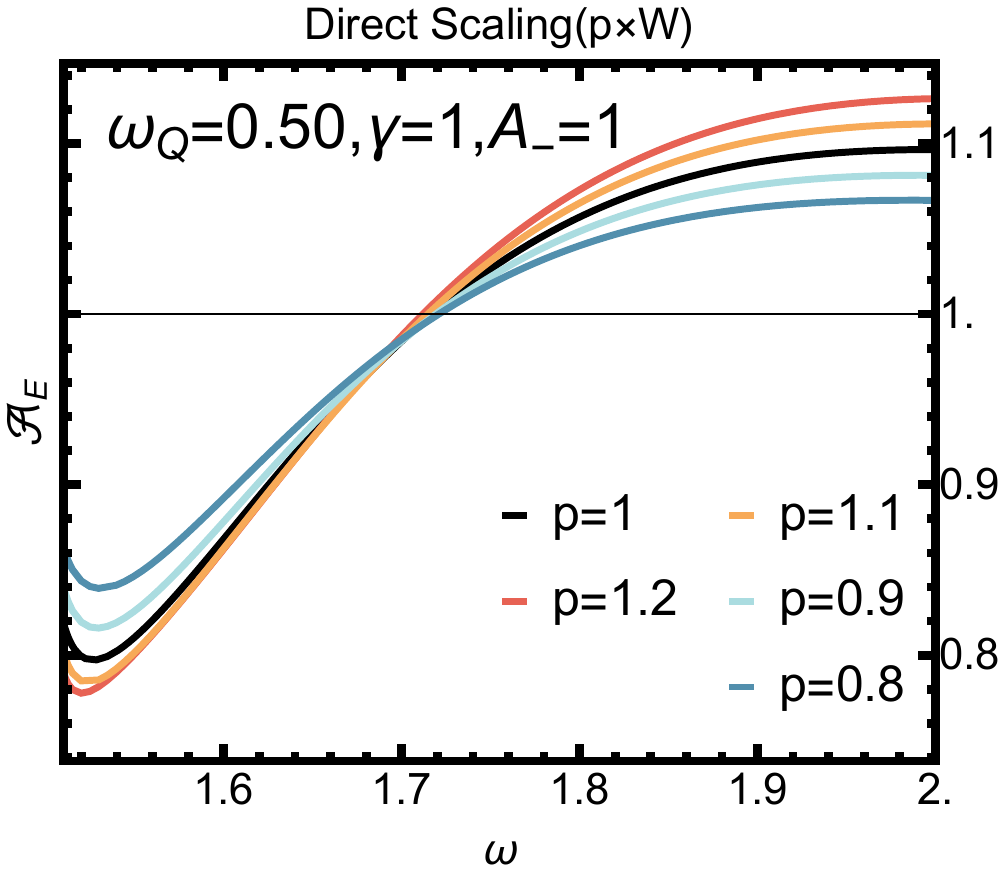}
     \includegraphics[height=3.7cm]{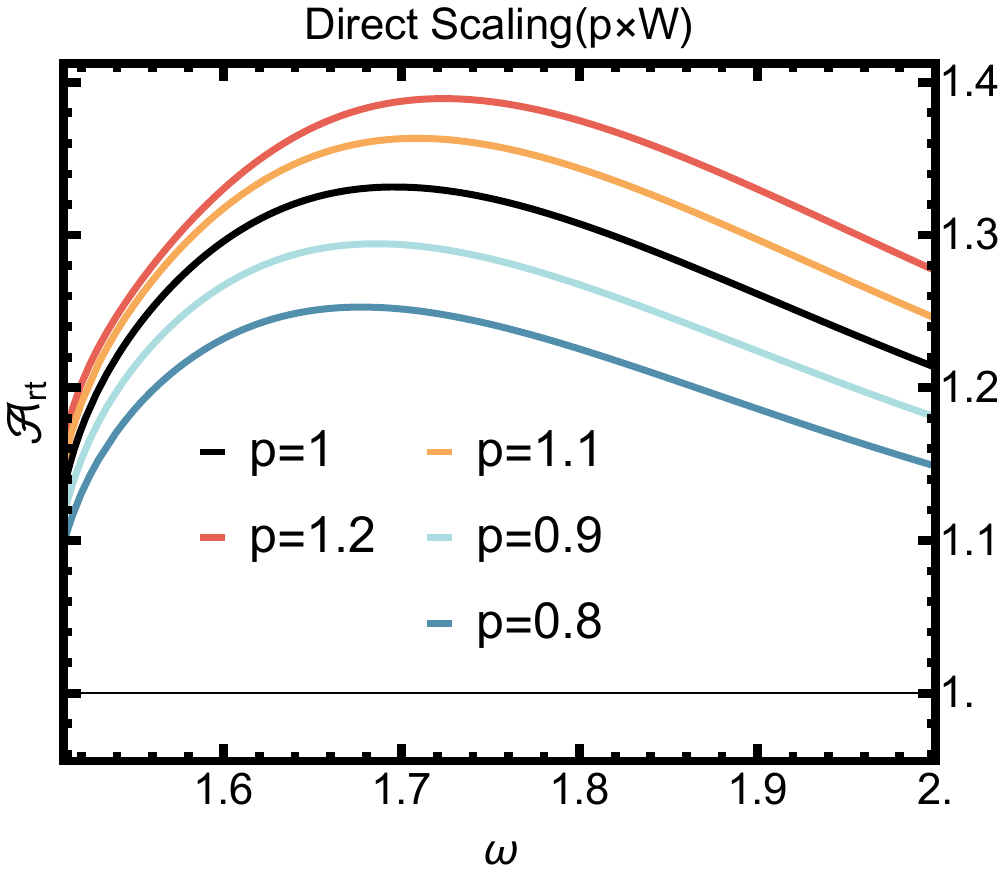}
		\includegraphics[height=3.7cm]{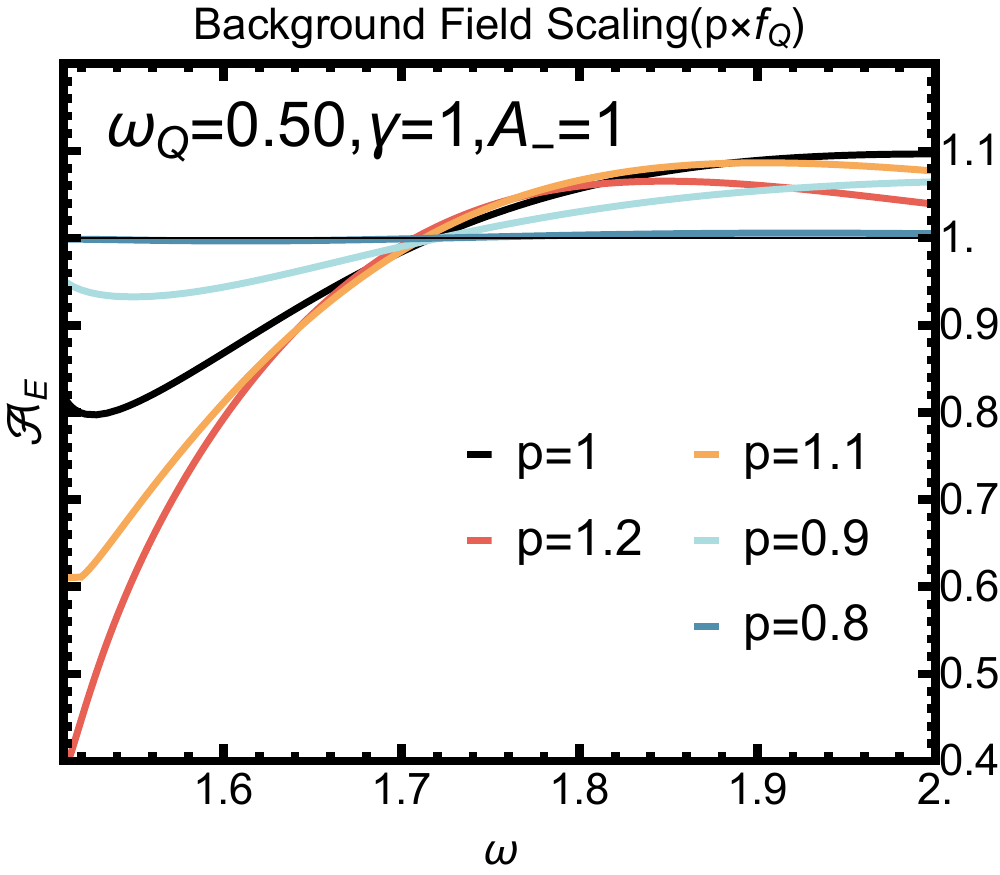}
     \includegraphics[height=3.7cm]{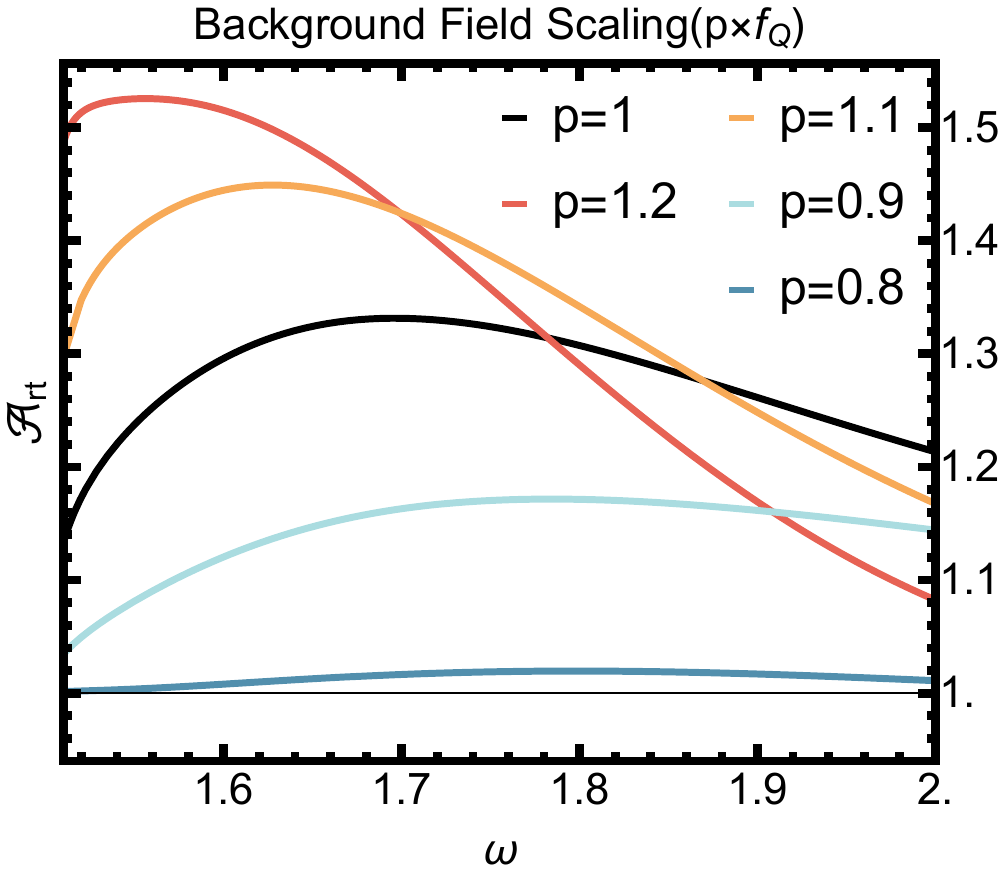}
	\caption{Effects of direct scaling of background coefficient $W$ and background field scaling on the energy and energy flux amplification factors. 
}
 \label{fig:bds}
\end{figure} 

\begin{figure}
	\centering
	 \includegraphics[height=5.0cm]{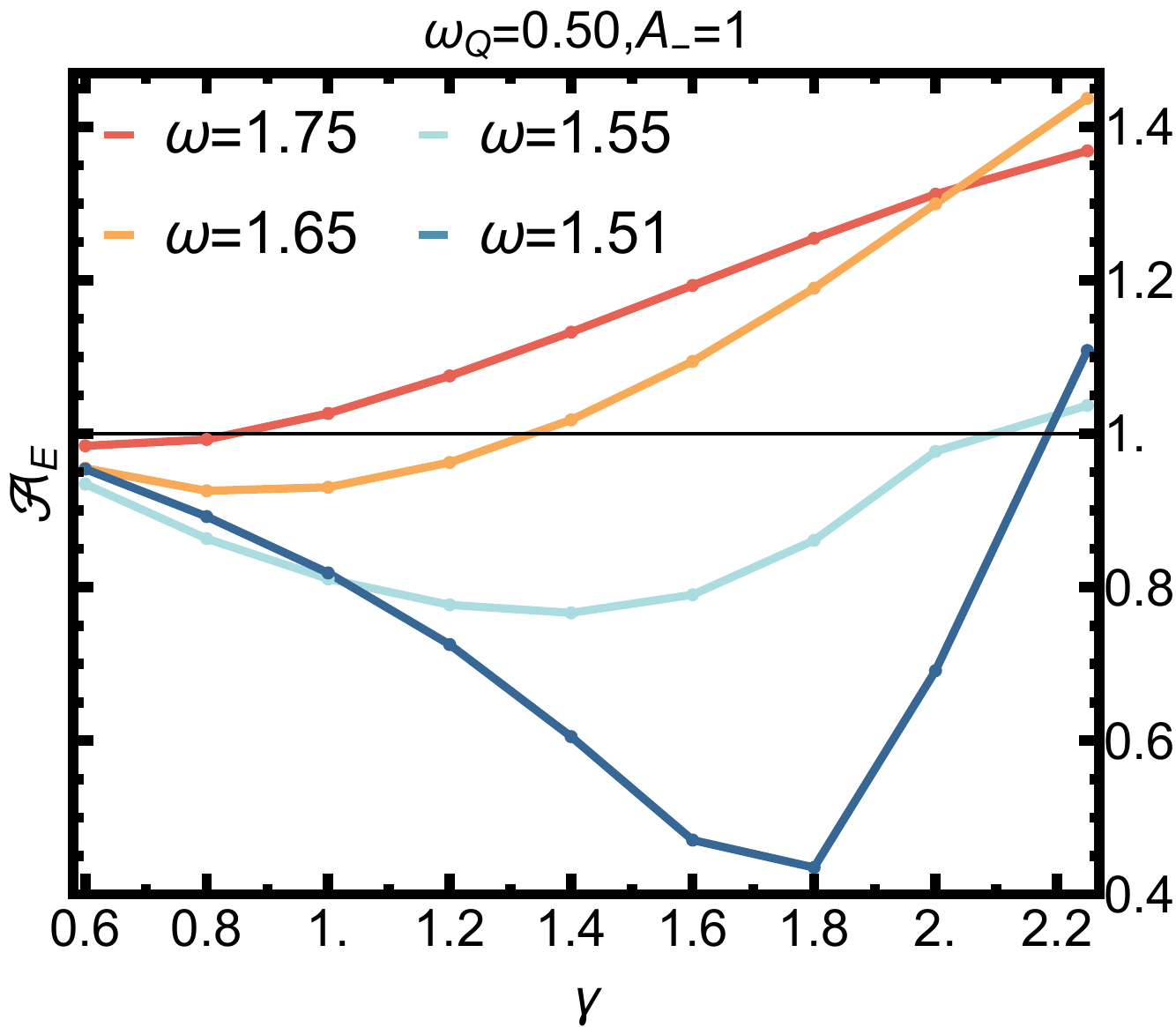}
      \includegraphics[height=5.0cm]{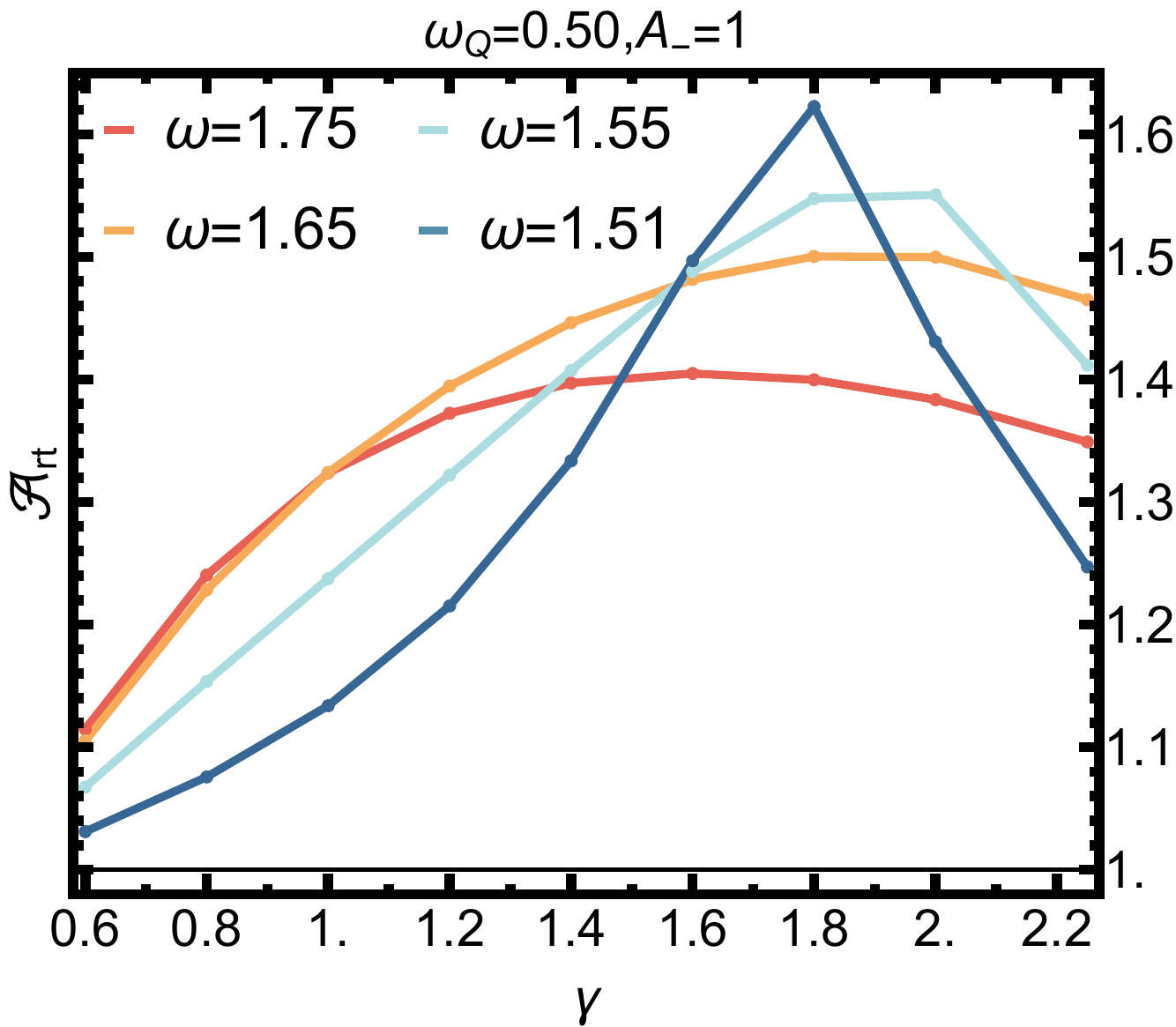}
	\caption{Energy and energy flux amplification factors for the single ingoing mode ($A_-=1$) for $\omega_Q=0.50$ and different mass ratios $\gamma$. 
}
 \label{fig:bds2}
\end{figure}

\subsection{Single ingoing mode}

The perturbative wave equations have been established in Eqs.~\eqref{equ::ptphi}, \eqref{equ::ptxi}, subject to the suitable boundary conditions given in Eqs.~\eqref{bc::1}, Eq.~\eqref{bc::2}. When employing the relaxation method to solve these equations, it is essential to ensure that both absolute and relative errors remain below $10^{-10}$ while conserving the particle flux, expressed as $|\mathcal{A}_N - 1| \to 0$. If the latter condition is not adequately satisfied, we can increase the number of computational points to achieve the required accuracy, though this will inevitably lead to longer computation times.

In Fig.~\ref{fig:sging}, we present the energy and energy flux amplification factors for a single ingoing mode, arranged in four columns. Note that increasing $\gamma$ (when $\gamma \le 1$) results in a wider range of amplification factors. Decreasing $\omega_Q$ leads to multiple peaks in the amplification factors. For the case where $\gamma = 10^{-4}$, the amplification factor is close to 1.

In Fig.~\ref{fig:rat}, we present the ratio of outgoing particle number for the parameters $\omega_Q=0.5$ and $\gamma=1$, with left, middle, and right panels corresponding to the different ingoing modes: $A_-=1$, $B_+=1$ and $D_+=1$, respectively. For each single ingoing mode configuration, the outgoing mode corresponds to a different energy (or energy flux) than the ingoing mode, making it easier for the amplification factor to reach its upper limit. Based on this understanding, to achieve a higher amplification factor, a higher outgoing ratio, distinct from the ingoing mode, is required, especially when the outgoing mode has a higher energy (or energy flux) than the ingoing mode. 

The figures reveal several intriguing results, which we summarize and attempt to explain in the following discussion: 
\begin{itemize}
\item A heavier real scalar field $\xi$, generally produces a larger amplification factor, except in the case of energy flux with $\omega_Q = 0.5$ and $\gamma=1$.

\item As $\gamma$ approaches smaller values, the amplification factor correspondingly decreases. In fact, for $\gamma =10^{-4}$ or $10^{-2}$, and $\omega_Q=0.95$, all amplification factors are nearly equal to $1$, and thus are not explicitly shown in the figures. 

\item As $\omega_Q$ approaches the lower bound of its range, the amplification factor exhibits more pronounced peaks. 
\end{itemize}

\subsubsection{Factors affecting amplification}

The system consists of coupled equations, with the background coefficient represented by $W$. If this background coefficient $W$ is neglected, the system reduces to a set of decoupled equations, which evidently cannot produce any superradiance. Conversely, as the background coefficient increases, the amplification factor may grow. Therefore, we investigate the impact of the background coefficient on the amplification factors through the following three approaches:
\begin{itemize}
\item {\it Direct Scaling}: Multiply $W$ directly by a parameter $p$. This method straightforwardly increases the background coefficient without affecting other components of the system. 

\item {\it Background Field Scaling:} Multiply the background field $f_Q$ by a parameter $p$. This adjustment not only influences the background coefficient $W$, but also modifies the term $U_\xi$. This approach is more physically realistic for larger soliton configurations, as the profile of $\chi_Q$ does not change significantly for $\gamma \ge 1$.

\item {\it Changing the Mass Ratio $\gamma$:} Vary the mass ratio $\gamma$, consider different background solutions, and compute the amplification factors for each case.

\end{itemize}

In Fig.~\ref{fig:W}, we present the background coefficient $W$ for different values of $\omega_Q$ and $\gamma$. It is evident that the background coefficient increases with the mass ratio $\gamma$. Specifically, for $\gamma=10^{-4}$ and $10^{-2}$, the background coefficient $W$ remains consistently small and can be considered negligible.

In Fig.~\ref{fig:bds}, the direct scaling of the background coefficient $W$ and background field scaling with the energy and energy flux amplification factors are presented. The top and bottom panels correspond to the direct scaling and background field scaling approaches, respectively, where the background solutions for both approaches are not the FLS solitons. However, these two approaches help us understand the relationship between the background coefficient $W$, the potential term $U_\xi$, and the amplification factors. It is clear that increasing the scaling parameter $p$ for direct scaling of the background coefficient in this case, leads to an increase in the energy flux amplification factor. In fact, the amplification factor is more likely to approach to its limits due to the conversion rate of outgoing modes with higher energy flux. For single ingoing mode $A_-$, the energy flux amplification factors are always great than 1. On the other hand, when both the background coefficient and potential term are increased, the amplification factors do not always increase. For FLS solitons, to achieve a larger background coefficient, a higher mass ratio is required, which in turn results in larger potentials.  

\begin{figure}
	\centering
		\includegraphics[height=5.0cm]{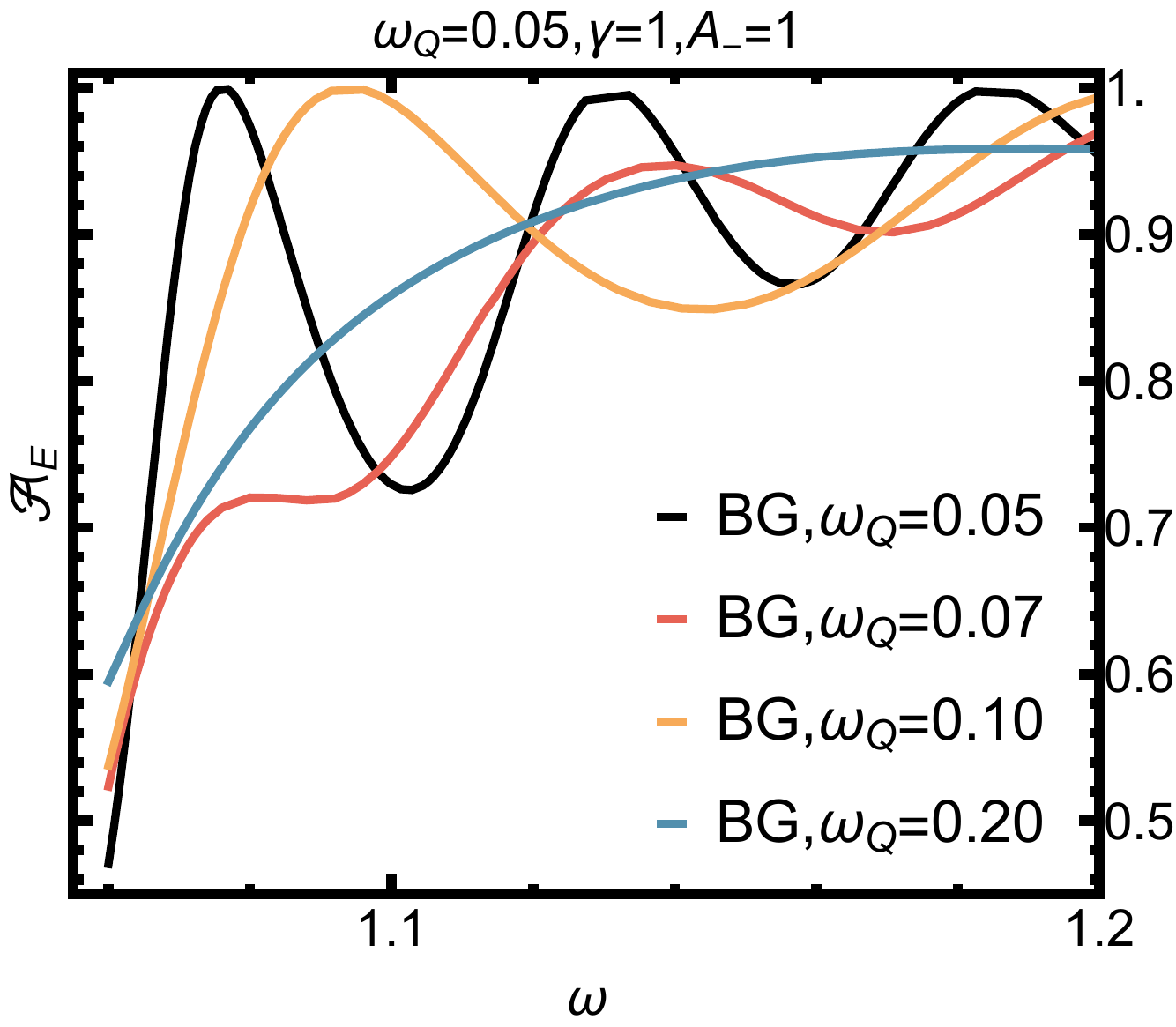}
            \includegraphics[height=5.0cm]{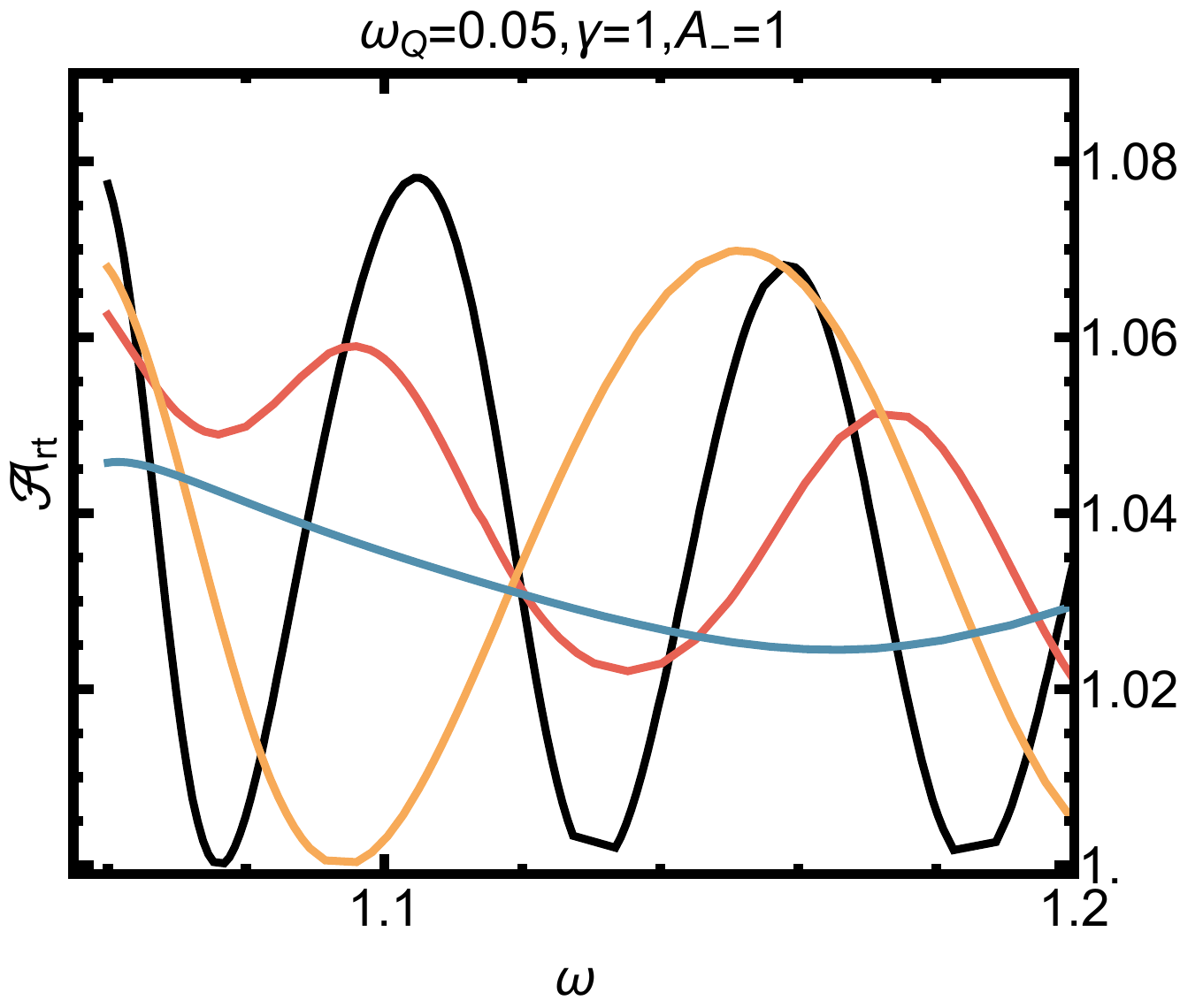}
	\caption{Energy and energy flux amplification factors for the single ingoing mode ($A_-=1$) with $\omega_Q=0.05$ and $\gamma=1$, for different background solutions. 
    The colors represent different background solutions with varying values of $\omega_Q$.
}
 \label{fig:bg}
\end{figure}

\begin{figure*}
	\centering
		\includegraphics[height=4.6cm]{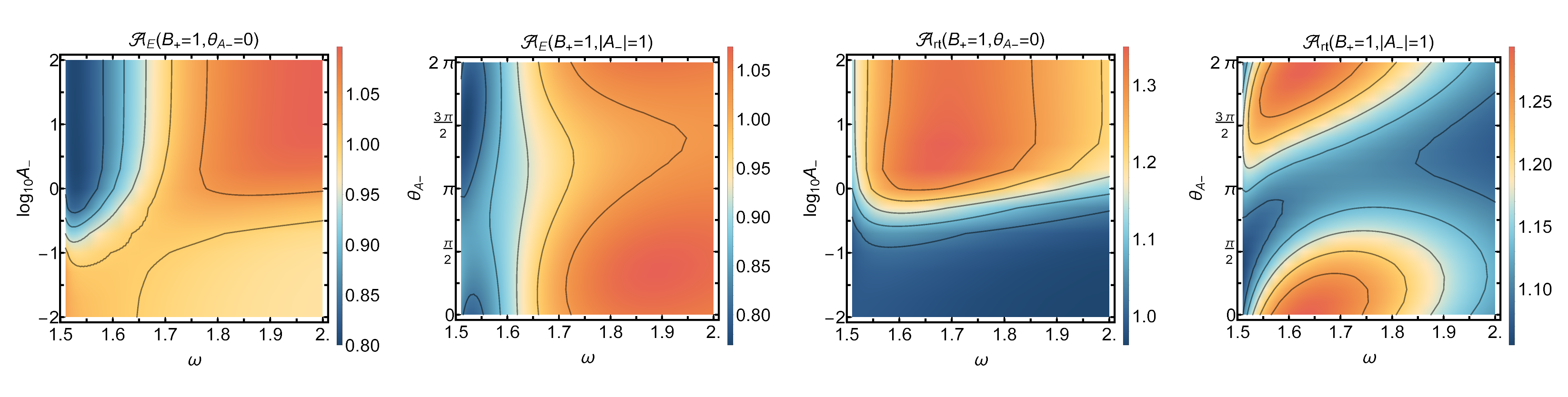}
            \includegraphics[height=4.6cm]{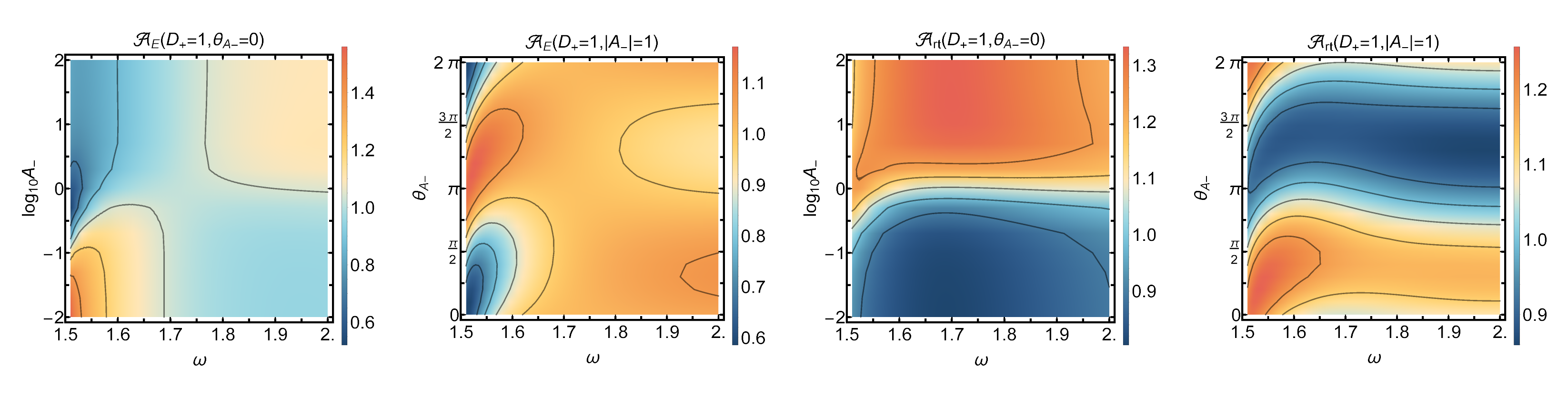}
		\includegraphics[height=4.5cm]{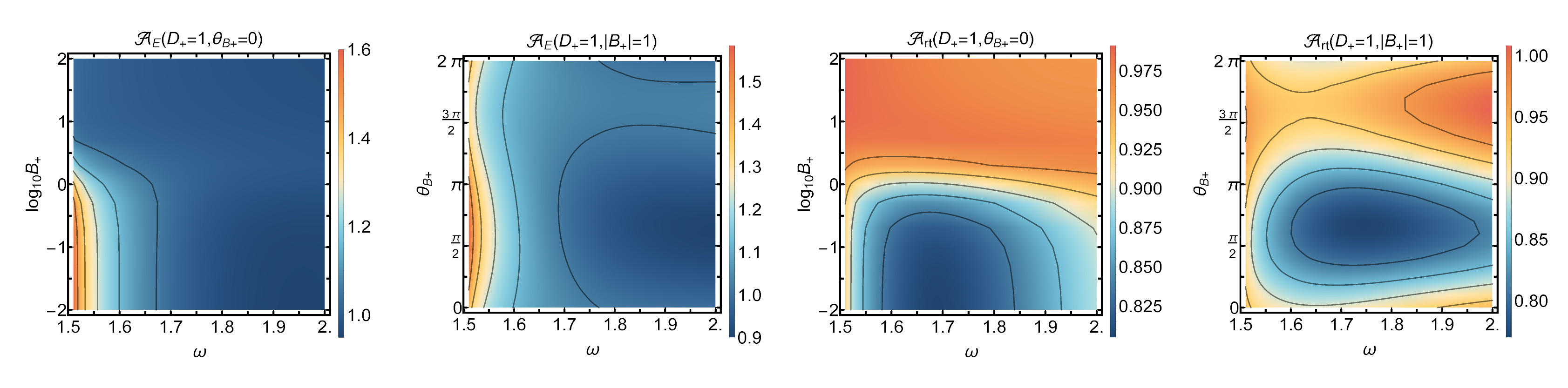}
	\caption{Amplification factors $\mathcal{A}_E$ and $\mathcal{A}_{rt}$ for double ingoing modes for $\omega_Q=0.50$ and $\gamma=1$. For example, in the top left figure, $B_+=1,\theta_{A-}=0$ means fixing the normalized amplitude $B_+$ and setting the phase $\theta_{A-}=0$ for $A_-=|A_-|\exp(i\theta_{A-})$, with $|A_-|$ adjustable.
}
 \label{fig:double}
\end{figure*}

Although increasing the background coefficient $W$, which can be achieved by increasing the mass ratio $\gamma$, may lead to lager amplification factors, we also need to test the real situation. In Fig.~\ref{fig:bds2}, we present the energy and energy flux amplification factors for the single ingoing mode $A_-=1$, with different mass ratios $\gamma$ and $\omega_Q=0.50$. It is clear that in the real situation, increasing the mass ratio $\gamma$ can give us larger amplification factors, but when $\gamma$ closes to $(1+\omega_Q)^2$, the energy flux amplification factors decrease, which is because more of the outgoing particles are taken up by the energy flux $\omega$ particles, due to the heavier scalar field.

We have shown that increasing the background coefficient $W$ leads the amplification factors to approach their absolute limits. This tendency suggests that more particles are scattered and converted into particles of different energy (or energy flux) states. However, it does not necessarily result in an increase in amplification; it can also lead to a decrease. The underlying reason is that a larger fraction of high-energy particles are being converted into low-energy particles, redistributing the overall energy. Nevertheless, it remains possible to selectively introduce only low-energy incident particles while increasing the coupling coefficient ({\it i.e.}, adopting a background solution with larger mass ratio $\gamma$) to achieve an enhanced amplification effect.

In the special case where $\omega_Q$ approaches its upper bound and for a light field $\xi$, specifically when $\omega_Q=0.95$ and $\gamma=10^{-4}$ or $10^{-2}$, the background coefficient becomes sufficiently small, where $W\to0$, and can be neglected. Under these conditions, the coupled perturbation equations can be well-approximated by decoupled ones, which, although incapable of producing significant superradiant amplification, still allow for a remarkably broad range of amplification factors.

When $\omega_Q$ approaches its lower bound, the allowed range of the energy flux amplification becames vary narrow, as discussed in Eq.~\eqref{range::Art}. In this regime, even if the background coefficient $W$ is larger, the amplification remains limited to a small energy flux amplification factor. By contrast, the energy amplification factors for $\omega_Q=0.95$ and $\gamma=10$, is significantly large, though this is not explicitly shown in the figures.

\subsubsection{Amplification peaks}

In Fig.~\ref{fig:bg}, we show the energy and energy flux amplification factors for the single ingoing mode $A_-=1$ with $\omega_Q=0.05$ and $\gamma=1$, with different background solutions. The up and down plots show the energy and energy flux amplification factors, respectively. The colors correspond to variations in the background solutions $(f_Q,\chi_Q)$ for different values of $\omega_Q$. Meanwhile, for $\omega_{\pm}$, the value of $\omega_Q = 0.05$ remains unchanged. It is clear that larger FLS solitons result in more peaks in the amplification factor spectra. This reason will be analyzed in detail in our following paper.

\begin{figure}
	\centering
		\includegraphics[height=8.8cm]{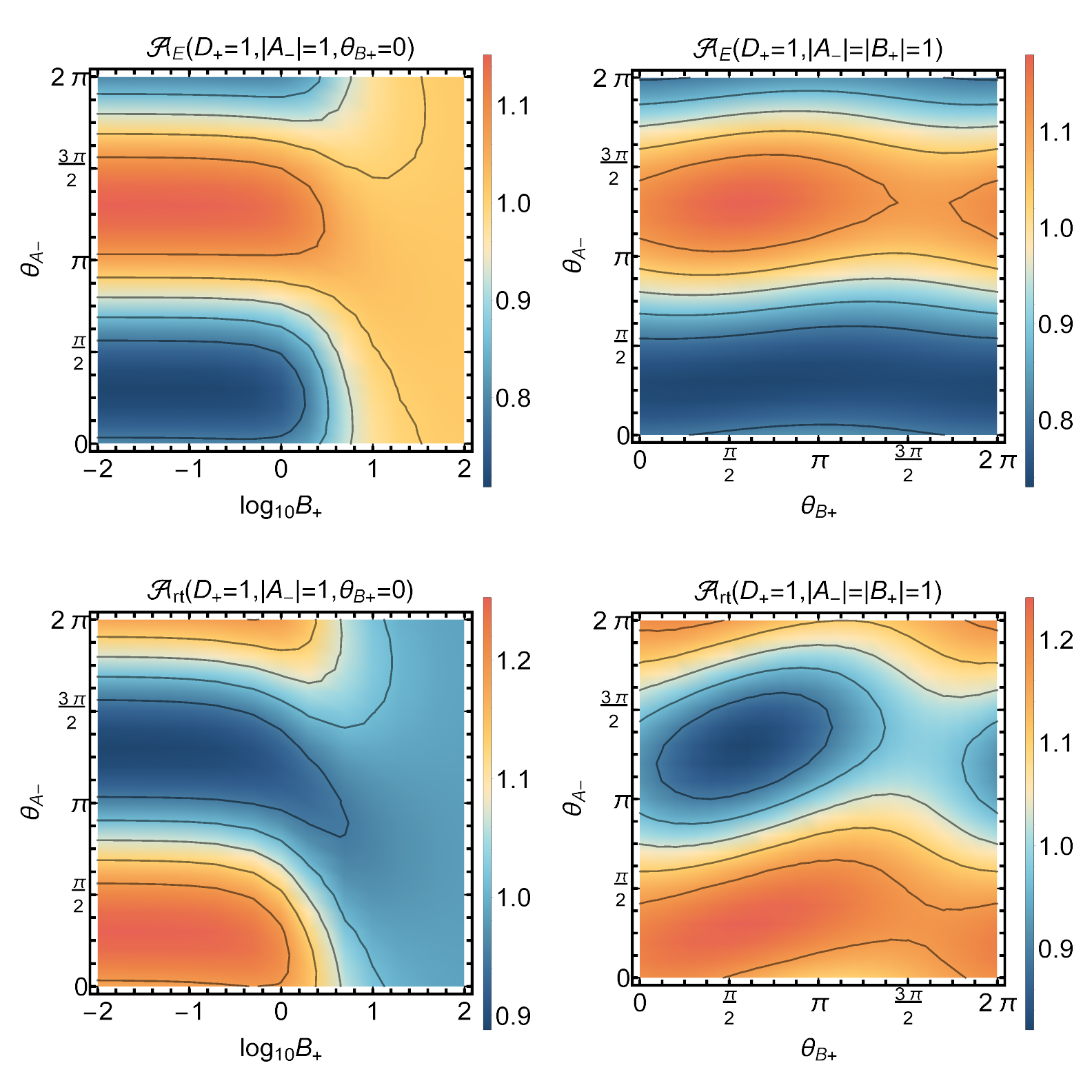}
	\caption{Amplification factors $\mathcal{A}_E$ and $\mathcal{A}_{rt}$ for triple ingoing modes for $\omega_Q=0.50,\gamma=1$, and $\omega=1.55$. Here, we fix $D_+=1$ and set $A_-=|A_-|\exp(i\theta_{A-})$ and $B_+=|B_+|\exp(i\theta_{B+})$.
}
 \label{fig:triple}
\end{figure}

\subsection{Multiple ingoing modes}

Although the single ingoing mode provides different information, it still contains valuable insights when compared to the double and triple ingoing modes. We present the case of double ingoing modes to illustrate the complex behavior of the amplification factor.

In Fig.~\ref{fig:double}, we present the energy and energy flux amplification factors for the double ingoing mode with parameters $\omega_Q=0.5$ and $\gamma=1$. For the double (or triple) ingoing modes, we need to consider the configuration of the ingoing modes, as there are two (or four) parameters to determine the amplitudes. Specifically, for a double ingoing mode, one mode can be normalized using the $U(1)$ and scaling symmetries, while the other mode requires determining both its modulus and phase. Therefore, each additional ingoing mode introduces two parameters to specify the configuration. For example, $B_+=1,\theta_{A-}=0$ means the case that $B_+=1,\theta_{A-}=0$ where $A_-=|A_-| \exp(i\theta_{A-})$, with $|A_-|$ adjustable. When the amplitudes of different ingoing modes have a large distinction, leading to a significant difference in the ingoing particle numbers, the smaller amplitude mode can be neglected and treated as a single ingoing mode approximately. In this case, the first and third columns can be interpreted as the changes in amplification factors during transitions between different single ingoing modes. The top, middle, and bottom rows can be considered as showing the transformations $B_+ \to A_-$, $D_+ \to A_-$, and $D_+ \to B_+$ for single ingoing mode, respectively. The second and fourth columns show identical amplitude magnitudes for the double ingoing modes, with the amplification factor varying according to the allowed phase difference. This illustrates the complex behavior of amplification in multi-mode scenarios. Here, we fix $B_+=1,|A_-|=1$ with $\theta_{A-}$ adjustable.
From Fig.~\ref{fig:double}, it is clear that by increasing the amplitude of $A_-$, the energy flux amplification factors can reach higher values, as argued in Section~\ref{sec:ampfcons}.

In Fig.~\ref{fig:triple}, we present the amplification spectra for the case of triple ingoing modes with parameters $\omega_Q=0.50$, $\gamma=1$, and $\omega=1.55$. We see that in this the behavior of the amplification factors is more complex, heavily depending on the different phases $\theta_{A-}$ and $\theta_{B+}$.

\section{Conclusion}
\label{sec:level5}

In this paper, we have studied superradiant amplifications of waves around non-topological solitions in the Friedberg-Lee-Sirlin model, which after re-formulation with dimensionless variables contains one theory parameter $\gamma$, the squared mass ratio between the two scalar fields. We first solve for the FLS soliton with the absolute and relative errors remaining below $10^{-10}$. It is found that as $\omega_Q$ decreases and $\gamma$ increases, both the charge and energy of the soliton increase.

We then investigated the perturbative scattering on the FLS soliton background and evaluate the amplification factors $\mathcal{A}_E$ and $\mathcal{A}_{rt}$ for various parameters and scattering scenarios. The solution profile of an FLS soliton is rather different from that of a $Q$-ball from the single scalar theory. As a result, for very large FLS solitons, the perturbative solution becomes highly sensitive to the conditions near the origin, and a standard shooting method fails to resolve it. We found that the relaxation method is effective to achieve sufficient accuracy in this case, which can be verified by checking the particle number conservation $\mathcal{A}_N = 1$. In our calculations, both the absolute and relative errors for the perturbative solutions were kept below $10^{-10}$, and the errors for all amplification factors below $10^{-4}$. In the other limit, as $\gamma$ approaches zero, the perturbative modes tend to decouple, and the amplification factors approach to 1. Furthermore, our numerical results showed that a background solution with a lower $\omega_Q$ exhibits more peaks in the amplification factors.

Generally, multiple ingoing modes lead to richer spectra of amplification factors. It is observed that for the case of double ingoing modes, increasing the amplitude of the low-energy complex-scalar mode tends to enhance energy and energy flux extraction. When the particle numbers of ingoing modes are fixed, the phase of the ingoing mode amplitudes can also influence the particle numbers of the outgoing modes, providing a means to achieve larger amplification factors. In the case of triple ingoing modes, with fixed particle numbers for the ingoing modes, the amplification factors depend on two phase parameters, resulting in a more intricate behavior.

The particle number conservation, coupled with a linear fractional optimization scheme, also allowed us to derive the bounds on the amplification factors for generic scenarios of ingoing modes. The same method also applies to the single-field $Q$-ball case, which is explicitly computed in Appendix~\ref{sec:levela} for the reader's convenience. We also analytically derived the bounds on the amplification factors near the mass gap.

In this paper, we focused on perturbative wave scattering off a spherically symmetric FLS soliton. A natural avenue for generalization is to consider the spinning case and to investigate the superradiant amplification of FLS solitons in the time domain, which is left for future work.

\acknowledgments

We would like to thank Victor Jaramillo and Meng-Fan Zhu for helpful discussions. SYZ acknowledges support from the National Natural Science Foundation of China under grant No.~12475074, 12075233 and 12247103. QXX acknowledges support from CSC (File No.~202406340173).

\appendix

\section{Amplification limits for single scalar models}
\label{sec:levela}

\begin{figure}
	\centering
		\includegraphics[height=5.0cm]{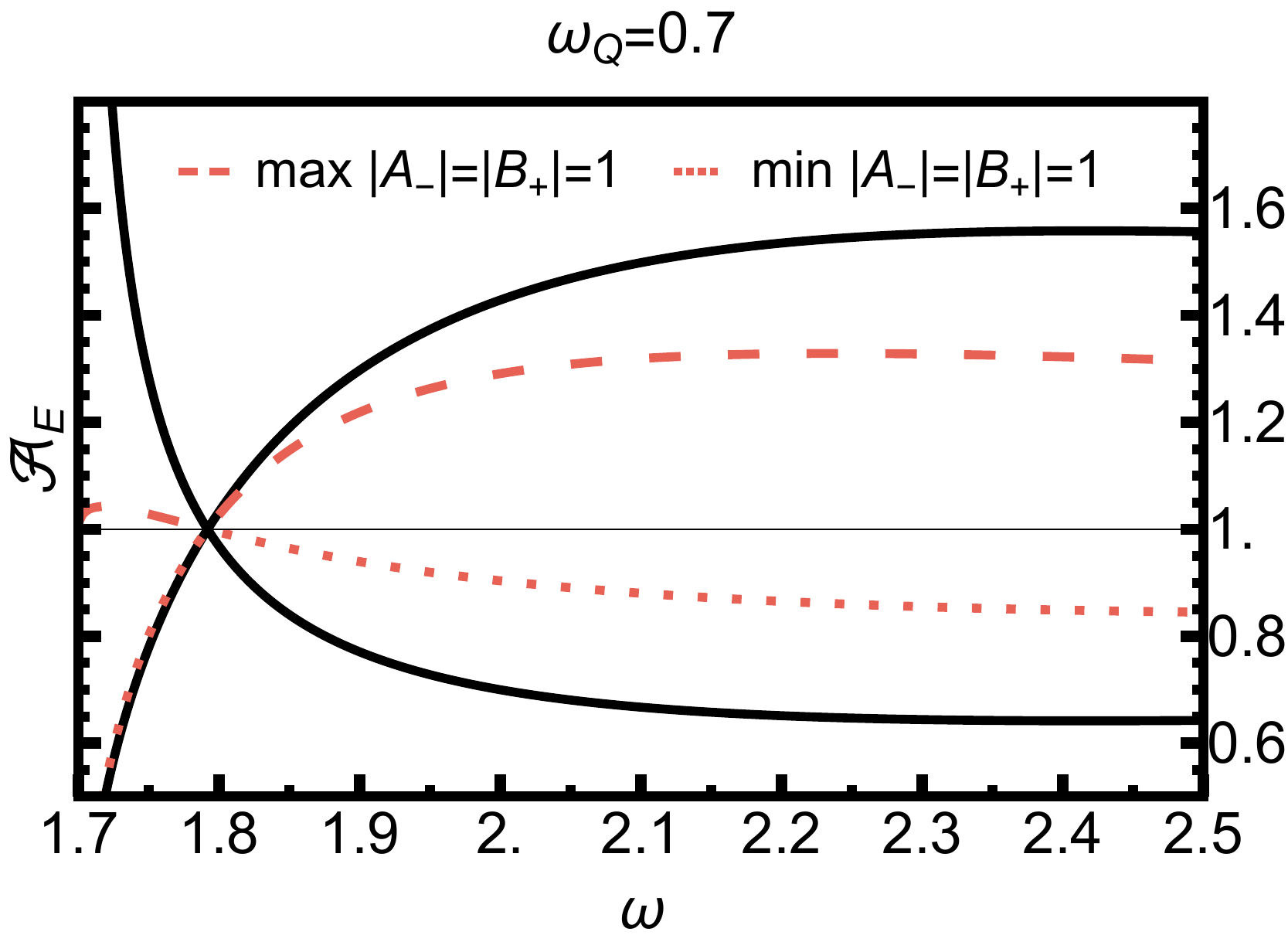}
		\includegraphics[height=5.0cm]{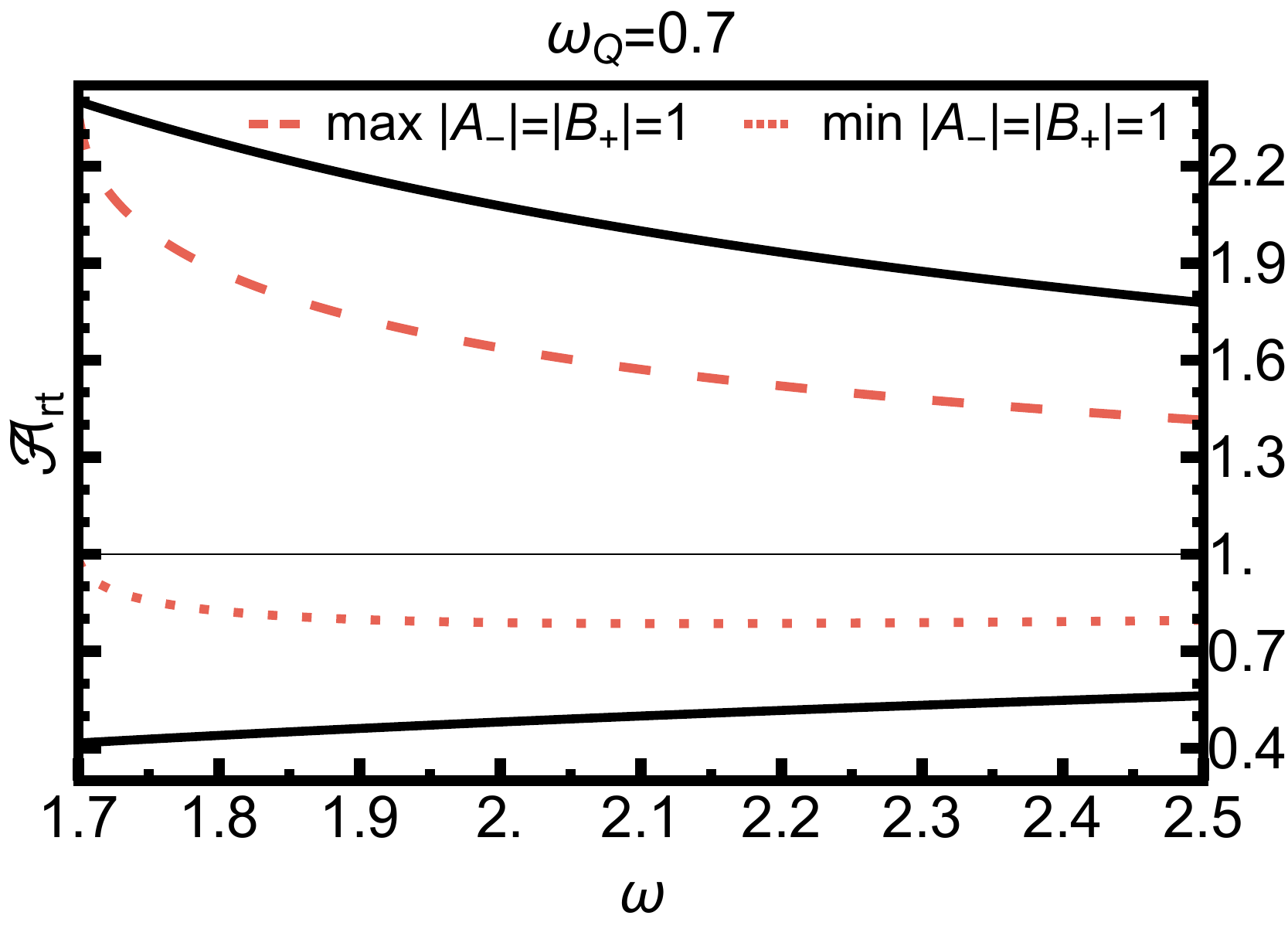}
	\caption{Absolute limits of $\mathcal{A}_E$ and $\mathcal{A}_{rt}$ with $\omega_Q=0.7$ for the single complex field theory. The black lines represent the amplification limits without amplitude constraints. The red lines correspond to $|A_-|=|B_+|=1$ for the double ingoing modes.  (The bounds for the case of single ingoing mode was derived in \cite{Zhang:2024ufh}.)}
 \label{fig:aert2}
\end{figure}

\begin{figure*}
	\centering
		\includegraphics[height=5.0cm]{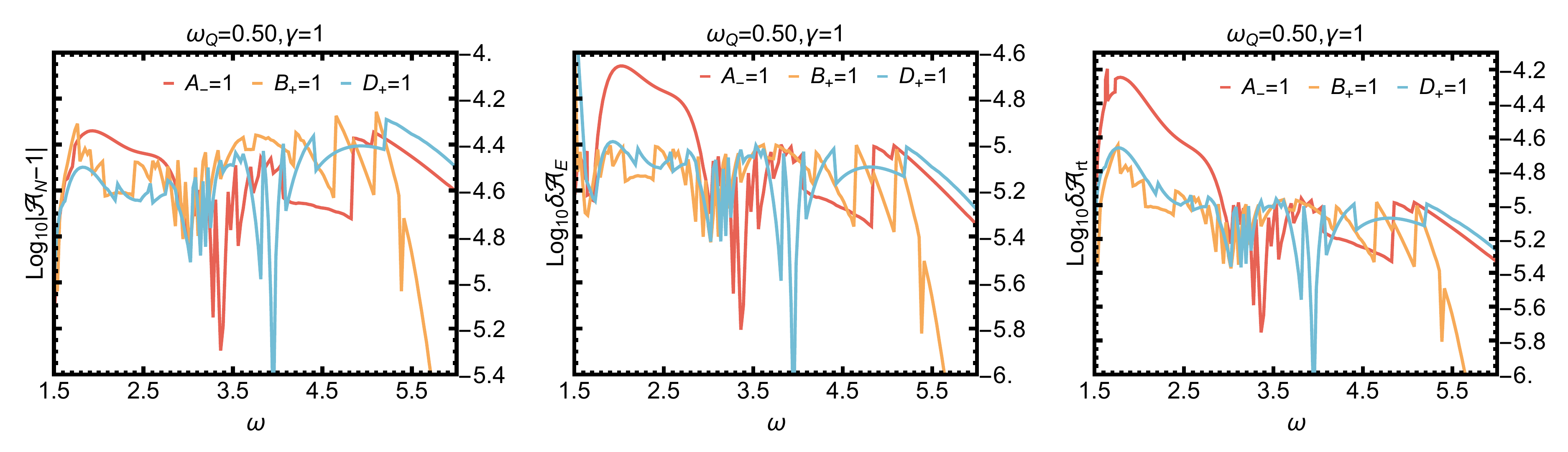}
	\caption{Errors of the amplification factors $\mathcal{A}_N$, $\mathcal{A}_E$, and $\mathcal{A}_{rt}$ with $\omega_Q=0.50$ and $\gamma=1$ for each single ingoing mode. 
    The middle and right subfigures display the errors of the energy and energy flux amplification factors, respectively, subject to the conservation of particle number. }
 \label{fig:error}
\end{figure*}

In this appendix, we also present the absolute bounds on the amplification factors for waves scattering around a $Q$-ball in a single scalar model.

Specifically, these bounds apply to a generic complex scalar model given by the following Lagrangian:
\begin{align}
    \mathcal{L} = - \p_\mu \Phi^\dagger \p^\mu \Phi - |\Phi|^2 + V_{\rm int}(|\Phi|).
\end{align}
Here the nonlinear potential $V_{\rm int}$ must allow for the existence of a $Q$-ball, but the absolute bounds are independent of its specific form, as their derivation only relies on the quadratic Lagrangian. The corresponding amplification factors $\mathcal{A}_N,\mathcal{A}_E,$ and $\mathcal{A}_{rt}$ are defined similar to those of Eqs.~\eqref{Ampc1}-\eqref{Ampc3}, but with $C_+$ and $D_+$ set to zero. Utilizing the LFO method described in Section~\ref{sec:amplim}, we can numerically derive the bounds for generic ingoing modes. 
For example, in Fig.~\ref{fig:aert2}, we plot the results for a $Q$-ball with internal frequency $\omega_Q=0.7$. 

From these figures, we observe that the bounds for the double ingoing modes exhibit a more intricate behavior, where the amplification factors can be either greater or smaller than $1$. Notably, these bounds are fully contained within the region defined by the scenario without any amplitude constraints and do not approach its boundaries. This suggests that the amplification factors achievable in multiple ingoing mode scenarios are, in theory, always limited by the bounds of the single ingoing mode case (or the case without amplitude constraints).

\section{Numerical accuracy}
\label{sec:levelb}

In this appendix, we briefly discuss the numerical parameters and accuracy in our calculations of the amplification factors. While the relaxation method yields convergent solutions, these solutions do not necessarily respect the constraint of the particle number conservation. However, by increasing the number of points in the solution domain in each iteration, we can improve the accuracy of the constraint. The relaxation method transforms the ODEs into a set of linear equations involving large matrices. Obviously, a larger number of domain points results in larger matrices in the relaxation evaluation, which in turn requires more computation time per iteration. For example, when solving for a background soliton, we implement each iteration of relaxation with a $5000 \times 5000$ matrix, which takes approximately one minute with parallel computation, corresponding to about 800 points in the solution domain. To solve the perturbative scattering equations, we typically implement the relaxation with around 3000 points, corresponding to a $20000 \times 20000$ matrix, and each requires approximately 20 iterations. This takes roughly 10 hours to compute one amplification data point, if we choose the absolute and relative errors to be below $10^{-10}$, which ensures that the errors in $|\mathcal{A}_N-1|$ are below $10^{-4}$. We implement the relaxation method mainly in Mathematica and also verify the results with an independent relaxation code in Python. 

As discussed in Section~\ref{sec:ampfcons}, the energy and energy flux amplification factors can be interpreted as the particle number, weighted by the particle energy and energy flux. To compute the errors in the other amplification factors, we require $\mathcal{A}_N = 1$ to adjust the amplitudes accordingly. In practice, we modify each amplitude to adjust the particle number and use the updated values to compute the energy and energy flux amplification factors. We define the error in these factors as the difference between the maximum and minimum values of the amplification factors, expressed as:
\begin{align}
\delta \mathcal{A}_E & \equiv \max(\mathcal{A}_E)_{\mathcal{A}_N=1} - \min(\mathcal{A}_E)_{\mathcal{A}_N=1}, \\
\delta \mathcal{A}_{rt} & \equiv \max(\mathcal{A}_{rt})_{\mathcal{A}_N=1} - \min(\mathcal{A}_{rt})_{\mathcal{A}_N=1}.
\end{align}
In Fig.~\ref{fig:error}, we present the errors of the amplification factors $\mathcal{A}_N$, $\mathcal{A}_E$, and $\mathcal{A}_{rt}$. The first subfigure illustrates the deviation of the particle number amplification factor, with all errors below $10^{-4}$. The second and third subfigures display the errors of the energy and energy flux amplification factors, respectively, while enforcing the conservation of particle number. We see that all errors in the amplification factors are also below $10^{-4}$.

\bibliographystyle{apsrev4-1}
\bibliography{zref}

\begin{thebibliography}{92}%
\makeatletter
\providecommand \@ifxundefined [1]{%
 \@ifx{#1\undefined}
}%
\providecommand \@ifnum [1]{%
 \ifnum #1\expandafter \@firstoftwo
 \else \expandafter \@secondoftwo
 \fi
}%
\providecommand \@ifx [1]{%
 \ifx #1\expandafter \@firstoftwo
 \else \expandafter \@secondoftwo
 \fi
}%
\providecommand \natexlab [1]{#1}%
\providecommand \enquote  [1]{``#1''}%
\providecommand \bibnamefont  [1]{#1}%
\providecommand \bibfnamefont [1]{#1}%
\providecommand \citenamefont [1]{#1}%
\providecommand \href@noop [0]{\@secondoftwo}%
\providecommand \href [0]{\begingroup \@sanitize@url \@href}%
\providecommand \@href[1]{\@@startlink{#1}\@@href}%
\providecommand \@@href[1]{\endgroup#1\@@endlink}%
\providecommand \@sanitize@url [0]{\catcode `\\12\catcode `\$12\catcode
  `\&12\catcode `\#12\catcode `\^12\catcode `\_12\catcode `\%12\relax}%
\providecommand \@@startlink[1]{}%
\providecommand \@@endlink[0]{}%
\providecommand \url  [0]{\begingroup\@sanitize@url \@url }%
\providecommand \@url [1]{\endgroup\@href {#1}{\urlprefix }}%
\providecommand \urlprefix  [0]{URL }%
\providecommand \Eprint [0]{\href }%
\providecommand \doibase [0]{http://dx.doi.org/}%
\providecommand \selectlanguage [0]{\@gobble}%
\providecommand \bibinfo  [0]{\@secondoftwo}%
\providecommand \bibfield  [0]{\@secondoftwo}%
\providecommand \translation [1]{[#1]}%
\providecommand \BibitemOpen [0]{}%
\providecommand \bibitemStop [0]{}%
\providecommand \bibitemNoStop [0]{.\EOS\space}%
\providecommand \EOS [0]{\spacefactor3000\relax}%
\providecommand \BibitemShut  [1]{\csname bibitem#1\endcsname}%
\let\auto@bib@innerbib\@empty
\bibitem [{\citenamefont {Friedberg}\ \emph
  {et~al.}(1976{\natexlab{a}})\citenamefont {Friedberg}, \citenamefont {Lee},\
  and\ \citenamefont {Sirlin}}]{Friedberg:1976me}%
  \BibitemOpen
  \bibfield  {author} {\bibinfo {author} {\bibfnamefont {R.}~\bibnamefont
  {Friedberg}}, \bibinfo {author} {\bibfnamefont {T.}~\bibnamefont {Lee}}, \
  and\ \bibinfo {author} {\bibfnamefont {A.}~\bibnamefont {Sirlin}},\ }\href
  {\doibase 10.1103/PhysRevD.13.2739} {\bibfield  {journal} {\bibinfo
  {journal} {Phys. Rev. D}\ }\textbf {\bibinfo {volume} {13}},\ \bibinfo
  {pages} {2739} (\bibinfo {year} {1976}{\natexlab{a}})}\BibitemShut {NoStop}%
\bibitem [{\citenamefont {Coleman}(1985)}]{Coleman:1985ki}%
  \BibitemOpen
  \bibfield  {author} {\bibinfo {author} {\bibfnamefont {S.~R.}\ \bibnamefont
  {Coleman}},\ }\href {\doibase 10.1016/0550-3213(86)90520-1} {\bibfield
  {journal} {\bibinfo  {journal} {Nucl. Phys. B}\ }\textbf {\bibinfo {volume}
  {262}},\ \bibinfo {pages} {263} (\bibinfo {year} {1985})},\ \bibinfo {note}
  {[Addendum: Nucl.Phys.B 269, 744 (1986)]}\BibitemShut {NoStop}%
\bibitem [{\citenamefont {Zhou}(2024)}]{Zhou:2024mea}%
  \BibitemOpen
  \bibfield  {author} {\bibinfo {author} {\bibfnamefont {S.-Y.}\ \bibnamefont
  {Zhou}},\ }\href@noop {} {\  (\bibinfo {year} {2024})},\ \Eprint
  {http://arxiv.org/abs/2411.16604} {arXiv:2411.16604 [hep-th]} \BibitemShut
  {NoStop}%
\bibitem [{\citenamefont {Volkov}\ and\ \citenamefont
  {Wohnert}(2002)}]{Volkov:2002aj}%
  \BibitemOpen
  \bibfield  {author} {\bibinfo {author} {\bibfnamefont {M.~S.}\ \bibnamefont
  {Volkov}}\ and\ \bibinfo {author} {\bibfnamefont {E.}~\bibnamefont
  {Wohnert}},\ }\href {\doibase 10.1103/PhysRevD.66.085003} {\bibfield
  {journal} {\bibinfo  {journal} {Phys. Rev. D}\ }\textbf {\bibinfo {volume}
  {66}},\ \bibinfo {pages} {085003} (\bibinfo {year} {2002})},\ \Eprint
  {http://arxiv.org/abs/hep-th/0205157} {arXiv:hep-th/0205157} \BibitemShut
  {NoStop}%
\bibitem [{\citenamefont {Kleihaus}\ \emph {et~al.}(2005)\citenamefont
  {Kleihaus}, \citenamefont {Kunz},\ and\ \citenamefont
  {List}}]{Kleihaus:2005me}%
  \BibitemOpen
  \bibfield  {author} {\bibinfo {author} {\bibfnamefont {B.}~\bibnamefont
  {Kleihaus}}, \bibinfo {author} {\bibfnamefont {J.}~\bibnamefont {Kunz}}, \
  and\ \bibinfo {author} {\bibfnamefont {M.}~\bibnamefont {List}},\ }\href
  {\doibase 10.1103/PhysRevD.72.064002} {\bibfield  {journal} {\bibinfo
  {journal} {Phys. Rev. D}\ }\textbf {\bibinfo {volume} {72}},\ \bibinfo
  {pages} {064002} (\bibinfo {year} {2005})},\ \Eprint
  {http://arxiv.org/abs/gr-qc/0505143} {arXiv:gr-qc/0505143} \BibitemShut
  {NoStop}%
\bibitem [{\citenamefont {Almumin}\ \emph {et~al.}(2024)\citenamefont
  {Almumin}, \citenamefont {Heeck}, \citenamefont {Rajaraman},\ and\
  \citenamefont {Verhaaren}}]{Almumin:2023wwi}%
  \BibitemOpen
  \bibfield  {author} {\bibinfo {author} {\bibfnamefont {Y.}~\bibnamefont
  {Almumin}}, \bibinfo {author} {\bibfnamefont {J.}~\bibnamefont {Heeck}},
  \bibinfo {author} {\bibfnamefont {A.}~\bibnamefont {Rajaraman}}, \ and\
  \bibinfo {author} {\bibfnamefont {C.~B.}\ \bibnamefont {Verhaaren}},\ }\href
  {\doibase 10.1140/epjc/s10052-024-12712-x} {\bibfield  {journal} {\bibinfo
  {journal} {Eur. Phys. J. C}\ }\textbf {\bibinfo {volume} {84}},\ \bibinfo
  {pages} {364} (\bibinfo {year} {2024})},\ \Eprint
  {http://arxiv.org/abs/2302.11589} {arXiv:2302.11589 [hep-th]} \BibitemShut
  {NoStop}%
\bibitem [{\citenamefont {Copeland}\ \emph {et~al.}(2014)\citenamefont
  {Copeland}, \citenamefont {Saffin},\ and\ \citenamefont
  {Zhou}}]{Copeland:2014qra}%
  \BibitemOpen
  \bibfield  {author} {\bibinfo {author} {\bibfnamefont {E.~J.}\ \bibnamefont
  {Copeland}}, \bibinfo {author} {\bibfnamefont {P.~M.}\ \bibnamefont
  {Saffin}}, \ and\ \bibinfo {author} {\bibfnamefont {S.-Y.}\ \bibnamefont
  {Zhou}},\ }\href {\doibase 10.1103/PhysRevLett.113.231603} {\bibfield
  {journal} {\bibinfo  {journal} {Phys. Rev. Lett.}\ }\textbf {\bibinfo
  {volume} {113}},\ \bibinfo {pages} {231603} (\bibinfo {year} {2014})},\
  \Eprint {http://arxiv.org/abs/1409.3232} {arXiv:1409.3232 [hep-th]}
  \BibitemShut {NoStop}%
\bibitem [{\citenamefont {Xie}\ \emph {et~al.}(2021)\citenamefont {Xie},
  \citenamefont {Saffin},\ and\ \citenamefont {Zhou}}]{Xie:2021glp}%
  \BibitemOpen
  \bibfield  {author} {\bibinfo {author} {\bibfnamefont {Q.-X.}\ \bibnamefont
  {Xie}}, \bibinfo {author} {\bibfnamefont {P.~M.}\ \bibnamefont {Saffin}}, \
  and\ \bibinfo {author} {\bibfnamefont {S.-Y.}\ \bibnamefont {Zhou}},\ }\href
  {\doibase 10.1007/JHEP07(2021)062} {\bibfield  {journal} {\bibinfo  {journal}
  {JHEP}\ }\textbf {\bibinfo {volume} {07}},\ \bibinfo {pages} {062} (\bibinfo
  {year} {2021})},\ \Eprint {http://arxiv.org/abs/2101.06988} {arXiv:2101.06988
  [hep-th]} \BibitemShut {NoStop}%
\bibitem [{\citenamefont {Hou}\ \emph {et~al.}(2022)\citenamefont {Hou},
  \citenamefont {Saffin}, \citenamefont {Xie},\ and\ \citenamefont
  {Zhou}}]{Hou:2022jcd}%
  \BibitemOpen
  \bibfield  {author} {\bibinfo {author} {\bibfnamefont {S.-Y.}\ \bibnamefont
  {Hou}}, \bibinfo {author} {\bibfnamefont {P.~M.}\ \bibnamefont {Saffin}},
  \bibinfo {author} {\bibfnamefont {Q.-X.}\ \bibnamefont {Xie}}, \ and\
  \bibinfo {author} {\bibfnamefont {S.-Y.}\ \bibnamefont {Zhou}},\ }\href
  {\doibase 10.1007/JHEP07(2022)060} {\bibfield  {journal} {\bibinfo  {journal}
  {JHEP}\ }\textbf {\bibinfo {volume} {07}},\ \bibinfo {pages} {060} (\bibinfo
  {year} {2022})},\ \Eprint {http://arxiv.org/abs/2202.08392} {arXiv:2202.08392
  [hep-ph]} \BibitemShut {NoStop}%
\bibitem [{\citenamefont {Xie}\ \emph {et~al.}(2024)\citenamefont {Xie},
  \citenamefont {Saffin}, \citenamefont {Tranberg},\ and\ \citenamefont
  {Zhou}}]{Xie:2023psz}%
  \BibitemOpen
  \bibfield  {author} {\bibinfo {author} {\bibfnamefont {Q.-X.}\ \bibnamefont
  {Xie}}, \bibinfo {author} {\bibfnamefont {P.~M.}\ \bibnamefont {Saffin}},
  \bibinfo {author} {\bibfnamefont {A.}~\bibnamefont {Tranberg}}, \ and\
  \bibinfo {author} {\bibfnamefont {S.-Y.}\ \bibnamefont {Zhou}},\ }\href
  {\doibase 10.1007/JHEP01(2024)165} {\bibfield  {journal} {\bibinfo  {journal}
  {JHEP}\ }\textbf {\bibinfo {volume} {01}},\ \bibinfo {pages} {165} (\bibinfo
  {year} {2024})},\ \Eprint {http://arxiv.org/abs/2312.01139} {arXiv:2312.01139
  [hep-th]} \BibitemShut {NoStop}%
\bibitem [{\citenamefont {Cohen}\ \emph {et~al.}(1986)\citenamefont {Cohen},
  \citenamefont {Coleman}, \citenamefont {Georgi},\ and\ \citenamefont
  {Manohar}}]{Cohen:1986ct}%
  \BibitemOpen
  \bibfield  {author} {\bibinfo {author} {\bibfnamefont {A.~G.}\ \bibnamefont
  {Cohen}}, \bibinfo {author} {\bibfnamefont {S.~R.}\ \bibnamefont {Coleman}},
  \bibinfo {author} {\bibfnamefont {H.}~\bibnamefont {Georgi}}, \ and\ \bibinfo
  {author} {\bibfnamefont {A.}~\bibnamefont {Manohar}},\ }\href {\doibase
  10.1016/0550-3213(86)90004-0} {\bibfield  {journal} {\bibinfo  {journal}
  {Nucl. Phys. B}\ }\textbf {\bibinfo {volume} {272}},\ \bibinfo {pages} {301}
  (\bibinfo {year} {1986})}\BibitemShut {NoStop}%
\bibitem [{\citenamefont {Tranberg}\ and\ \citenamefont
  {Weir}(2014)}]{Tranberg:2013cka}%
  \BibitemOpen
  \bibfield  {author} {\bibinfo {author} {\bibfnamefont {A.}~\bibnamefont
  {Tranberg}}\ and\ \bibinfo {author} {\bibfnamefont {D.~J.}\ \bibnamefont
  {Weir}},\ }\href {\doibase 10.1007/JHEP04(2014)184} {\bibfield  {journal}
  {\bibinfo  {journal} {JHEP}\ }\textbf {\bibinfo {volume} {04}},\ \bibinfo
  {pages} {184} (\bibinfo {year} {2014})},\ \Eprint
  {http://arxiv.org/abs/1310.7487} {arXiv:1310.7487 [hep-ph]} \BibitemShut
  {NoStop}%
\bibitem [{\citenamefont {Kovtun}\ and\ \citenamefont
  {Zantedeschi}(2022)}]{Kovtun:2020udn}%
  \BibitemOpen
  \bibfield  {author} {\bibinfo {author} {\bibfnamefont {A.}~\bibnamefont
  {Kovtun}}\ and\ \bibinfo {author} {\bibfnamefont {M.}~\bibnamefont
  {Zantedeschi}},\ }\href {\doibase 10.1103/PhysRevD.105.085019} {\bibfield
  {journal} {\bibinfo  {journal} {Phys. Rev. D}\ }\textbf {\bibinfo {volume}
  {105}},\ \bibinfo {pages} {085019} (\bibinfo {year} {2022})},\ \Eprint
  {http://arxiv.org/abs/2008.02187} {arXiv:2008.02187 [hep-th]} \BibitemShut
  {NoStop}%
\bibitem [{\citenamefont {Alonso-Izquierdo}\ \emph {et~al.}(2024)\citenamefont
  {Alonso-Izquierdo}, \citenamefont {Martinez}, \citenamefont {Sanchez},\ and\
  \citenamefont {Gonzalez~Leon}}]{Alonso-Izquierdo:2023hrr}%
  \BibitemOpen
  \bibfield  {author} {\bibinfo {author} {\bibfnamefont {A.}~\bibnamefont
  {Alonso-Izquierdo}}, \bibinfo {author} {\bibfnamefont {D.~C.}\ \bibnamefont
  {Martinez}}, \bibinfo {author} {\bibfnamefont {C.~G.}\ \bibnamefont
  {Sanchez}}, \ and\ \bibinfo {author} {\bibfnamefont {M.~A.}\ \bibnamefont
  {Gonzalez~Leon}},\ }\href {\doibase 10.1016/j.chaos.2024.114732} {\bibfield
  {journal} {\bibinfo  {journal} {Chaos Solitons Fractals}\ }\textbf {\bibinfo
  {volume} {181}},\ \bibinfo {pages} {114732} (\bibinfo {year} {2024})},\
  \Eprint {http://arxiv.org/abs/2311.12728} {arXiv:2311.12728 [hep-th]}
  \BibitemShut {NoStop}%
\bibitem [{\citenamefont {Friedberg}\ \emph
  {et~al.}(1976{\natexlab{b}})\citenamefont {Friedberg}, \citenamefont {Lee},\
  and\ \citenamefont {Sirlin}}]{Friedberg:1976az}%
  \BibitemOpen
  \bibfield  {author} {\bibinfo {author} {\bibfnamefont {R.}~\bibnamefont
  {Friedberg}}, \bibinfo {author} {\bibfnamefont {T.~D.}\ \bibnamefont {Lee}},
  \ and\ \bibinfo {author} {\bibfnamefont {A.}~\bibnamefont {Sirlin}},\ }\href
  {\doibase 10.1016/0550-3213(76)90274-1} {\bibfield  {journal} {\bibinfo
  {journal} {Nucl. Phys. B}\ }\textbf {\bibinfo {volume} {115}},\ \bibinfo
  {pages} {1} (\bibinfo {year} {1976}{\natexlab{b}})}\BibitemShut {NoStop}%
\bibitem [{\citenamefont {Friedberg}\ \emph
  {et~al.}(1976{\natexlab{c}})\citenamefont {Friedberg}, \citenamefont {Lee},\
  and\ \citenamefont {Sirlin}}]{Friedberg:1976ay}%
  \BibitemOpen
  \bibfield  {author} {\bibinfo {author} {\bibfnamefont {R.}~\bibnamefont
  {Friedberg}}, \bibinfo {author} {\bibfnamefont {T.~D.}\ \bibnamefont {Lee}},
  \ and\ \bibinfo {author} {\bibfnamefont {A.}~\bibnamefont {Sirlin}},\ }\href
  {\doibase 10.1016/0550-3213(76)90275-3} {\bibfield  {journal} {\bibinfo
  {journal} {Nucl. Phys. B}\ }\textbf {\bibinfo {volume} {115}},\ \bibinfo
  {pages} {32} (\bibinfo {year} {1976}{\natexlab{c}})}\BibitemShut {NoStop}%
\bibitem [{\citenamefont {Lee}\ \emph {et~al.}(1989)\citenamefont {Lee},
  \citenamefont {Stein-Schabes}, \citenamefont {Watkins},\ and\ \citenamefont
  {Widrow}}]{Lee:1988ag}%
  \BibitemOpen
  \bibfield  {author} {\bibinfo {author} {\bibfnamefont {K.-M.}\ \bibnamefont
  {Lee}}, \bibinfo {author} {\bibfnamefont {J.~A.}\ \bibnamefont
  {Stein-Schabes}}, \bibinfo {author} {\bibfnamefont {R.}~\bibnamefont
  {Watkins}}, \ and\ \bibinfo {author} {\bibfnamefont {L.~M.}\ \bibnamefont
  {Widrow}},\ }\href {\doibase 10.1103/PhysRevD.39.1665} {\bibfield  {journal}
  {\bibinfo  {journal} {Phys. Rev. D}\ }\textbf {\bibinfo {volume} {39}},\
  \bibinfo {pages} {1665} (\bibinfo {year} {1989})}\BibitemShut {NoStop}%
\bibitem [{\citenamefont {Kusenko}\ \emph {et~al.}(1998)\citenamefont
  {Kusenko}, \citenamefont {Shaposhnikov},\ and\ \citenamefont
  {Tinyakov}}]{Kusenko:1997vi}%
  \BibitemOpen
  \bibfield  {author} {\bibinfo {author} {\bibfnamefont {A.}~\bibnamefont
  {Kusenko}}, \bibinfo {author} {\bibfnamefont {M.~E.}\ \bibnamefont
  {Shaposhnikov}}, \ and\ \bibinfo {author} {\bibfnamefont {P.~G.}\
  \bibnamefont {Tinyakov}},\ }\href {\doibase 10.1134/1.567658} {\bibfield
  {journal} {\bibinfo  {journal} {Pisma Zh. Eksp. Teor. Fiz.}\ }\textbf
  {\bibinfo {volume} {67}},\ \bibinfo {pages} {229} (\bibinfo {year} {1998})},\
  \Eprint {http://arxiv.org/abs/hep-th/9801041} {arXiv:hep-th/9801041}
  \BibitemShut {NoStop}%
\bibitem [{\citenamefont {Benci}\ and\ \citenamefont
  {Fortunato}(2011)}]{Benci:2010cs}%
  \BibitemOpen
  \bibfield  {author} {\bibinfo {author} {\bibfnamefont {V.}~\bibnamefont
  {Benci}}\ and\ \bibinfo {author} {\bibfnamefont {D.}~\bibnamefont
  {Fortunato}},\ }\href {\doibase 10.1063/1.3629848} {\bibfield  {journal}
  {\bibinfo  {journal} {J. Math. Phys.}\ }\textbf {\bibinfo {volume} {52}},\
  \bibinfo {pages} {093701} (\bibinfo {year} {2011})},\ \Eprint
  {http://arxiv.org/abs/1011.5044} {arXiv:1011.5044 [math-ph]} \BibitemShut
  {NoStop}%
\bibitem [{\citenamefont {Gulamov}\ \emph {et~al.}(2014)\citenamefont
  {Gulamov}, \citenamefont {Nugaev},\ and\ \citenamefont
  {Smolyakov}}]{Gulamov:2013cra}%
  \BibitemOpen
  \bibfield  {author} {\bibinfo {author} {\bibfnamefont {I.~E.}\ \bibnamefont
  {Gulamov}}, \bibinfo {author} {\bibfnamefont {E.~Y.}\ \bibnamefont {Nugaev}},
  \ and\ \bibinfo {author} {\bibfnamefont {M.~N.}\ \bibnamefont {Smolyakov}},\
  }\href {\doibase 10.1103/PhysRevD.89.085006} {\bibfield  {journal} {\bibinfo
  {journal} {Phys. Rev. D}\ }\textbf {\bibinfo {volume} {89}},\ \bibinfo
  {pages} {085006} (\bibinfo {year} {2014})},\ \Eprint
  {http://arxiv.org/abs/1311.0325} {arXiv:1311.0325 [hep-th]} \BibitemShut
  {NoStop}%
\bibitem [{\citenamefont {Gulamov}\ \emph {et~al.}(2015)\citenamefont
  {Gulamov}, \citenamefont {Nugaev}, \citenamefont {Panin},\ and\ \citenamefont
  {Smolyakov}}]{Gulamov:2015fya}%
  \BibitemOpen
  \bibfield  {author} {\bibinfo {author} {\bibfnamefont {I.~E.}\ \bibnamefont
  {Gulamov}}, \bibinfo {author} {\bibfnamefont {E.~Y.}\ \bibnamefont {Nugaev}},
  \bibinfo {author} {\bibfnamefont {A.~G.}\ \bibnamefont {Panin}}, \ and\
  \bibinfo {author} {\bibfnamefont {M.~N.}\ \bibnamefont {Smolyakov}},\ }\href
  {\doibase 10.1103/PhysRevD.92.045011} {\bibfield  {journal} {\bibinfo
  {journal} {Phys. Rev. D}\ }\textbf {\bibinfo {volume} {92}},\ \bibinfo
  {pages} {045011} (\bibinfo {year} {2015})},\ \Eprint
  {http://arxiv.org/abs/1506.05786} {arXiv:1506.05786 [hep-th]} \BibitemShut
  {NoStop}%
\bibitem [{\citenamefont {Loiko}\ and\ \citenamefont
  {Shnir}(2019)}]{Loiko:2019gwk}%
  \BibitemOpen
  \bibfield  {author} {\bibinfo {author} {\bibfnamefont {V.}~\bibnamefont
  {Loiko}}\ and\ \bibinfo {author} {\bibfnamefont {Y.}~\bibnamefont {Shnir}},\
  }\href {\doibase 10.1016/j.physletb.2019.134810} {\bibfield  {journal}
  {\bibinfo  {journal} {Phys. Lett. B}\ }\textbf {\bibinfo {volume} {797}},\
  \bibinfo {pages} {134810} (\bibinfo {year} {2019})},\ \Eprint
  {http://arxiv.org/abs/1906.01943} {arXiv:1906.01943 [hep-th]} \BibitemShut
  {NoStop}%
\bibitem [{\citenamefont {Nugaev}\ and\ \citenamefont
  {Shkerin}(2020)}]{Nugaev:2019vru}%
  \BibitemOpen
  \bibfield  {author} {\bibinfo {author} {\bibfnamefont {E.~Y.}\ \bibnamefont
  {Nugaev}}\ and\ \bibinfo {author} {\bibfnamefont {A.~V.}\ \bibnamefont
  {Shkerin}},\ }\href {\doibase 10.1134/S1063776120020077} {\bibfield
  {journal} {\bibinfo  {journal} {J. Exp. Theor. Phys.}\ }\textbf {\bibinfo
  {volume} {130}},\ \bibinfo {pages} {301} (\bibinfo {year} {2020})},\ \Eprint
  {http://arxiv.org/abs/1905.05146} {arXiv:1905.05146 [hep-th]} \BibitemShut
  {NoStop}%
\bibitem [{\citenamefont {Heeck}\ \emph
  {et~al.}(2021{\natexlab{a}})\citenamefont {Heeck}, \citenamefont {Rajaraman},
  \citenamefont {Riley},\ and\ \citenamefont {Verhaaren}}]{Heeck:2021zvk}%
  \BibitemOpen
  \bibfield  {author} {\bibinfo {author} {\bibfnamefont {J.}~\bibnamefont
  {Heeck}}, \bibinfo {author} {\bibfnamefont {A.}~\bibnamefont {Rajaraman}},
  \bibinfo {author} {\bibfnamefont {R.}~\bibnamefont {Riley}}, \ and\ \bibinfo
  {author} {\bibfnamefont {C.~B.}\ \bibnamefont {Verhaaren}},\ }\href {\doibase
  10.1103/PhysRevD.103.116004} {\bibfield  {journal} {\bibinfo  {journal}
  {Phys. Rev. D}\ }\textbf {\bibinfo {volume} {103}},\ \bibinfo {pages}
  {116004} (\bibinfo {year} {2021}{\natexlab{a}})},\ \Eprint
  {http://arxiv.org/abs/2103.06905} {arXiv:2103.06905 [hep-th]} \BibitemShut
  {NoStop}%
\bibitem [{\citenamefont {Heeck}\ \emph
  {et~al.}(2021{\natexlab{b}})\citenamefont {Heeck}, \citenamefont {Rajaraman},
  \citenamefont {Riley},\ and\ \citenamefont {Verhaaren}}]{Heeck:2021bce}%
  \BibitemOpen
  \bibfield  {author} {\bibinfo {author} {\bibfnamefont {J.}~\bibnamefont
  {Heeck}}, \bibinfo {author} {\bibfnamefont {A.}~\bibnamefont {Rajaraman}},
  \bibinfo {author} {\bibfnamefont {R.}~\bibnamefont {Riley}}, \ and\ \bibinfo
  {author} {\bibfnamefont {C.~B.}\ \bibnamefont {Verhaaren}},\ }\href {\doibase
  10.1007/JHEP10(2021)103} {\bibfield  {journal} {\bibinfo  {journal} {JHEP}\
  }\textbf {\bibinfo {volume} {10}},\ \bibinfo {pages} {103} (\bibinfo {year}
  {2021}{\natexlab{b}})},\ \Eprint {http://arxiv.org/abs/2107.10280}
  {arXiv:2107.10280 [hep-th]} \BibitemShut {NoStop}%
\bibitem [{\citenamefont {Kinach}\ and\ \citenamefont
  {Choptuik}(2024)}]{Kinach:2024qzc}%
  \BibitemOpen
  \bibfield  {author} {\bibinfo {author} {\bibfnamefont {M.~P.}\ \bibnamefont
  {Kinach}}\ and\ \bibinfo {author} {\bibfnamefont {M.~W.}\ \bibnamefont
  {Choptuik}},\ }\href {\doibase 10.1103/PhysRevD.110.075033} {\bibfield
  {journal} {\bibinfo  {journal} {Phys. Rev. D}\ }\textbf {\bibinfo {volume}
  {110}},\ \bibinfo {pages} {075033} (\bibinfo {year} {2024})},\ \Eprint
  {http://arxiv.org/abs/2408.07561} {arXiv:2408.07561 [hep-th]} \BibitemShut
  {NoStop}%
\bibitem [{\citenamefont {Kusenko}\ and\ \citenamefont
  {Shaposhnikov}(1998)}]{Kusenko:1997si}%
  \BibitemOpen
  \bibfield  {author} {\bibinfo {author} {\bibfnamefont {A.}~\bibnamefont
  {Kusenko}}\ and\ \bibinfo {author} {\bibfnamefont {M.~E.}\ \bibnamefont
  {Shaposhnikov}},\ }\href {\doibase 10.1016/S0370-2693(97)01375-0} {\bibfield
  {journal} {\bibinfo  {journal} {Phys. Lett. B}\ }\textbf {\bibinfo {volume}
  {418}},\ \bibinfo {pages} {46} (\bibinfo {year} {1998})},\ \Eprint
  {http://arxiv.org/abs/hep-ph/9709492} {arXiv:hep-ph/9709492} \BibitemShut
  {NoStop}%
\bibitem [{\citenamefont {Enqvist}\ and\ \citenamefont
  {McDonald}(1998)}]{Enqvist:1997si}%
  \BibitemOpen
  \bibfield  {author} {\bibinfo {author} {\bibfnamefont {K.}~\bibnamefont
  {Enqvist}}\ and\ \bibinfo {author} {\bibfnamefont {J.}~\bibnamefont
  {McDonald}},\ }\href {\doibase 10.1016/S0370-2693(98)00271-8} {\bibfield
  {journal} {\bibinfo  {journal} {Phys. Lett. B}\ }\textbf {\bibinfo {volume}
  {425}},\ \bibinfo {pages} {309} (\bibinfo {year} {1998})},\ \Eprint
  {http://arxiv.org/abs/hep-ph/9711514} {arXiv:hep-ph/9711514} \BibitemShut
  {NoStop}%
\bibitem [{\citenamefont {Fujii}\ and\ \citenamefont
  {Hamaguchi}(2002)}]{Fujii:2002kr}%
  \BibitemOpen
  \bibfield  {author} {\bibinfo {author} {\bibfnamefont {M.}~\bibnamefont
  {Fujii}}\ and\ \bibinfo {author} {\bibfnamefont {K.}~\bibnamefont
  {Hamaguchi}},\ }\href {\doibase 10.1103/PhysRevD.66.083501} {\bibfield
  {journal} {\bibinfo  {journal} {Phys. Rev. D}\ }\textbf {\bibinfo {volume}
  {66}},\ \bibinfo {pages} {083501} (\bibinfo {year} {2002})},\ \Eprint
  {http://arxiv.org/abs/hep-ph/0205044} {arXiv:hep-ph/0205044} \BibitemShut
  {NoStop}%
\bibitem [{\citenamefont {Enqvist}\ and\ \citenamefont
  {Mazumdar}(2003)}]{Enqvist:2003gh}%
  \BibitemOpen
  \bibfield  {author} {\bibinfo {author} {\bibfnamefont {K.}~\bibnamefont
  {Enqvist}}\ and\ \bibinfo {author} {\bibfnamefont {A.}~\bibnamefont
  {Mazumdar}},\ }\href {\doibase 10.1016/S0370-1573(03)00119-4} {\bibfield
  {journal} {\bibinfo  {journal} {Phys. Rept.}\ }\textbf {\bibinfo {volume}
  {380}},\ \bibinfo {pages} {99} (\bibinfo {year} {2003})},\ \Eprint
  {http://arxiv.org/abs/hep-ph/0209244} {arXiv:hep-ph/0209244} \BibitemShut
  {NoStop}%
\bibitem [{\citenamefont {Roszkowski}\ and\ \citenamefont
  {Seto}(2007)}]{Roszkowski:2006kw}%
  \BibitemOpen
  \bibfield  {author} {\bibinfo {author} {\bibfnamefont {L.}~\bibnamefont
  {Roszkowski}}\ and\ \bibinfo {author} {\bibfnamefont {O.}~\bibnamefont
  {Seto}},\ }\href {\doibase 10.1103/PhysRevLett.98.161304} {\bibfield
  {journal} {\bibinfo  {journal} {Phys. Rev. Lett.}\ }\textbf {\bibinfo
  {volume} {98}},\ \bibinfo {pages} {161304} (\bibinfo {year} {2007})},\
  \Eprint {http://arxiv.org/abs/hep-ph/0608013} {arXiv:hep-ph/0608013}
  \BibitemShut {NoStop}%
\bibitem [{\citenamefont {Shoemaker}\ and\ \citenamefont
  {Kusenko}(2009)}]{Shoemaker:2009kg}%
  \BibitemOpen
  \bibfield  {author} {\bibinfo {author} {\bibfnamefont {I.~M.}\ \bibnamefont
  {Shoemaker}}\ and\ \bibinfo {author} {\bibfnamefont {A.}~\bibnamefont
  {Kusenko}},\ }\href {\doibase 10.1103/PhysRevD.80.075021} {\bibfield
  {journal} {\bibinfo  {journal} {Phys. Rev. D}\ }\textbf {\bibinfo {volume}
  {80}},\ \bibinfo {pages} {075021} (\bibinfo {year} {2009})},\ \Eprint
  {http://arxiv.org/abs/0909.3334} {arXiv:0909.3334 [hep-ph]} \BibitemShut
  {NoStop}%
\bibitem [{\citenamefont {Zhou}(2015)}]{Zhou:2015yfa}%
  \BibitemOpen
  \bibfield  {author} {\bibinfo {author} {\bibfnamefont {S.-Y.}\ \bibnamefont
  {Zhou}},\ }\href {\doibase 10.1088/1475-7516/2015/06/033} {\bibfield
  {journal} {\bibinfo  {journal} {JCAP}\ }\textbf {\bibinfo {volume} {06}},\
  \bibinfo {pages} {033} (\bibinfo {year} {2015})},\ \Eprint
  {http://arxiv.org/abs/1501.01217} {arXiv:1501.01217 [astro-ph.CO]}
  \BibitemShut {NoStop}%
\bibitem [{\citenamefont {Kawasaki}\ and\ \citenamefont
  {Nakatsuka}(2020)}]{Kawasaki:2019ywz}%
  \BibitemOpen
  \bibfield  {author} {\bibinfo {author} {\bibfnamefont {M.}~\bibnamefont
  {Kawasaki}}\ and\ \bibinfo {author} {\bibfnamefont {H.}~\bibnamefont
  {Nakatsuka}},\ }\href {\doibase 10.1088/1475-7516/2020/04/017} {\bibfield
  {journal} {\bibinfo  {journal} {JCAP}\ }\textbf {\bibinfo {volume} {04}},\
  \bibinfo {pages} {017} (\bibinfo {year} {2020})},\ \Eprint
  {http://arxiv.org/abs/1912.06993} {arXiv:1912.06993 [hep-ph]} \BibitemShut
  {NoStop}%
\bibitem [{\citenamefont {Gouttenoire}\ \emph {et~al.}(2021)\citenamefont
  {Gouttenoire}, \citenamefont {Servant},\ and\ \citenamefont
  {Simakachorn}}]{Gouttenoire:2021jhk}%
  \BibitemOpen
  \bibfield  {author} {\bibinfo {author} {\bibfnamefont {Y.}~\bibnamefont
  {Gouttenoire}}, \bibinfo {author} {\bibfnamefont {G.}~\bibnamefont
  {Servant}}, \ and\ \bibinfo {author} {\bibfnamefont {P.}~\bibnamefont
  {Simakachorn}},\ }\href@noop {} {\  (\bibinfo {year} {2021})},\ \Eprint
  {http://arxiv.org/abs/2111.01150} {arXiv:2111.01150 [hep-ph]} \BibitemShut
  {NoStop}%
\bibitem [{\citenamefont {Kasai}\ \emph {et~al.}(2022)\citenamefont {Kasai},
  \citenamefont {Kawasaki},\ and\ \citenamefont {Murai}}]{Kasai:2022vhq}%
  \BibitemOpen
  \bibfield  {author} {\bibinfo {author} {\bibfnamefont {K.}~\bibnamefont
  {Kasai}}, \bibinfo {author} {\bibfnamefont {M.}~\bibnamefont {Kawasaki}}, \
  and\ \bibinfo {author} {\bibfnamefont {K.}~\bibnamefont {Murai}},\ }\href
  {\doibase 10.1088/1475-7516/2022/10/048} {\bibfield  {journal} {\bibinfo
  {journal} {JCAP}\ }\textbf {\bibinfo {volume} {10}},\ \bibinfo {pages} {048}
  (\bibinfo {year} {2022})},\ \Eprint {http://arxiv.org/abs/2205.10148}
  {arXiv:2205.10148 [astro-ph.CO]} \BibitemShut {NoStop}%
\bibitem [{\citenamefont {El~Bourakadi}\ \emph {et~al.}(2023)\citenamefont
  {El~Bourakadi}, \citenamefont {Ferricha-Alami}, \citenamefont {Sakhi},
  \citenamefont {Bennai},\ and\ \citenamefont {Chakir}}]{ElBourakadi:2023pue}%
  \BibitemOpen
  \bibfield  {author} {\bibinfo {author} {\bibfnamefont {K.}~\bibnamefont
  {El~Bourakadi}}, \bibinfo {author} {\bibfnamefont {M.}~\bibnamefont
  {Ferricha-Alami}}, \bibinfo {author} {\bibfnamefont {Z.}~\bibnamefont
  {Sakhi}}, \bibinfo {author} {\bibfnamefont {M.}~\bibnamefont {Bennai}}, \
  and\ \bibinfo {author} {\bibfnamefont {H.}~\bibnamefont {Chakir}},\
  }\href@noop {} {\  (\bibinfo {year} {2023})},\ \Eprint
  {http://arxiv.org/abs/2307.15541} {arXiv:2307.15541 [hep-ph]} \BibitemShut
  {NoStop}%
\bibitem [{\citenamefont {Blaschke}\ \emph {et~al.}(2025)\citenamefont
  {Blaschke}, \citenamefont {Roma\'nczukiewicz}, \citenamefont
  {S\l{}awi\'nska},\ and\ \citenamefont {Wereszczy\'nski}}]{Blaschke:2024dlt}%
  \BibitemOpen
  \bibfield  {author} {\bibinfo {author} {\bibfnamefont {F.}~\bibnamefont
  {Blaschke}}, \bibinfo {author} {\bibfnamefont {T.}~\bibnamefont
  {Roma\'nczukiewicz}}, \bibinfo {author} {\bibfnamefont {K.}~\bibnamefont
  {S\l{}awi\'nska}}, \ and\ \bibinfo {author} {\bibfnamefont {A.}~\bibnamefont
  {Wereszczy\'nski}},\ }\href {\doibase 10.1103/PhysRevLett.134.081601}
  {\bibfield  {journal} {\bibinfo  {journal} {Phys. Rev. Lett.}\ }\textbf
  {\bibinfo {volume} {134}},\ \bibinfo {pages} {081601} (\bibinfo {year}
  {2025})},\ \Eprint {http://arxiv.org/abs/2410.24109} {arXiv:2410.24109
  [hep-th]} \BibitemShut {NoStop}%
\bibitem [{\citenamefont {Saffin}\ \emph {et~al.}(2023)\citenamefont {Saffin},
  \citenamefont {Xie},\ and\ \citenamefont {Zhou}}]{Saffin:2022tub}%
  \BibitemOpen
  \bibfield  {author} {\bibinfo {author} {\bibfnamefont {P.~M.}\ \bibnamefont
  {Saffin}}, \bibinfo {author} {\bibfnamefont {Q.-X.}\ \bibnamefont {Xie}}, \
  and\ \bibinfo {author} {\bibfnamefont {S.-Y.}\ \bibnamefont {Zhou}},\ }\href
  {\doibase 10.1103/PhysRevLett.131.111601} {\bibfield  {journal} {\bibinfo
  {journal} {Phys. Rev. Lett.}\ }\textbf {\bibinfo {volume} {131}},\ \bibinfo
  {pages} {111601} (\bibinfo {year} {2023})},\ \Eprint
  {http://arxiv.org/abs/2212.03269} {arXiv:2212.03269 [hep-th]} \BibitemShut
  {NoStop}%
\bibitem [{\citenamefont {Dicke}(1954)}]{Dicke:1954zz}%
  \BibitemOpen
  \bibfield  {author} {\bibinfo {author} {\bibfnamefont {R.~H.}\ \bibnamefont
  {Dicke}},\ }\href {\doibase 10.1103/PhysRev.93.99} {\bibfield  {journal}
  {\bibinfo  {journal} {Phys. Rev.}\ }\textbf {\bibinfo {volume} {93}},\
  \bibinfo {pages} {99} (\bibinfo {year} {1954})}\BibitemShut {NoStop}%
\bibitem [{\citenamefont {Bekenstein}\ and\ \citenamefont
  {Schiffer}(1998)}]{Bekenstein:1998nt}%
  \BibitemOpen
  \bibfield  {author} {\bibinfo {author} {\bibfnamefont {J.~D.}\ \bibnamefont
  {Bekenstein}}\ and\ \bibinfo {author} {\bibfnamefont {M.}~\bibnamefont
  {Schiffer}},\ }\href {\doibase 10.1103/PhysRevD.58.064014} {\bibfield
  {journal} {\bibinfo  {journal} {Phys. Rev. D}\ }\textbf {\bibinfo {volume}
  {58}},\ \bibinfo {pages} {064014} (\bibinfo {year} {1998})},\ \Eprint
  {http://arxiv.org/abs/gr-qc/9803033} {arXiv:gr-qc/9803033} \BibitemShut
  {NoStop}%
\bibitem [{\citenamefont {Brito}\ \emph
  {et~al.}(2015{\natexlab{a}})\citenamefont {Brito}, \citenamefont {Cardoso},\
  and\ \citenamefont {Pani}}]{Brito:2015oca}%
  \BibitemOpen
  \bibfield  {author} {\bibinfo {author} {\bibfnamefont {R.}~\bibnamefont
  {Brito}}, \bibinfo {author} {\bibfnamefont {V.}~\bibnamefont {Cardoso}}, \
  and\ \bibinfo {author} {\bibfnamefont {P.}~\bibnamefont {Pani}},\ }\href
  {\doibase 10.1007/978-3-319-19000-6} {\bibfield  {journal} {\bibinfo
  {journal} {Lect. Notes Phys.}\ }\textbf {\bibinfo {volume} {906}},\ \bibinfo
  {pages} {pp.1} (\bibinfo {year} {2015}{\natexlab{a}})},\ \Eprint
  {http://arxiv.org/abs/1501.06570} {arXiv:1501.06570 [gr-qc]} \BibitemShut
  {NoStop}%
\bibitem [{\citenamefont {Teukolsky}\ and\ \citenamefont
  {Press}(1974)}]{Teukolsky:1974yv}%
  \BibitemOpen
  \bibfield  {author} {\bibinfo {author} {\bibfnamefont {S.~A.}\ \bibnamefont
  {Teukolsky}}\ and\ \bibinfo {author} {\bibfnamefont {W.~H.}\ \bibnamefont
  {Press}},\ }\href {\doibase 10.1086/153180} {\bibfield  {journal} {\bibinfo
  {journal} {Astrophys. J.}\ }\textbf {\bibinfo {volume} {193}},\ \bibinfo
  {pages} {443} (\bibinfo {year} {1974})}\BibitemShut {NoStop}%
\bibitem [{\citenamefont {Cardoso}\ and\ \citenamefont
  {Dias}(2004)}]{Cardoso:2004hs}%
  \BibitemOpen
  \bibfield  {author} {\bibinfo {author} {\bibfnamefont {V.}~\bibnamefont
  {Cardoso}}\ and\ \bibinfo {author} {\bibfnamefont {O.~J.~C.}\ \bibnamefont
  {Dias}},\ }\href {\doibase 10.1103/PhysRevD.70.084011} {\bibfield  {journal}
  {\bibinfo  {journal} {Phys. Rev. D}\ }\textbf {\bibinfo {volume} {70}},\
  \bibinfo {pages} {084011} (\bibinfo {year} {2004})},\ \Eprint
  {http://arxiv.org/abs/hep-th/0405006} {arXiv:hep-th/0405006} \BibitemShut
  {NoStop}%
\bibitem [{\citenamefont {Dolan}(2007)}]{Dolan:2007mj}%
  \BibitemOpen
  \bibfield  {author} {\bibinfo {author} {\bibfnamefont {S.~R.}\ \bibnamefont
  {Dolan}},\ }\href {\doibase 10.1103/PhysRevD.76.084001} {\bibfield  {journal}
  {\bibinfo  {journal} {Phys. Rev. D}\ }\textbf {\bibinfo {volume} {76}},\
  \bibinfo {pages} {084001} (\bibinfo {year} {2007})},\ \Eprint
  {http://arxiv.org/abs/0705.2880} {arXiv:0705.2880 [gr-qc]} \BibitemShut
  {NoStop}%
\bibitem [{\citenamefont {Arvanitaki}\ \emph {et~al.}(2010)\citenamefont
  {Arvanitaki}, \citenamefont {Dimopoulos}, \citenamefont {Dubovsky},
  \citenamefont {Kaloper},\ and\ \citenamefont
  {March-Russell}}]{Arvanitaki:2009fg}%
  \BibitemOpen
  \bibfield  {author} {\bibinfo {author} {\bibfnamefont {A.}~\bibnamefont
  {Arvanitaki}}, \bibinfo {author} {\bibfnamefont {S.}~\bibnamefont
  {Dimopoulos}}, \bibinfo {author} {\bibfnamefont {S.}~\bibnamefont
  {Dubovsky}}, \bibinfo {author} {\bibfnamefont {N.}~\bibnamefont {Kaloper}}, \
  and\ \bibinfo {author} {\bibfnamefont {J.}~\bibnamefont {March-Russell}},\
  }\href {\doibase 10.1103/PhysRevD.81.123530} {\bibfield  {journal} {\bibinfo
  {journal} {Phys. Rev. D}\ }\textbf {\bibinfo {volume} {81}},\ \bibinfo
  {pages} {123530} (\bibinfo {year} {2010})},\ \Eprint
  {http://arxiv.org/abs/0905.4720} {arXiv:0905.4720 [hep-th]} \BibitemShut
  {NoStop}%
\bibitem [{\citenamefont {Bredberg}\ \emph {et~al.}(2010)\citenamefont
  {Bredberg}, \citenamefont {Hartman}, \citenamefont {Song},\ and\
  \citenamefont {Strominger}}]{Bredberg:2009pv}%
  \BibitemOpen
  \bibfield  {author} {\bibinfo {author} {\bibfnamefont {I.}~\bibnamefont
  {Bredberg}}, \bibinfo {author} {\bibfnamefont {T.}~\bibnamefont {Hartman}},
  \bibinfo {author} {\bibfnamefont {W.}~\bibnamefont {Song}}, \ and\ \bibinfo
  {author} {\bibfnamefont {A.}~\bibnamefont {Strominger}},\ }\href {\doibase
  10.1007/JHEP04(2010)019} {\bibfield  {journal} {\bibinfo  {journal} {JHEP}\
  }\textbf {\bibinfo {volume} {04}},\ \bibinfo {pages} {019} (\bibinfo {year}
  {2010})},\ \Eprint {http://arxiv.org/abs/0907.3477} {arXiv:0907.3477
  [hep-th]} \BibitemShut {NoStop}%
\bibitem [{\citenamefont {Arvanitaki}\ and\ \citenamefont
  {Dubovsky}(2011)}]{Arvanitaki:2010sy}%
  \BibitemOpen
  \bibfield  {author} {\bibinfo {author} {\bibfnamefont {A.}~\bibnamefont
  {Arvanitaki}}\ and\ \bibinfo {author} {\bibfnamefont {S.}~\bibnamefont
  {Dubovsky}},\ }\href {\doibase 10.1103/PhysRevD.83.044026} {\bibfield
  {journal} {\bibinfo  {journal} {Phys. Rev. D}\ }\textbf {\bibinfo {volume}
  {83}},\ \bibinfo {pages} {044026} (\bibinfo {year} {2011})},\ \Eprint
  {http://arxiv.org/abs/1004.3558} {arXiv:1004.3558 [hep-th]} \BibitemShut
  {NoStop}%
\bibitem [{\citenamefont {Pani}\ \emph {et~al.}(2012)\citenamefont {Pani},
  \citenamefont {Cardoso}, \citenamefont {Gualtieri}, \citenamefont {Berti},\
  and\ \citenamefont {Ishibashi}}]{Pani:2012vp}%
  \BibitemOpen
  \bibfield  {author} {\bibinfo {author} {\bibfnamefont {P.}~\bibnamefont
  {Pani}}, \bibinfo {author} {\bibfnamefont {V.}~\bibnamefont {Cardoso}},
  \bibinfo {author} {\bibfnamefont {L.}~\bibnamefont {Gualtieri}}, \bibinfo
  {author} {\bibfnamefont {E.}~\bibnamefont {Berti}}, \ and\ \bibinfo {author}
  {\bibfnamefont {A.}~\bibnamefont {Ishibashi}},\ }\href {\doibase
  10.1103/PhysRevLett.109.131102} {\bibfield  {journal} {\bibinfo  {journal}
  {Phys. Rev. Lett.}\ }\textbf {\bibinfo {volume} {109}},\ \bibinfo {pages}
  {131102} (\bibinfo {year} {2012})},\ \Eprint {http://arxiv.org/abs/1209.0465}
  {arXiv:1209.0465 [gr-qc]} \BibitemShut {NoStop}%
\bibitem [{\citenamefont {Witek}\ \emph {et~al.}(2013)\citenamefont {Witek},
  \citenamefont {Cardoso}, \citenamefont {Ishibashi},\ and\ \citenamefont
  {Sperhake}}]{Witek:2012tr}%
  \BibitemOpen
  \bibfield  {author} {\bibinfo {author} {\bibfnamefont {H.}~\bibnamefont
  {Witek}}, \bibinfo {author} {\bibfnamefont {V.}~\bibnamefont {Cardoso}},
  \bibinfo {author} {\bibfnamefont {A.}~\bibnamefont {Ishibashi}}, \ and\
  \bibinfo {author} {\bibfnamefont {U.}~\bibnamefont {Sperhake}},\ }\href
  {\doibase 10.1103/PhysRevD.87.043513} {\bibfield  {journal} {\bibinfo
  {journal} {Phys. Rev. D}\ }\textbf {\bibinfo {volume} {87}},\ \bibinfo
  {pages} {043513} (\bibinfo {year} {2013})},\ \Eprint
  {http://arxiv.org/abs/1212.0551} {arXiv:1212.0551 [gr-qc]} \BibitemShut
  {NoStop}%
\bibitem [{\citenamefont {Brito}\ \emph {et~al.}(2013)\citenamefont {Brito},
  \citenamefont {Cardoso},\ and\ \citenamefont {Pani}}]{Brito:2013wya}%
  \BibitemOpen
  \bibfield  {author} {\bibinfo {author} {\bibfnamefont {R.}~\bibnamefont
  {Brito}}, \bibinfo {author} {\bibfnamefont {V.}~\bibnamefont {Cardoso}}, \
  and\ \bibinfo {author} {\bibfnamefont {P.}~\bibnamefont {Pani}},\ }\href
  {\doibase 10.1103/PhysRevD.88.023514} {\bibfield  {journal} {\bibinfo
  {journal} {Phys. Rev. D}\ }\textbf {\bibinfo {volume} {88}},\ \bibinfo
  {pages} {023514} (\bibinfo {year} {2013})},\ \Eprint
  {http://arxiv.org/abs/1304.6725} {arXiv:1304.6725 [gr-qc]} \BibitemShut
  {NoStop}%
\bibitem [{\citenamefont {Brito}\ \emph
  {et~al.}(2015{\natexlab{b}})\citenamefont {Brito}, \citenamefont {Cardoso},\
  and\ \citenamefont {Pani}}]{Brito:2014wla}%
  \BibitemOpen
  \bibfield  {author} {\bibinfo {author} {\bibfnamefont {R.}~\bibnamefont
  {Brito}}, \bibinfo {author} {\bibfnamefont {V.}~\bibnamefont {Cardoso}}, \
  and\ \bibinfo {author} {\bibfnamefont {P.}~\bibnamefont {Pani}},\ }\href
  {\doibase 10.1088/0264-9381/32/13/134001} {\bibfield  {journal} {\bibinfo
  {journal} {Class. Quant. Grav.}\ }\textbf {\bibinfo {volume} {32}},\ \bibinfo
  {pages} {134001} (\bibinfo {year} {2015}{\natexlab{b}})},\ \Eprint
  {http://arxiv.org/abs/1411.0686} {arXiv:1411.0686 [gr-qc]} \BibitemShut
  {NoStop}%
\bibitem [{\citenamefont {Berti}\ \emph {et~al.}(2015)\citenamefont {Berti}
  \emph {et~al.}}]{Berti:2015itd}%
  \BibitemOpen
  \bibfield  {author} {\bibinfo {author} {\bibfnamefont {E.}~\bibnamefont
  {Berti}} \emph {et~al.},\ }\href {\doibase 10.1088/0264-9381/32/24/243001}
  {\bibfield  {journal} {\bibinfo  {journal} {Class. Quant. Grav.}\ }\textbf
  {\bibinfo {volume} {32}},\ \bibinfo {pages} {243001} (\bibinfo {year}
  {2015})},\ \Eprint {http://arxiv.org/abs/1501.07274} {arXiv:1501.07274
  [gr-qc]} \BibitemShut {NoStop}%
\bibitem [{\citenamefont {Marsh}(2016)}]{Marsh:2015xka}%
  \BibitemOpen
  \bibfield  {author} {\bibinfo {author} {\bibfnamefont {D.~J.~E.}\
  \bibnamefont {Marsh}},\ }\href {\doibase 10.1016/j.physrep.2016.06.005}
  {\bibfield  {journal} {\bibinfo  {journal} {Phys. Rept.}\ }\textbf {\bibinfo
  {volume} {643}},\ \bibinfo {pages} {1} (\bibinfo {year} {2016})},\ \Eprint
  {http://arxiv.org/abs/1510.07633} {arXiv:1510.07633 [astro-ph.CO]}
  \BibitemShut {NoStop}%
\bibitem [{\citenamefont {East}\ and\ \citenamefont
  {Pretorius}(2017)}]{East:2017ovw}%
  \BibitemOpen
  \bibfield  {author} {\bibinfo {author} {\bibfnamefont {W.~E.}\ \bibnamefont
  {East}}\ and\ \bibinfo {author} {\bibfnamefont {F.}~\bibnamefont
  {Pretorius}},\ }\href {\doibase 10.1103/PhysRevLett.119.041101} {\bibfield
  {journal} {\bibinfo  {journal} {Phys. Rev. Lett.}\ }\textbf {\bibinfo
  {volume} {119}},\ \bibinfo {pages} {041101} (\bibinfo {year} {2017})},\
  \Eprint {http://arxiv.org/abs/1704.04791} {arXiv:1704.04791 [gr-qc]}
  \BibitemShut {NoStop}%
\bibitem [{\citenamefont {Baryakhtar}\ \emph {et~al.}(2017)\citenamefont
  {Baryakhtar}, \citenamefont {Lasenby},\ and\ \citenamefont
  {Teo}}]{Baryakhtar:2017ngi}%
  \BibitemOpen
  \bibfield  {author} {\bibinfo {author} {\bibfnamefont {M.}~\bibnamefont
  {Baryakhtar}}, \bibinfo {author} {\bibfnamefont {R.}~\bibnamefont {Lasenby}},
  \ and\ \bibinfo {author} {\bibfnamefont {M.}~\bibnamefont {Teo}},\ }\href
  {\doibase 10.1103/PhysRevD.96.035019} {\bibfield  {journal} {\bibinfo
  {journal} {Phys. Rev. D}\ }\textbf {\bibinfo {volume} {96}},\ \bibinfo
  {pages} {035019} (\bibinfo {year} {2017})},\ \Eprint
  {http://arxiv.org/abs/1704.05081} {arXiv:1704.05081 [hep-ph]} \BibitemShut
  {NoStop}%
\bibitem [{\citenamefont {Baumann}\ \emph {et~al.}(2019)\citenamefont
  {Baumann}, \citenamefont {Chia},\ and\ \citenamefont
  {Porto}}]{Baumann:2018vus}%
  \BibitemOpen
  \bibfield  {author} {\bibinfo {author} {\bibfnamefont {D.}~\bibnamefont
  {Baumann}}, \bibinfo {author} {\bibfnamefont {H.~S.}\ \bibnamefont {Chia}}, \
  and\ \bibinfo {author} {\bibfnamefont {R.~A.}\ \bibnamefont {Porto}},\ }\href
  {\doibase 10.1103/PhysRevD.99.044001} {\bibfield  {journal} {\bibinfo
  {journal} {Phys. Rev. D}\ }\textbf {\bibinfo {volume} {99}},\ \bibinfo
  {pages} {044001} (\bibinfo {year} {2019})},\ \Eprint
  {http://arxiv.org/abs/1804.03208} {arXiv:1804.03208 [gr-qc]} \BibitemShut
  {NoStop}%
\bibitem [{\citenamefont {Zhu}\ \emph {et~al.}(2020)\citenamefont {Zhu},
  \citenamefont {Baryakhtar}, \citenamefont {Papa}, \citenamefont {Tsuna},
  \citenamefont {Kawanaka},\ and\ \citenamefont {Eggenstein}}]{Zhu:2020tht}%
  \BibitemOpen
  \bibfield  {author} {\bibinfo {author} {\bibfnamefont {S.~J.}\ \bibnamefont
  {Zhu}}, \bibinfo {author} {\bibfnamefont {M.}~\bibnamefont {Baryakhtar}},
  \bibinfo {author} {\bibfnamefont {M.~A.}\ \bibnamefont {Papa}}, \bibinfo
  {author} {\bibfnamefont {D.}~\bibnamefont {Tsuna}}, \bibinfo {author}
  {\bibfnamefont {N.}~\bibnamefont {Kawanaka}}, \ and\ \bibinfo {author}
  {\bibfnamefont {H.-B.}\ \bibnamefont {Eggenstein}},\ }\href {\doibase
  10.1103/PhysRevD.102.063020} {\bibfield  {journal} {\bibinfo  {journal}
  {Phys. Rev. D}\ }\textbf {\bibinfo {volume} {102}},\ \bibinfo {pages}
  {063020} (\bibinfo {year} {2020})},\ \Eprint
  {http://arxiv.org/abs/2003.03359} {arXiv:2003.03359 [gr-qc]} \BibitemShut
  {NoStop}%
\bibitem [{\citenamefont {Zhang}\ \emph {et~al.}(2020)\citenamefont {Zhang},
  \citenamefont {Zhang}, \citenamefont {Li},\ and\ \citenamefont
  {Guo}}]{Zhang:2020sjh}%
  \BibitemOpen
  \bibfield  {author} {\bibinfo {author} {\bibfnamefont {C.-Y.}\ \bibnamefont
  {Zhang}}, \bibinfo {author} {\bibfnamefont {S.-J.}\ \bibnamefont {Zhang}},
  \bibinfo {author} {\bibfnamefont {P.-C.}\ \bibnamefont {Li}}, \ and\ \bibinfo
  {author} {\bibfnamefont {M.}~\bibnamefont {Guo}},\ }\href {\doibase
  10.1007/JHEP08(2020)105} {\bibfield  {journal} {\bibinfo  {journal} {JHEP}\
  }\textbf {\bibinfo {volume} {08}},\ \bibinfo {pages} {105} (\bibinfo {year}
  {2020})},\ \Eprint {http://arxiv.org/abs/2004.03141} {arXiv:2004.03141
  [gr-qc]} \BibitemShut {NoStop}%
\bibitem [{\citenamefont {Stott}(2020)}]{Stott:2020gjj}%
  \BibitemOpen
  \bibfield  {author} {\bibinfo {author} {\bibfnamefont {M.~J.}\ \bibnamefont
  {Stott}},\ }\href@noop {} {\  (\bibinfo {year} {2020})},\ \Eprint
  {http://arxiv.org/abs/2009.07206} {arXiv:2009.07206 [hep-ph]} \BibitemShut
  {NoStop}%
\bibitem [{\citenamefont {Baryakhtar}\ \emph {et~al.}(2021)\citenamefont
  {Baryakhtar}, \citenamefont {Galanis}, \citenamefont {Lasenby},\ and\
  \citenamefont {Simon}}]{Baryakhtar:2020gao}%
  \BibitemOpen
  \bibfield  {author} {\bibinfo {author} {\bibfnamefont {M.}~\bibnamefont
  {Baryakhtar}}, \bibinfo {author} {\bibfnamefont {M.}~\bibnamefont {Galanis}},
  \bibinfo {author} {\bibfnamefont {R.}~\bibnamefont {Lasenby}}, \ and\
  \bibinfo {author} {\bibfnamefont {O.}~\bibnamefont {Simon}},\ }\href
  {\doibase 10.1103/PhysRevD.103.095019} {\bibfield  {journal} {\bibinfo
  {journal} {Phys. Rev. D}\ }\textbf {\bibinfo {volume} {103}},\ \bibinfo
  {pages} {095019} (\bibinfo {year} {2021})},\ \Eprint
  {http://arxiv.org/abs/2011.11646} {arXiv:2011.11646 [hep-ph]} \BibitemShut
  {NoStop}%
\bibitem [{\citenamefont {Mehta}\ \emph {et~al.}(2021)\citenamefont {Mehta},
  \citenamefont {Demirtas}, \citenamefont {Long}, \citenamefont {Marsh},
  \citenamefont {McAllister},\ and\ \citenamefont {Stott}}]{Mehta:2021pwf}%
  \BibitemOpen
  \bibfield  {author} {\bibinfo {author} {\bibfnamefont {V.~M.}\ \bibnamefont
  {Mehta}}, \bibinfo {author} {\bibfnamefont {M.}~\bibnamefont {Demirtas}},
  \bibinfo {author} {\bibfnamefont {C.}~\bibnamefont {Long}}, \bibinfo {author}
  {\bibfnamefont {D.~J.~E.}\ \bibnamefont {Marsh}}, \bibinfo {author}
  {\bibfnamefont {L.}~\bibnamefont {McAllister}}, \ and\ \bibinfo {author}
  {\bibfnamefont {M.~J.}\ \bibnamefont {Stott}},\ }\href {\doibase
  10.1088/1475-7516/2021/07/033} {\bibfield  {journal} {\bibinfo  {journal}
  {JCAP}\ }\textbf {\bibinfo {volume} {07}},\ \bibinfo {pages} {033} (\bibinfo
  {year} {2021})},\ \Eprint {http://arxiv.org/abs/2103.06812} {arXiv:2103.06812
  [hep-th]} \BibitemShut {NoStop}%
\bibitem [{\citenamefont {Roy}\ \emph {et~al.}(2022)\citenamefont {Roy},
  \citenamefont {Vagnozzi},\ and\ \citenamefont {Visinelli}}]{Roy:2021uye}%
  \BibitemOpen
  \bibfield  {author} {\bibinfo {author} {\bibfnamefont {R.}~\bibnamefont
  {Roy}}, \bibinfo {author} {\bibfnamefont {S.}~\bibnamefont {Vagnozzi}}, \
  and\ \bibinfo {author} {\bibfnamefont {L.}~\bibnamefont {Visinelli}},\ }\href
  {\doibase 10.1103/PhysRevD.105.083002} {\bibfield  {journal} {\bibinfo
  {journal} {Phys. Rev. D}\ }\textbf {\bibinfo {volume} {105}},\ \bibinfo
  {pages} {083002} (\bibinfo {year} {2022})},\ \Eprint
  {http://arxiv.org/abs/2112.06932} {arXiv:2112.06932 [astro-ph.HE]}
  \BibitemShut {NoStop}%
\bibitem [{\citenamefont {Chen}\ \emph {et~al.}(2022)\citenamefont {Chen},
  \citenamefont {Roy}, \citenamefont {Vagnozzi},\ and\ \citenamefont
  {Visinelli}}]{Chen:2022nbb}%
  \BibitemOpen
  \bibfield  {author} {\bibinfo {author} {\bibfnamefont {Y.}~\bibnamefont
  {Chen}}, \bibinfo {author} {\bibfnamefont {R.}~\bibnamefont {Roy}}, \bibinfo
  {author} {\bibfnamefont {S.}~\bibnamefont {Vagnozzi}}, \ and\ \bibinfo
  {author} {\bibfnamefont {L.}~\bibnamefont {Visinelli}},\ }\href {\doibase
  10.1103/PhysRevD.106.043021} {\bibfield  {journal} {\bibinfo  {journal}
  {Phys. Rev. D}\ }\textbf {\bibinfo {volume} {106}},\ \bibinfo {pages}
  {043021} (\bibinfo {year} {2022})},\ \Eprint
  {http://arxiv.org/abs/2205.06238} {arXiv:2205.06238 [astro-ph.HE]}
  \BibitemShut {NoStop}%
\bibitem [{\citenamefont {Siemonsen}\ \emph {et~al.}(2023)\citenamefont
  {Siemonsen}, \citenamefont {May},\ and\ \citenamefont
  {East}}]{Siemonsen:2022yyf}%
  \BibitemOpen
  \bibfield  {author} {\bibinfo {author} {\bibfnamefont {N.}~\bibnamefont
  {Siemonsen}}, \bibinfo {author} {\bibfnamefont {T.}~\bibnamefont {May}}, \
  and\ \bibinfo {author} {\bibfnamefont {W.~E.}\ \bibnamefont {East}},\ }\href
  {\doibase 10.1103/PhysRevD.107.104003} {\bibfield  {journal} {\bibinfo
  {journal} {Phys. Rev. D}\ }\textbf {\bibinfo {volume} {107}},\ \bibinfo
  {pages} {104003} (\bibinfo {year} {2023})},\ \Eprint
  {http://arxiv.org/abs/2211.03845} {arXiv:2211.03845 [gr-qc]} \BibitemShut
  {NoStop}%
\bibitem [{\citenamefont {Zel'dovich}({\natexlab{a}})}]{Zeld1}%
  \BibitemOpen
  \bibfield  {author} {\bibinfo {author} {\bibfnamefont {Y.~B.}\ \bibnamefont
  {Zel'dovich}},\ }\href@noop {} {\bibfield  {journal} {\bibinfo  {journal}
  {Zh. Eksp. Teor. Fiz. Pis'ma {\bf 14}, 270 (1971) [JETP Letters {\bf 14}, 180
  (1971)]}\ } ({\natexlab{a}})}\BibitemShut {NoStop}%
\bibitem [{\citenamefont {Zel'dovich}({\natexlab{b}})}]{Zeld2}%
  \BibitemOpen
  \bibfield  {author} {\bibinfo {author} {\bibfnamefont {Y.~B.}\ \bibnamefont
  {Zel'dovich}},\ }\href@noop {} {\bibfield  {journal} {\bibinfo  {journal}
  {Zh. Eksp. Teor. Fiz. {\bf 62}, 2076 (1971) [JETP {\bf 35}, 1085 (1971)]}\ }
  ({\natexlab{b}})}\BibitemShut {NoStop}%
\bibitem [{\citenamefont {Cardoso}\ \emph {et~al.}(2023)\citenamefont
  {Cardoso}, \citenamefont {Vicente},\ and\ \citenamefont
  {Zhong}}]{Cardoso:2023dtm}%
  \BibitemOpen
  \bibfield  {author} {\bibinfo {author} {\bibfnamefont {V.}~\bibnamefont
  {Cardoso}}, \bibinfo {author} {\bibfnamefont {R.}~\bibnamefont {Vicente}}, \
  and\ \bibinfo {author} {\bibfnamefont {Z.}~\bibnamefont {Zhong}},\ }\href
  {\doibase 10.1103/PhysRevLett.131.111602} {\bibfield  {journal} {\bibinfo
  {journal} {Phys. Rev. Lett.}\ }\textbf {\bibinfo {volume} {131}},\ \bibinfo
  {pages} {111602} (\bibinfo {year} {2023})},\ \Eprint
  {http://arxiv.org/abs/2307.13734} {arXiv:2307.13734 [hep-th]} \BibitemShut
  {NoStop}%
\bibitem [{\citenamefont {Zhang}\ \emph {et~al.}(2024)\citenamefont {Zhang},
  \citenamefont {Chang}, \citenamefont {Saffin}, \citenamefont {Xie},\ and\
  \citenamefont {Zhou}}]{Zhang:2024ufh}%
  \BibitemOpen
  \bibfield  {author} {\bibinfo {author} {\bibfnamefont {G.-D.}\ \bibnamefont
  {Zhang}}, \bibinfo {author} {\bibfnamefont {F.-M.}\ \bibnamefont {Chang}},
  \bibinfo {author} {\bibfnamefont {P.~M.}\ \bibnamefont {Saffin}}, \bibinfo
  {author} {\bibfnamefont {Q.-X.}\ \bibnamefont {Xie}}, \ and\ \bibinfo
  {author} {\bibfnamefont {S.-Y.}\ \bibnamefont {Zhou}},\ }\href {\doibase
  10.1103/PhysRevD.110.043504} {\bibfield  {journal} {\bibinfo  {journal}
  {Phys. Rev. D}\ }\textbf {\bibinfo {volume} {110}},\ \bibinfo {pages}
  {043504} (\bibinfo {year} {2024})},\ \Eprint
  {http://arxiv.org/abs/2402.03193} {arXiv:2402.03193 [hep-th]} \BibitemShut
  {NoStop}%
\bibitem [{\citenamefont {Gao}\ \emph {et~al.}(2024)\citenamefont {Gao},
  \citenamefont {Saffin}, \citenamefont {Wang}, \citenamefont {Xie},\ and\
  \citenamefont {Zhou}}]{Gao:2023gof}%
  \BibitemOpen
  \bibfield  {author} {\bibinfo {author} {\bibfnamefont {H.-Y.}\ \bibnamefont
  {Gao}}, \bibinfo {author} {\bibfnamefont {P.~M.}\ \bibnamefont {Saffin}},
  \bibinfo {author} {\bibfnamefont {Y.-J.}\ \bibnamefont {Wang}}, \bibinfo
  {author} {\bibfnamefont {Q.-X.}\ \bibnamefont {Xie}}, \ and\ \bibinfo
  {author} {\bibfnamefont {S.-Y.}\ \bibnamefont {Zhou}},\ }\href {\doibase
  10.1007/s11433-023-2357-4} {\bibfield  {journal} {\bibinfo  {journal} {Sci.
  China Phys. Mech. Astron.}\ }\textbf {\bibinfo {volume} {67}},\ \bibinfo
  {pages} {260413} (\bibinfo {year} {2024})},\ \Eprint
  {http://arxiv.org/abs/2306.01868} {arXiv:2306.01868 [gr-qc]} \BibitemShut
  {NoStop}%
\bibitem [{\citenamefont {Chang}\ \emph {et~al.}(2025)\citenamefont {Chang},
  \citenamefont {Gao}, \citenamefont {Jaramillo}, \citenamefont {Meng},\ and\
  \citenamefont {Zhou}}]{Chang:2024xjp}%
  \BibitemOpen
  \bibfield  {author} {\bibinfo {author} {\bibfnamefont {F.-M.}\ \bibnamefont
  {Chang}}, \bibinfo {author} {\bibfnamefont {H.-Y.}\ \bibnamefont {Gao}},
  \bibinfo {author} {\bibfnamefont {V.}~\bibnamefont {Jaramillo}}, \bibinfo
  {author} {\bibfnamefont {X.}~\bibnamefont {Meng}}, \ and\ \bibinfo {author}
  {\bibfnamefont {S.-Y.}\ \bibnamefont {Zhou}},\ }\href {\doibase
  10.1103/PhysRevD.111.044053} {\bibfield  {journal} {\bibinfo  {journal}
  {Phys. Rev. D}\ }\textbf {\bibinfo {volume} {111}},\ \bibinfo {pages}
  {044053} (\bibinfo {year} {2025})},\ \Eprint
  {http://arxiv.org/abs/2412.01894} {arXiv:2412.01894 [gr-qc]} \BibitemShut
  {NoStop}%
\bibitem [{\citenamefont {Lee}\ and\ \citenamefont {Pang}(1992)}]{Lee:1991ax}%
  \BibitemOpen
  \bibfield  {author} {\bibinfo {author} {\bibfnamefont {T.~D.}\ \bibnamefont
  {Lee}}\ and\ \bibinfo {author} {\bibfnamefont {Y.}~\bibnamefont {Pang}},\
  }\href {\doibase 10.1016/0370-1573(92)90064-7} {\bibfield  {journal}
  {\bibinfo  {journal} {Phys. Rept.}\ }\textbf {\bibinfo {volume} {221}},\
  \bibinfo {pages} {251} (\bibinfo {year} {1992})}\BibitemShut {NoStop}%
\bibitem [{\citenamefont {Lee}(1978)}]{Lee:1978yu}%
  \BibitemOpen
  \bibfield  {author} {\bibinfo {author} {\bibfnamefont {T.~D.}\ \bibnamefont
  {Lee}},\ }\href@noop {} {\bibfield  {journal} {\bibinfo  {journal} {Comments
  Nucl. Part. Phys.}\ }\textbf {\bibinfo {volume} {7}},\ \bibinfo {pages} {165}
  (\bibinfo {year} {1978})}\BibitemShut {NoStop}%
\bibitem [{\citenamefont {Friedberg}\ and\ \citenamefont
  {Lee}(1978)}]{Friedberg:1978sc}%
  \BibitemOpen
  \bibfield  {author} {\bibinfo {author} {\bibfnamefont {R.}~\bibnamefont
  {Friedberg}}\ and\ \bibinfo {author} {\bibfnamefont {T.~D.}\ \bibnamefont
  {Lee}},\ }\href {\doibase 10.1103/PhysRevD.18.2623} {\bibfield  {journal}
  {\bibinfo  {journal} {Phys. Rev. D}\ }\textbf {\bibinfo {volume} {18}},\
  \bibinfo {pages} {2623} (\bibinfo {year} {1978})}\BibitemShut {NoStop}%
\bibitem [{\citenamefont {Goldflam}\ and\ \citenamefont
  {Wilets}(1982)}]{Goldflam:1981tg}%
  \BibitemOpen
  \bibfield  {author} {\bibinfo {author} {\bibfnamefont {R.}~\bibnamefont
  {Goldflam}}\ and\ \bibinfo {author} {\bibfnamefont {L.}~\bibnamefont
  {Wilets}},\ }\href {\doibase 10.1103/PhysRevD.25.1951} {\bibfield  {journal}
  {\bibinfo  {journal} {Phys. Rev. D}\ }\textbf {\bibinfo {volume} {25}},\
  \bibinfo {pages} {1951} (\bibinfo {year} {1982})}\BibitemShut {NoStop}%
\bibitem [{\citenamefont {Cahill}\ and\ \citenamefont
  {Roberts}(1985)}]{Cahill:1985mh}%
  \BibitemOpen
  \bibfield  {author} {\bibinfo {author} {\bibfnamefont {R.~T.}\ \bibnamefont
  {Cahill}}\ and\ \bibinfo {author} {\bibfnamefont {C.~D.}\ \bibnamefont
  {Roberts}},\ }\href {\doibase 10.1103/PhysRevD.32.2419} {\bibfield  {journal}
  {\bibinfo  {journal} {Phys. Rev. D}\ }\textbf {\bibinfo {volume} {32}},\
  \bibinfo {pages} {2419} (\bibinfo {year} {1985})}\BibitemShut {NoStop}%
\bibitem [{\citenamefont {Loiko}\ \emph {et~al.}(2018)\citenamefont {Loiko},
  \citenamefont {Perapechka},\ and\ \citenamefont {Shnir}}]{Loiko:2018mhb}%
  \BibitemOpen
  \bibfield  {author} {\bibinfo {author} {\bibfnamefont {V.}~\bibnamefont
  {Loiko}}, \bibinfo {author} {\bibfnamefont {I.}~\bibnamefont {Perapechka}}, \
  and\ \bibinfo {author} {\bibfnamefont {Y.}~\bibnamefont {Shnir}},\ }\href
  {\doibase 10.1103/PhysRevD.98.045018} {\bibfield  {journal} {\bibinfo
  {journal} {Phys. Rev. D}\ }\textbf {\bibinfo {volume} {98}},\ \bibinfo
  {pages} {045018} (\bibinfo {year} {2018})},\ \Eprint
  {http://arxiv.org/abs/1805.11929} {arXiv:1805.11929 [hep-th]} \BibitemShut
  {NoStop}%
\bibitem [{\citenamefont {Zhong}\ and\ \citenamefont
  {Cheng}(2019)}]{Zhong:2018hwm}%
  \BibitemOpen
  \bibfield  {author} {\bibinfo {author} {\bibfnamefont {Y.}~\bibnamefont
  {Zhong}}\ and\ \bibinfo {author} {\bibfnamefont {H.}~\bibnamefont {Cheng}},\
  }\href {\doibase 10.1007/s10773-019-04117-4} {\bibfield  {journal} {\bibinfo
  {journal} {Int. J. Theor. Phys.}\ }\textbf {\bibinfo {volume} {58}},\
  \bibinfo {pages} {2251} (\bibinfo {year} {2019})},\ \Eprint
  {http://arxiv.org/abs/1807.03695} {arXiv:1807.03695 [hep-th]} \BibitemShut
  {NoStop}%
\bibitem [{\citenamefont {Loiko}\ and\ \citenamefont
  {Shnir}(2022)}]{Loiko:2022noq}%
  \BibitemOpen
  \bibfield  {author} {\bibinfo {author} {\bibfnamefont {V.}~\bibnamefont
  {Loiko}}\ and\ \bibinfo {author} {\bibfnamefont {Y.}~\bibnamefont {Shnir}},\
  }\href {\doibase 10.1103/PhysRevD.106.045021} {\bibfield  {journal} {\bibinfo
   {journal} {Phys. Rev. D}\ }\textbf {\bibinfo {volume} {106}},\ \bibinfo
  {pages} {045021} (\bibinfo {year} {2022})},\ \Eprint
  {http://arxiv.org/abs/2207.02646} {arXiv:2207.02646 [hep-th]} \BibitemShut
  {NoStop}%
\bibitem [{\citenamefont {Heeck}\ and\ \citenamefont
  {Sokhashvili}(2023)}]{Heeck:2023idx}%
  \BibitemOpen
  \bibfield  {author} {\bibinfo {author} {\bibfnamefont {J.}~\bibnamefont
  {Heeck}}\ and\ \bibinfo {author} {\bibfnamefont {M.}~\bibnamefont
  {Sokhashvili}},\ }\href {\doibase 10.1140/epjc/s10052-023-11710-9} {\bibfield
   {journal} {\bibinfo  {journal} {Eur. Phys. J. C}\ }\textbf {\bibinfo
  {volume} {83}},\ \bibinfo {pages} {526} (\bibinfo {year} {2023})},\ \Eprint
  {http://arxiv.org/abs/2303.09566} {arXiv:2303.09566 [hep-ph]} \BibitemShut
  {NoStop}%
\bibitem [{\citenamefont {Kim}\ and\ \citenamefont
  {Nugaev}(2024)}]{Kim:2023zvf}%
  \BibitemOpen
  \bibfield  {author} {\bibinfo {author} {\bibfnamefont {E.}~\bibnamefont
  {Kim}}\ and\ \bibinfo {author} {\bibfnamefont {E.}~\bibnamefont {Nugaev}},\
  }\href {\doibase 10.1140/epjc/s10052-024-13167-w} {\bibfield  {journal}
  {\bibinfo  {journal} {Eur. Phys. J. C}\ }\textbf {\bibinfo {volume} {84}},\
  \bibinfo {pages} {797} (\bibinfo {year} {2024})},\ \Eprint
  {http://arxiv.org/abs/2309.09661} {arXiv:2309.09661 [hep-ph]} \BibitemShut
  {NoStop}%
\bibitem [{\citenamefont {Kim}\ \emph {et~al.}(2024)\citenamefont {Kim},
  \citenamefont {Nugaev},\ and\ \citenamefont {Shnir}}]{Kim:2024vam}%
  \BibitemOpen
  \bibfield  {author} {\bibinfo {author} {\bibfnamefont {E.}~\bibnamefont
  {Kim}}, \bibinfo {author} {\bibfnamefont {E.}~\bibnamefont {Nugaev}}, \ and\
  \bibinfo {author} {\bibfnamefont {Y.}~\bibnamefont {Shnir}},\ }\href
  {\doibase 10.1016/j.physletb.2024.138881} {\bibfield  {journal} {\bibinfo
  {journal} {Phys. Lett. B}\ }\textbf {\bibinfo {volume} {856}},\ \bibinfo
  {pages} {138881} (\bibinfo {year} {2024})},\ \Eprint
  {http://arxiv.org/abs/2405.09262} {arXiv:2405.09262 [hep-ph]} \BibitemShut
  {NoStop}%
\bibitem [{\citenamefont {Hamada}\ \emph {et~al.}(2024)\citenamefont {Hamada},
  \citenamefont {Kawana}, \citenamefont {Kim},\ and\ \citenamefont
  {Lu}}]{Hamada:2024pbs}%
  \BibitemOpen
  \bibfield  {author} {\bibinfo {author} {\bibfnamefont {Y.}~\bibnamefont
  {Hamada}}, \bibinfo {author} {\bibfnamefont {K.}~\bibnamefont {Kawana}},
  \bibinfo {author} {\bibfnamefont {T.}~\bibnamefont {Kim}}, \ and\ \bibinfo
  {author} {\bibfnamefont {P.}~\bibnamefont {Lu}},\ }\href {\doibase
  10.1007/JHEP08(2024)242} {\bibfield  {journal} {\bibinfo  {journal} {JHEP}\
  }\textbf {\bibinfo {volume} {08}},\ \bibinfo {pages} {242} (\bibinfo {year}
  {2024})},\ \Eprint {http://arxiv.org/abs/2407.11115} {arXiv:2407.11115
  [hep-ph]} \BibitemShut {NoStop}%
\bibitem [{\citenamefont {Kunz}\ \emph {et~al.}(2019)\citenamefont {Kunz},
  \citenamefont {Perapechka},\ and\ \citenamefont {Shnir}}]{Kunz:2019sgn}%
  \BibitemOpen
  \bibfield  {author} {\bibinfo {author} {\bibfnamefont {J.}~\bibnamefont
  {Kunz}}, \bibinfo {author} {\bibfnamefont {I.}~\bibnamefont {Perapechka}}, \
  and\ \bibinfo {author} {\bibfnamefont {Y.}~\bibnamefont {Shnir}},\ }\href
  {\doibase 10.1007/JHEP07(2019)109} {\bibfield  {journal} {\bibinfo  {journal}
  {JHEP}\ }\textbf {\bibinfo {volume} {07}},\ \bibinfo {pages} {109} (\bibinfo
  {year} {2019})},\ \Eprint {http://arxiv.org/abs/1904.13379} {arXiv:1904.13379
  [gr-qc]} \BibitemShut {NoStop}%
\bibitem [{\citenamefont {Kunz}\ \emph {et~al.}(2022)\citenamefont {Kunz},
  \citenamefont {Loiko},\ and\ \citenamefont {Shnir}}]{Kunz:2021mbm}%
  \BibitemOpen
  \bibfield  {author} {\bibinfo {author} {\bibfnamefont {J.}~\bibnamefont
  {Kunz}}, \bibinfo {author} {\bibfnamefont {V.}~\bibnamefont {Loiko}}, \ and\
  \bibinfo {author} {\bibfnamefont {Y.}~\bibnamefont {Shnir}},\ }\href
  {\doibase 10.1103/PhysRevD.105.085013} {\bibfield  {journal} {\bibinfo
  {journal} {Phys. Rev. D}\ }\textbf {\bibinfo {volume} {105}},\ \bibinfo
  {pages} {085013} (\bibinfo {year} {2022})},\ \Eprint
  {http://arxiv.org/abs/2112.06626} {arXiv:2112.06626 [gr-qc]} \BibitemShut
  {NoStop}%
\bibitem [{\citenamefont {Herdeiro}\ \emph {et~al.}(2023)\citenamefont
  {Herdeiro}, \citenamefont {Radu},\ and\ \citenamefont {dos Santos
  Costa~Filho}}]{Herdeiro:2023lze}%
  \BibitemOpen
  \bibfield  {author} {\bibinfo {author} {\bibfnamefont {C.}~\bibnamefont
  {Herdeiro}}, \bibinfo {author} {\bibfnamefont {E.}~\bibnamefont {Radu}}, \
  and\ \bibinfo {author} {\bibfnamefont {E.}~\bibnamefont {dos Santos
  Costa~Filho}},\ }\href {\doibase 10.1088/1475-7516/2023/05/022} {\bibfield
  {journal} {\bibinfo  {journal} {JCAP}\ }\textbf {\bibinfo {volume} {05}},\
  \bibinfo {pages} {022} (\bibinfo {year} {2023})},\ \Eprint
  {http://arxiv.org/abs/2301.04172} {arXiv:2301.04172 [gr-qc]} \BibitemShut
  {NoStop}%
\bibitem [{\citenamefont {Kunz}\ and\ \citenamefont
  {Shnir}(2023)}]{Kunz:2023qfg}%
  \BibitemOpen
  \bibfield  {author} {\bibinfo {author} {\bibfnamefont {J.}~\bibnamefont
  {Kunz}}\ and\ \bibinfo {author} {\bibfnamefont {Y.}~\bibnamefont {Shnir}},\
  }\href {\doibase 10.1103/PhysRevD.107.104062} {\bibfield  {journal} {\bibinfo
   {journal} {Phys. Rev. D}\ }\textbf {\bibinfo {volume} {107}},\ \bibinfo
  {pages} {104062} (\bibinfo {year} {2023})},\ \Eprint
  {http://arxiv.org/abs/2303.16562} {arXiv:2303.16562 [hep-th]} \BibitemShut
  {NoStop}%
\bibitem [{\citenamefont {deSa}\ \emph {et~al.}(2024)\citenamefont {deSa},
  \citenamefont {Lima}, \citenamefont {Herdeiro},\ and\ \citenamefont
  {Crispino}}]{deSa:2024dhj}%
  \BibitemOpen
  \bibfield  {author} {\bibinfo {author} {\bibfnamefont {P.~L.~B.}\
  \bibnamefont {deSa}}, \bibinfo {author} {\bibfnamefont {H.~C.~D.}\
  \bibnamefont {Lima}, \bibfnamefont {Jr.}}, \bibinfo {author} {\bibfnamefont
  {C.~A.~R.}\ \bibnamefont {Herdeiro}}, \ and\ \bibinfo {author} {\bibfnamefont
  {L.~C.~B.}\ \bibnamefont {Crispino}},\ }\href {\doibase
  10.1103/PhysRevD.110.104047} {\bibfield  {journal} {\bibinfo  {journal}
  {Phys. Rev. D}\ }\textbf {\bibinfo {volume} {110}},\ \bibinfo {pages}
  {104047} (\bibinfo {year} {2024})},\ \Eprint
  {http://arxiv.org/abs/2406.02695} {arXiv:2406.02695 [gr-qc]} \BibitemShut
  {NoStop}%
\bibitem [{\citenamefont {Kunz}\ \emph {et~al.}(2024)\citenamefont {Kunz},
  \citenamefont {Loiko},\ and\ \citenamefont {Shnir}}]{Kunz:2024uux}%
  \BibitemOpen
  \bibfield  {author} {\bibinfo {author} {\bibfnamefont {J.}~\bibnamefont
  {Kunz}}, \bibinfo {author} {\bibfnamefont {V.}~\bibnamefont {Loiko}}, \ and\
  \bibinfo {author} {\bibfnamefont {Y.}~\bibnamefont {Shnir}},\ }\href
  {\doibase 10.1103/PhysRevD.110.125020} {\bibfield  {journal} {\bibinfo
  {journal} {Phys. Rev. D}\ }\textbf {\bibinfo {volume} {110}},\ \bibinfo
  {pages} {125020} (\bibinfo {year} {2024})},\ \Eprint
  {http://arxiv.org/abs/2407.21463} {arXiv:2407.21463 [hep-th]} \BibitemShut
  {NoStop}%
\bibitem [{\citenamefont {Jaramillo}\ and\ \citenamefont
  {Zhou}(2025)}]{Jaramillo:2024cus}%
  \BibitemOpen
  \bibfield  {author} {\bibinfo {author} {\bibfnamefont {V.}~\bibnamefont
  {Jaramillo}}\ and\ \bibinfo {author} {\bibfnamefont {S.-Y.}\ \bibnamefont
  {Zhou}},\ }\href {\doibase 10.1103/PhysRevD.111.024027} {\bibfield  {journal}
  {\bibinfo  {journal} {Phys. Rev. D}\ }\textbf {\bibinfo {volume} {111}},\
  \bibinfo {pages} {024027} (\bibinfo {year} {2025})},\ \Eprint
  {http://arxiv.org/abs/2411.08985} {arXiv:2411.08985 [gr-qc]} \BibitemShut
  {NoStop}%
\bibitem [{\citenamefont {Azatov}\ \emph {et~al.}(2024)\citenamefont {Azatov},
  \citenamefont {Ho},\ and\ \citenamefont {Khalil}}]{Azatov:2024npx}%
  \BibitemOpen
  \bibfield  {author} {\bibinfo {author} {\bibfnamefont {A.}~\bibnamefont
  {Azatov}}, \bibinfo {author} {\bibfnamefont {Q.~T.}\ \bibnamefont {Ho}}, \
  and\ \bibinfo {author} {\bibfnamefont {M.~M.}\ \bibnamefont {Khalil}},\
  }\href@noop {} {\  (\bibinfo {year} {2024})},\ \Eprint
  {http://arxiv.org/abs/2412.13885} {arXiv:2412.13885 [hep-ph]} \BibitemShut
  {NoStop}%
\bibitem [{\citenamefont {Press}\ \emph {et~al.}(1992)\citenamefont {Press},
  \citenamefont {Teukolsky}, \citenamefont {Vetterling},\ and\ \citenamefont
  {Flannery}}]{NRC}%
  \BibitemOpen
  \bibfield  {author} {\bibinfo {author} {\bibfnamefont {W.~H.}\ \bibnamefont
  {Press}}, \bibinfo {author} {\bibfnamefont {S.~A.}\ \bibnamefont
  {Teukolsky}}, \bibinfo {author} {\bibfnamefont {W.~T.}\ \bibnamefont
  {Vetterling}}, \ and\ \bibinfo {author} {\bibfnamefont {B.~P.}\ \bibnamefont
  {Flannery}},\ }\href@noop {} {\emph {\bibinfo {title} {Numerical Recipes in
  C: The Art of Scientific Computing Second Edition}}}\ (\bibinfo  {publisher}
  {Cambridge University Press},\ \bibinfo {year} {1992})\BibitemShut {NoStop}%
\end{thebibliography}%

\end{document}